\documentclass[a4paper, 11pt]{article}
\pdfoutput=1 

\usepackage{jheppub} 

\usepackage[T1]{fontenc} 
\usepackage{blkarray}
\usepackage{relsize}

\usepackage{filecontents}

\usepackage[bottom]{footmisc}
\usepackage[utf8]{inputenc}
\usepackage{amssymb, amsthm, tabto, epigraph, mathtools, dsfont, verbatim, slashed, mathrsfs}
\usepackage{graphicx, wrapfig}
\usepackage{enumitem}
\usepackage{relsize}
\usepackage{subcaption}
\usepackage{csquotes}
\usepackage{hyperref}
\usepackage{physics}
\usepackage{tabularx}
\usepackage[dvipsnames]{xcolor}
\usepackage{color}

\usepackage{amsmath}
\usepackage{empheq}
 
\setlist{nolistsep}
\hypersetup{
    colorlinks=true,
    linkcolor=Mahogany,
    filecolor=magenta,      
    urlcolor=blue,
}

\newlength\dlf  

\allowdisplaybreaks

\setlist{nolistsep}
\hypersetup{
    colorlinks=true,
    linkcolor=Maroon,
    filecolor=Maroon,      
    urlcolor=Maroon,
    citecolor=Maroon
}

\theoremstyle{remark}

\theoremstyle{definition}
\newtheorem{definition}{Definition}[section]

\usepackage{amssymb}
\usepackage{tikz}
\usetikzlibrary{shapes,arrows,cd,chains,decorations.markings,decorations.pathmorphing,calc}
\tikzset{
->-/.style args={#1rotate#2}{decoration={markings, mark=at position #1 with {\arrow[scale=1.5,rotate = #2 ]{stealth}}}, postaction={decorate}}
}
\tikzset{curve/.style={settings={#1},to path={(\tikztostart)
    .. controls ($(\tikztostart)!\pv{pos}!(\tikztotarget)!\pv{height}!270:(\tikztotarget)$)
    and ($(\tikztostart)!1-\pv{pos}!(\tikztotarget)!\pv{height}!270:(\tikztotarget)$)
    .. (\tikztotarget)\tikztonodes}},
    settings/.code={\tikzset{quiver/.cd,#1}
        \def\pv##1{\pgfkeysvalueof{/tikz/quiver/##1}}},
    quiver/.cd,pos/.initial=0.35,height/.initial=0}
\tikzset{tail reversed/.code={\pgfsetarrowsstart{tikzcd to}}}
\tikzset{2tail/.code={\pgfsetarrowsstart{Implies[reversed]}}}
\tikzset{2tail reversed/.code={\pgfsetarrowsstart{Implies}}}

\newmuskip\pFqskip
\mathchardef\pFcomma=\mathcode`, 

\graphicspath{{./images/}}

\renewcommand{\bar}{\overline}

\usepackage{braket}

\usepackage[export]{adjustbox}

\title{\center{Topological Cosets via Anyon Condensation and Applications to Gapped $\mathrm{\bf{QCD_{2}}}$}}

\abstract{The coset construction of two-dimensional conformal field theory (2D CFT) defines a 2D CFT by taking the quotient of two previously known chiral algebras. In this work, we use the methods of non-abelian (non-invertible) anyon condensation to describe 2D topological cosets, defined by the special case where the quotient of chiral algebras is a conformal embedding. In this case, the coset has zero central charge, and the coset theory is thus purely topological. Using 
non-abelian anyon condensation we describe in general the spectrum of line and local operators as well as their fusion, operator product expansion, and the action of the lines on local operators. An important application of our results is to QCD$_{2}$ with massless fermions in any representation that leads to a gapped phase, where topological cosets (conjecturally) describe the infrared fixed point.  We discuss several such examples in detail. For instance, we find that the $Spin(8)_{1}/SU(3)_{3}$ and $Spin(16)_{1}/Spin(9)_{2}$ topological cosets appearing at the infrared fixed point of appropriate QCD$_{2}$ theories are described by $\mathbb{Z}_{2} \times \mathbb{Z}_{2}$ triality and $\mathbb{Z}_{2} \times \mathrm{Rep(S_{3})}$ fusion categories respectively. Additionally, using this setup, we argue that chiral $Spin(8)$ QCD$_{2}$ with massless chiral fermions in the vectorial and spinorial representations is not only gapped, but moreover trivially gapped, with a unique ground state.}

\author[*]{Clay C\'ordova,}
\author[*]{and Diego Garc\'ia-Sep\'ulveda}

\affiliation[*]{Kadanoff Center for Theoretical Physics \& Enrico Fermi Institute, University of Chicago}

\emailAdd{clayc@uchicago.edu}
\emailAdd{dgarciasepulveda@uchicago.edu}

\begin{document}

\maketitle

\section{Introduction}

For the past half-century, the Standard Model of particle physics has been our best description of the dynamics of elementary particles.  However, despite its tremendous success, we still lack a clear analytical understanding of the low-energy regime of the QCD sector, where strong interactions take over, and we observe a gapped energy spectrum with no long-range topological degrees of freedom.
This enduring mystery is one of the main motivations for developing new general tools to understand strongly coupled dynamics in quantum field theory. Directly addressing these questions is a formidable long-term challenge.  Nevertheless, analogs of QCD in lower dimensions are under better technical control and provide a valuable window into the strongly interacting regime of quantum field theory.  

In this context, in recent years there has been significant progress in our understanding of the strongly coupled regime of 2D QCD \cite{Cherman:2019hbq, Komargodski:2020mxz, Dempsey:2021xpf, Anand:2021qnd, Delmastro:2021otj, Popov:2022vud, Dempsey:2022uie, Delmastro:2022prj, Dempsey:2023fvm, Ambrosino:2023dik, Dempsey:2024ofo, Ambrosino:2024prz, Cherman:2024onj, Damia:2024kyt,Dempsey:2024alw}. For instance, an elegant explanation for (de)confinement in massless $SU(N)$ adjoint 2D QCD was derived in \cite{Komargodski:2020mxz} based on the presence of non-invertible topological line defects \cite{Frohlich:2009gb, Bhardwaj:2017xup, Chang:2018iay, Gaiotto:2014kfa} in these systems. Meanwhile, \cite{Delmastro:2021otj} established criteria for a 2D QCD theory (with vanishing bare quark masses) to be gapped or gapless at long distances. Specifically, if we take a gauge theory with gauge group $F$ with left-moving fermions in some representation $R_{\ell}$ and right-moving fermions in some representation $R_{r}$, the corresponding QCD theory is gapped if and only if the following operator equations hold:
\begin{align} 
    T_{SO(\mathrm{dim}(R_{\ell}))_{1}} - T_{F_{I(R_{\ell})}} = 0, \label{gaplessorgapped1} \\
    \bar{T}_{SO(\mathrm{dim}(R_{r}))_{1}} - \bar{T}_{F_{I(R_{r})}} = 0, \label{gaplessorgapped2}
\end{align}
where $I(R)$ is the Dynkin index of the representation $R$ and $T_{F_{k}}$ ($\bar{T}_{F_{k}}$) is the holomorphic (antiholomorphic) part of the stress-energy tensor of the WZW theory with symmetry group $F$ and level $k$ (denoted $F_{k}$). To simplify the discussion, we will assume from now on that the gauge theory we are working with is non-chiral $R_{\ell} = R_{r} = R$, unless otherwise specified. Equivalently, a non-chiral 2D QCD theory is gapped if and only if the coset
\begin{equation} \label{2DQCD-TopologicalCoset}
    \frac{SO ( \mathrm{dim}(R) )_{1}}{F_{I(R)}}
\end{equation}
is a conformal embedding, i.e.\ it has vanishing central charge:\footnote{Notice that $SO(\mathrm{dim}(R))_{1}$ is a fermionic theory. Since the criterion for a mass gap concerns the central charge, it applies to either the fermionic or bosonic version of the coset theory.} 
\begin{equation}
    c_{SO (\mathrm{dim}(R))_{1}/F_{I(R)}} = c_{SO(\mathrm{dim}(R))_{1}} - c_{F_{I(R)}} = 0.
\end{equation}
As in \cite{Delmastro:2021otj}, we call cosets satisfying this condition \textit{topological cosets}. A complete list of gapped 2D QCD theories can be found in \cite{Delmastro:2021otj}, and a list of all conformal embeddings can be found in \cite{davydov2013witt}.

An intuitive way to understand this result is to bosonize the fermionic theory and cast 2D QCD as a gauged WZW model with a kinetic term for the gauge fields with coupling $g_{\mathrm{YM}}$ (for details on bosonization and fermionization, see Appendix \ref{Fermionization}). 
Here, the WZW model captures the global symmetry of the free UV fermions and is therefore given by $Spin(\mathrm{dim}(R))_{1}$.  However, the following discussion holds for any conformal embedding. Assuming that the infrared fixed point is obtained upon taking the limit $g_{\mathrm{YM}} \to \infty$:
\begin{align}
    \lim_{g_{\mathrm{YM}} \to \infty} &\int \mathcal{D}g \, \mathcal{D}A \ \mathrm{exp} \bigg[ -S_{\mathrm{WZW}}[g,A] + \frac{1}{4g_{\mathrm{YM}}^{2}}\int_{\Sigma} d^{2}x \ \mathrm{Tr}(F^{2}) \bigg] \nonumber \\  & = \int \mathcal{D}g \, \mathcal{D}A \ \mathrm{exp} \bigg[ -S_{\mathrm{WZW}}[g,A] \bigg], \label{gaugedWZW}
\end{align}
the theory is gapless or gapped if and only if the gauged WZW model is gapless or gapped, reproducing the rigorous statement obtained in \cite{Delmastro:2021otj} based on Eqns. \eqref{gaplessorgapped1} and \eqref{gaplessorgapped2}.

It is textbook conformal field theory (CFT) that for a generic coset (i.e., non-zero central charge) there exist standard algebraic methods to characterize the associated coset CFT, meaning (roughly) that we can recover, e.g., the spectrum of local operators and their fusion rules. For an overview, see \cite{DiFrancesco:1997nk}. In this sense, describing explicitly \eqref{gaugedWZW} in the gapless case is straightforward. For concrete examples in the context of 2D QCD, see \cite{Delmastro:2021otj,Delmastro:2022prj}. Not as well known, however, is that historically the standard coset construction has largely ignored the existence of additional topological sectors, with multiple vacua and topological line defects that are important to fully characterize the theory \eqref{gaugedWZW}. This observation is particularly significant in the case that the gauged WZW model is given by a conformal embedding (corresponding to a gapped 2D QCD theory), as in this case the topological sectors are all the information that is contained in \eqref{gaugedWZW}. This is the situation we will be interested in below.

\begin{figure}[!b]
        \centering
        \includegraphics[scale=0.14]{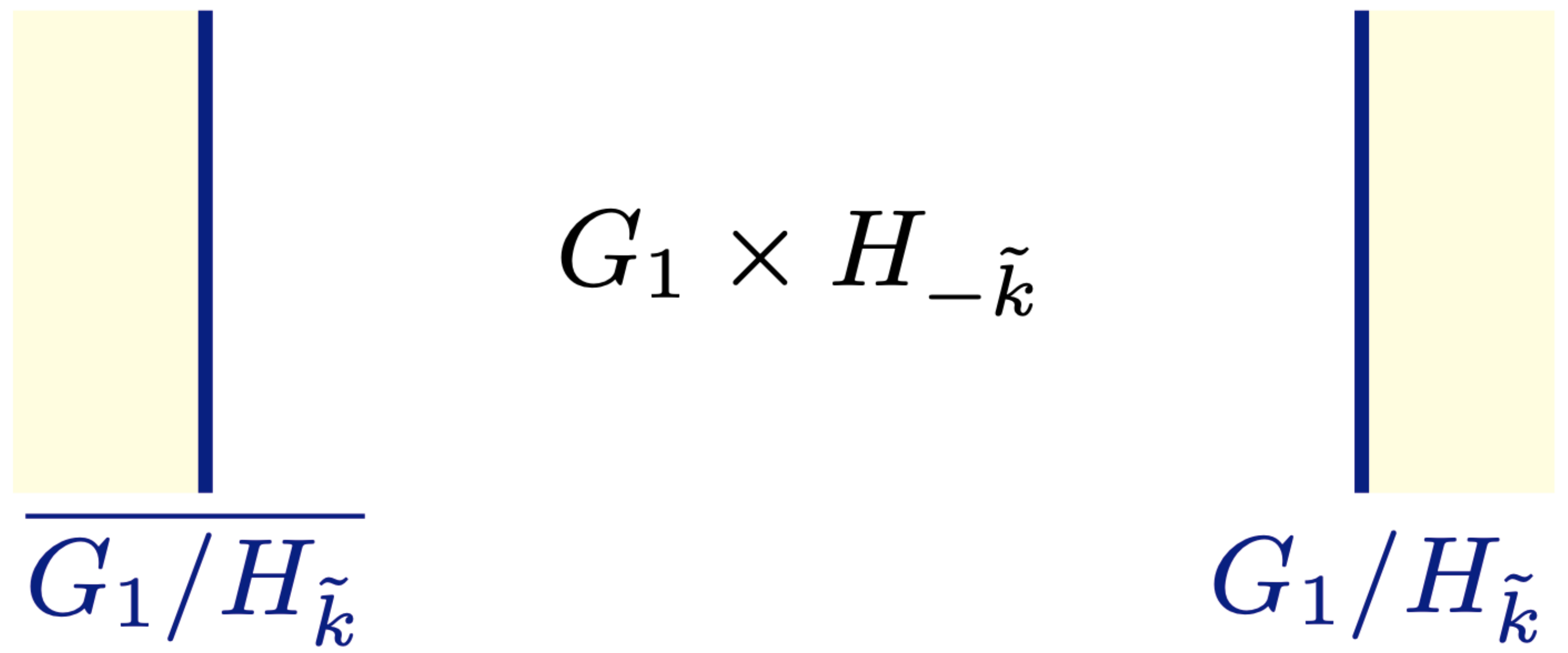} 
        \caption{Finding the full IR description of gapped 2D QCD is tantamount to describing the topological local operators and topological line operators of the 2D theory obtained upon interval compactification of a $G_{1} \times H_{-\tilde{k}}$ topological order with topological coset boundary conditions. The latter boundary conditions exist when $G_{1} / H_{\tilde{k}}$ is a topological coset, i.e., when the embedding of chiral algebras is conformal.} \label{GoverKCoset}
\end{figure}

A fruitful way to describe the topological coset \eqref{gaugedWZW} makes use of a suitable 3D construction. More precisely, the 2D theory can be obtained after interval compactification of the well-known 3D construction of gauged WZW models of \cite{Moore:1989yh}. Indeed, in this construction, if the topological coset consists of a WZW model $G_{1}$\footnote{As reviewed in \cite{DiFrancesco:1997nk}, unless $H_{\tilde{k}} = G_{k}$, conformal embeddings can only occur when the numerator has level $k=1$, which we assume in the rest of this discussion.} with gauge group $H$ and index of embedding $\tilde{k}$, we can construct the topological coset $G_{1}/H_{\tilde{k}}$ by starting with the three-dimensional Chern-Simons theory $G_{1} \times H_{-\tilde{k}}$ and setting coset boundary conditions describing the embedding of $H_{\tilde{k}}$ into $G_{1}$ on the left and right boundaries of the interval. See Figure \ref{GoverKCoset} and  Appendix \ref{CircleInterval2DQCD} for additional details.

Thus far, the construction described is not unique to conformal embeddings. The crucial difference between the coset boundary conditions for a conformal embedding and that of a generic coset at non-zero central charge is that in the former case the coset boundary conditions are \textit{topological} (also called \textit{gapped}) boundary conditions.\footnote{Mathematically, this is the statement that conformal embeddings belong to the trivial Witt  class \cite{davydov2013witt}.} Gapped boundary conditions are special in the sense that, unlike a generic boundary condition, they can always be described by a gauging operation for topological defect lines with non-necessarily-invertible fusion rules known as \textit{non-abelian anyon condensation}, or \textit{non-invertible anyon condensation}. For a collection of references and applications in this subject, see \cite{Eliens:2013epa, kong2014anyon, Lan:2014uaa, Hung:2015hfa, neupert2016boson, Burnell_2018, Cong:2017ffh, Kaidi:2021gbs, Yu:2021zmu, Cordova:2023jip, Zhang:2024bye}.

Indeed, topological defects in quantum field theory have seen significant applications in recent years. Their interpretation as ``non-invertible symmetries'' was originally developed in \cite{Gaiotto:2014kfa, Bhardwaj:2017xup, Chang:2018iay, Thorngren:2019iar, Thorngren:2021yso, Choi:2021kmx, Kaidi:2021xfk, Chang:2022hud}.  See also \cite{Fuchs:2002cm,Fuchs:2003id, Fuchs:2004dz,Fuchs:2004xi,Frohlich:2004ef,Frohlich:2006ch} for earlier pioneering work in the subject.  Most relevant to our analysis below, generalized symmetries have proved useful in characterizing phases of general quantum field theories \cite{Bhardwaj:2023fca, Bhardwaj:2023idu, Bhardwaj:2023bbf, Bhardwaj:2024qrf}, and in signaling phases of 2D gauge theory in particular in \cite{Komargodski:2020mxz}.  These symmetries can be gauged \cite{Bhardwaj:2017xup, Gaiotto:2019xmp,Kaidi:2021gbs, Roumpedakis:2022aik, Perez-Lona:2023djo, Diatlyk:2023fwf, Cordova:2023jip} and can constrain strongly coupled RG flows through anomalies \cite{Thorngren:2019iar, Choi:2021kmx, Choi:2022zal, Apte:2022xtu, Kaidi:2023maf, Choi:2023xjw, Zhang:2023wlu, Cordova:2023bja, Antinucci:2023ezl, Putrov:2024uor}. They have also recently seen applications directly relevant to particle physics in 2D including scattering theory \cite{Copetti:2024dcz, Copetti:2024rqj}, and representation theory of generalized multiplets of particles \cite{Cordova:2024vsq, Cordova:2024iti} and operators \cite{Kitaev:2011dxc, Lin:2022dhv, Bartsch:2022mpm, Bartsch:2022ytj, Bhardwaj:2023ayw, Bartsch:2023wvv, Bhardwaj:2024igy, Choi:2024tri, Choi:2024wfm, Dimofte:2024bwe}.

In this paper, we will use the technology of non-invertible anyon condensation to provide a explicit description of topological cosets --and thus in particular the IR fixed point of (bosonized) gapped 2D QCD-- explaining how to obtain the spectrum of line operators and their fusion ring, the spectrum of local operators and their topological OPE, and action of line operators on local operators.  For previous work in this direction, see \cite{Gaiotto:2020iye, Komargodski:2020mxz, Delmastro:2021otj, Delmastro:2022prj}. Below we will concentrate on the bosonic version of the topological cosets, and will leave the study of their fermionic analogs (crucial for a proper description of gapped fermionic 2D QCD) for future investigations, perhaps along the lines of \cite{Chang:2022hud, Inamura:2022lun, Bhardwaj:2024ydc}.

In mathematical terms, the problem of solving for the line operators has a rather succinct statement. In general, an arbitrary gapped boundary condition for a 3D TQFT always supports a set of topological line defects described by some fusion category $\mathcal{F}$. In turn, full knowledge of the boundary fusion category $\mathcal{F}$ allows us to recover the bulk 3D TQFT (see e.g. \cite{PhysRevB.71.045110, PhysRevB.103.195155, Turaev:1992hq}). Indeed, the corresponding bulk topological order is the Drinfeld center $\mathcal{Z}(\mathcal{F})$ of the fusion category $\mathcal{F}$. Because conformal embeddings always admit a particular gapped boundary describing the embedding of $H_{\tilde{k}}$ into $G_{1}$, this means that the Chern-Simons theory $G_{1} \times H_{-\tilde{k}}$ is the Drinfeld center $\mathcal{Z}(\mathcal{F}_{G_{1}/H_{\tilde{k}}})$ of a particular boundary fusion category $\mathcal{F}_{G_{1}/H_{\tilde{k}}}$ associated to this specific gapped boundary. The problem we must solve then is the inverse of the boundary to bulk correspondence outlined above. Namely, we must determine the boundary fusion category $\mathcal{F}_{G_{1}/H_{\tilde{k}}}$ given that we know $G_{1} \times H_{-\tilde{k}}$ allows for the specific gapped boundary associated to the conformal embedding $H_{\tilde{k}} \hookrightarrow G_{1}$. As we will describe with more precision in the main text, since we are picking the same boundary condition to the left and right of the interval, this data will fully characterize the 2D theory we are interested in after interval compactification. 

This presentation also clarifies the key role of non-invertible symmetry in characterizing the IR of gapped QCD$_{2}$.  Indeed, the boundary fusion category $\mathcal{F}_{G_{1}/H_{\tilde{k}}}$ defined above is not the full symmetry of the RG flow, but rather precisely the symmetry along the RG flow that is spontaneously broken at long distances. See Appendix \ref{CircleInterval2DQCD}. In particular, the vacua of the IR are in one-to-one correspondence with simple objects in $\mathcal{F}_{G_{1}/H_{\tilde{k}}}$. Due to its importance, we sometimes refer to this symmetry as the \emph{coset zero-form symmetry}. Mathematically, the IR TQFT furnishes a regular module of this coset zero-form symmetry $\mathcal{F}_{G_{1}/H_{\tilde{k}}}$. This relationship between the bulk 3D TQFT and the boundary symmetry is a particular incidence of the symmetry TQFT construction utilized e.g.\ in \cite{Gaiotto:2014kfa, Gaiotto:2020iye, Apruzzi:2021nmk, Freed:2022qnc, Chatterjee:2022kxb, Inamura:2023ldn, Kaidi:2022cpf, Kaidi:2023maf, Bhardwaj:2023bbf, Zhang:2023wlu, Bhardwaj:2024qrf, Brennan:2024fgj, Antinucci:2024zjp, Bonetti:2024cjk, Apruzzi:2024htg, Apruzzi:2023uma, DelZotto:2024tae}.

A topological coset also consists of a set of local operators. In the context of 2D QCD, these operators are the ones that survive the RG flow and do not decouple at long distances. For works studying the flow of these local operators, see \cite{Delmastro:2021otj, Delmastro:2022prj}. Below, we will also explain precisely how the gapped boundary conditions in the 3D construction determine the set of local operators, and furthermore, their OPE.\footnote{In mathematical terms, the OPE and the sphere one-point functions of the IR local operators furnish a commutative Frobenius algebra \cite{Sawin:1995rh, Abrams:1996ty}.} See Section \ref{TopologicalCosets} for more details. In particular, this allows us to go beyond the counting of local operators/vacua, and recover the leading contribution of the OPE of local operators of 2D QCD as we approach the IR fixed point:
\begin{equation}
   \phi^{UV}_{1}(x_{1}) \phi^{UV}_{2}(x_{2}) \to \phi^{IR}_{1} \phi^{IR}_{2}.
\end{equation}
In particular, since the IR theory is gapped, the right-hand side above is independent of position and non-singular in the limit $x_{1}\rightarrow x_{2}.$

As an example illustrating our results, we will find that in $SU(3)$ gauge theory with a single adjoint fermion, the topological coset $Spin(8)_{1}/SU(3)_{3}$ is described by a $\mathbb{Z}_{2} \times \mathbb{Z}_{2}$ triality fusion category:
\begin{table}[h]
\centering
\begin{tabular}{ |c|c|c|c|c|c| } 
\hline
$\times$  & $\mathcal{N}$ & $\bar{\mathcal{N}}$ & $\mathbf{v}$ & $\mathbf{s}$ & $\mathbf{c}$ \\
\hline
$\mathcal{N}$ & $2 \, \bar{\mathcal{N}}$ & $0 + \mathbf{v} + \mathbf{s} + \mathbf{c}$  & $\mathcal{N}$  & $\mathcal{N}$  & $\mathcal{N}$ \\ 
\hline
$\bar{\mathcal{N}}$ & $0 + \mathbf{v} + \mathbf{s} + \mathbf{c}$ & $2 \, \mathcal{N}$  & $\bar{\mathcal{N}}$  & $\bar{\mathcal{N}}$  & $\bar{\mathcal{N}}$ \\ 
\hline
$\mathbf{v}$ & $\mathcal{N}$ & $\bar{\mathcal{N}}$  & $0$  & $\mathbf{c}$  & $\mathbf{s}$ \\ 
\hline
$\mathbf{s}$ & $\mathcal{N}$ & $\bar{\mathcal{N}}$  & $\mathbf{c}$  & $0$  & $\mathbf{v}$ \\ 
\hline
$\mathbf{c}$ & $\mathcal{N}$ & $\bar{\mathcal{N}}$  & $\mathbf{s}$  & $\mathbf{v}$  & $0$ \\ 
\hline
\end{tabular}
\end{table}

\noindent and the local operators, in the basis that diagonalizes the above fusion ring (the one determined by the branching rules of the conformal embedding), satisfy the IR multiplication table:
\begin{table}[h] 
\hspace{-2.1cm} 
\begin{tabular}{|c|c|c|c|c|c|c|} 
\hline
$\boldsymbol{\cdot}$  & $\phi_{(0, \mathbf{10})}$ & $\phi_{(0, \overline{\mathbf{10}})}$ & $\phi_{(\mathbf{v}, \mathbf{8})}$ & $\phi_{(\mathbf{s}, \mathbf{8})}$ & $\phi_{(\mathbf{c}, \mathbf{8})}$ \\
\hline
$\phi_{(0, \mathbf{10})}$  & $\phi_{(0, \overline{\mathbf{10}})}$  & $\phi_{(0, \mathbf{1})}$  & $\phi_{(\mathbf{v}, \mathbf{8})}$ & $\phi_{(\mathbf{s}, \mathbf{8})}$  & $\phi_{(\mathbf{c}, \mathbf{8})}$  \\ 
\hline
$\phi_{(0, \overline{\mathbf{10}})}$ & $\phi_{(0, \mathbf{1})}$  & $\phi_{(0, \mathbf{10})}$  & $\phi_{(\mathbf{v}, \mathbf{8})}$  & $\phi_{(\mathbf{s}, \mathbf{8})}$ & $\phi_{(\mathbf{c}, \mathbf{8})}$ \\ 
\hline
$\phi_{(\mathbf{v}, \mathbf{8})}$ & $\phi_{(\mathbf{v}, \mathbf{8})}$  & $\phi_{(\mathbf{v}, \mathbf{8})}$  & $3\phi_{(0, \mathbf{1})} + 3\phi_{(0, \mathbf{10})} + 3\phi_{(0, \bar{\mathbf{10}})}$   & $3\phi_{(\mathbf{c}, \mathbf{8})}$ & $3\phi_{(\mathbf{s}, \mathbf{8})}$ \\ 
\hline
$\phi_{(\mathbf{s}, \mathbf{8})}$  & $\phi_{(\mathbf{s}, \mathbf{8})}$  & $\phi_{(\mathbf{s}, \mathbf{8})}$  & $3\phi_{(\mathbf{c}, \mathbf{8})}$  & $3\phi_{(0, \mathbf{1})} + 3\phi_{(0, \mathbf{10})} + 3 \phi_{(0, \bar{\mathbf{10}})}$ & $3\phi_{(\mathbf{v}, \mathbf{8})}$ \\ 
\hline
$\phi_{(\mathbf{c}, \mathbf{8})}$ & $\phi_{(\mathbf{c}, \mathbf{8})}$  & $\phi_{(\mathbf{c}, \mathbf{8})}$  & $3\phi_{(\mathbf{s}, \mathbf{8})}$   & $3\phi_{(\mathbf{v}, \mathbf{8})}$ & $3\phi_{(0, \mathbf{1})} + 3\phi_{(0, \mathbf{10})} +3\phi_{(0, \bar{\mathbf{10}})}$ \\ 
\hline
\end{tabular}
\end{table}

See Section \ref{examples} for details in the notation.

\begin{figure}[t] \hspace{0.cm}
\begin{subfigure}{.45\textwidth}
  \includegraphics[scale=1.2]{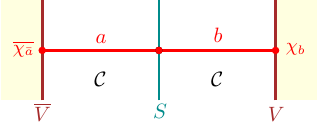}
\end{subfigure}%
\hspace{0.85cm}
\begin{subfigure}{.45\textwidth}
  \includegraphics[scale=0.145]{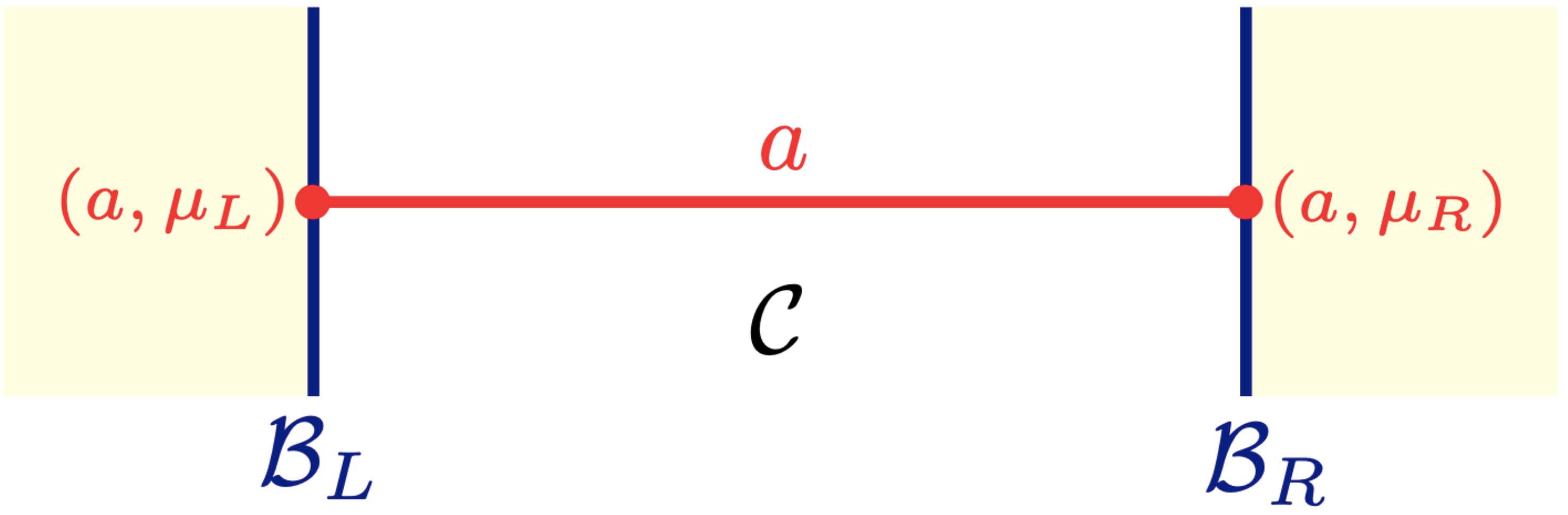}
\end{subfigure}
\caption{On the left: A modular invariant for a chiral algebra $V$ is obtained starting with an appropriate 3D bulk MTC $\mathcal{C}$, possibly with an insertion of a surface operator $S$, and summing over all the possible junctions of anyons stretching perpendicularly between the left and right boundaries, both with canonical boundary conditions for $V$. The different modular invariants are given by different choices of bulk surface operators $S$. On the right: If the bulk MTC $\mathcal{C}$ allows for (several) topological boundary conditions, then the spectrum of topological local operators of the 2D theory obtained after interval compactification is obtained by stretching an anyon between the left and right boundaries. The total partition function is obtained by summing over all such insertions, each with a unit contribution.}
\label{BoundaryGluing} 
\end{figure}

In the spirit of understanding the IR limit of gauge theories like the standard model, it is also interesting to consider the possibility of gapped QCD theories that are trivially gapped, i.e.\ those where all particle spectra are massive and where there is a unique vacuum state. To see how this possibility is addressed from the 2D-3D correspondence point of view, it is useful to recall the more standard case of counting states of a 2D RCFT $\mathcal{R}$ with a single vacuum from the 3D point of view and how the chiral and antichiral sides of $\mathcal{R}$ are glued together to furnish a full 2D RCFT. \cite{Fuchs:2002cm, Kapustin:2010if} (see also \cite{Komargodski:2020mxz, Gaiotto:2020iye}). This will be useful background for Section \ref{chiral2DQCDSection} (see also Appendix \ref{CircleInterval2DQCD}). Indeed, if $\mathcal{R}$ is described by some chiral algebra $V$, then we can construct $\mathcal{R}$ from the 3D perspective by taking a bulk TQFT described by an appropriate MTC $\mathcal{C}$\footnote{The ``appropriate'' MTC in this context is the so-called category of $V$-modules of the chiral algebra $V$.} on an interval with canonical boundary conditions respect to such chiral algebra (for references on gapless boundaries see the seminal work \cite{Elitzur:1989nr} or more recently \cite{Kong:2017etd, Kong:2019byq, Kong:2019cuu}, specially remark 5.4 in \cite{Kong:2019byq}). An arbitrary modular invariant giving the partition function for $\mathcal{R}$ is then constructed by studying all the allowed endpoints of anyons of the bulk MTC on the two boundary conditions, possibly with some surface operator inserted in the middle region \cite{Kapustin:2010if} dictating how the chiral and antichiral sides are glued together, as shown in the left in Figure \ref{BoundaryGluing}. When the surface defect $S$ is invertible the construction leads to a permutation modular invariant, and in the case where the modular invariant is of pure extension type the surface defect $S$ is non-invertible, as in the higher-gauging defects of \cite{Roumpedakis:2022aik}.

The situation when the boundaries are topological is similar, with the difference that characterizing topological boundaries is more subtle technically than the gapless boundaries describing (non-necessarily diagonal) RCFTs with single vacuum, and with the difference that an allowed insertion of an anyon in between two boundaries does not contribute a pair of holomorphic and antiholomorphic characters to the (torus) partition function, but rather just a unit contribution. Allowed insertions of anyons in between topological boundaries then count the number of vacua in the corresponding 2D theory, as shown in the right in Figure \ref{BoundaryGluing}. Thus, if the unique anyon that can be stretched in between two different topological boundaries is the identity anyon, then we have found a trivially gapped IR fixed point. Indeed, this is the essential idea advocated in \cite{Zhang:2023wlu} to explore anomalies of non-invertible symmetries in two spacetime dimensions. Here, we will take this construction and apply it on our context to argue in Section \ref{chiral2DQCDSection} that the chiral $Spin(8)$ QCD$_{2}$ with massless fermions in the vectorial and spinorial representations is trivially gapped.

\section{Lagrangian Algebras and Gapped Boundaries} \label{GappedBoundariesandTopologicalCosets}

In this section, we describe the algebraic theory of topological boundaries in terms of non-abelian (non-invertible) anyon condensation and Lagrangian algebras. In the next section we put the theory to use to describe topological cosets. In Appendix \ref{MTCsection} we recall standard material on the algebraic formulation of anyons via MTCs that is useful in the following.

\subsection{Lagrangian Algebras for Gapped Boundaries}

\subsubsection{Abelian Case}

To understand the general situation, we start by briefly recalling the theory of gapped boundaries for abelian MTCs. These are MTCs whose simple anyons are all abelian, and their fusion rules are always those of some abelian group $G$. In this case, we can efficiently describe topological boundaries in terms of the gauging of abelian higher-form symmetries \cite{Gaiotto:2014kfa}.

\begin{figure}[t] \hspace{2.7cm}
\begin{subfigure}{.45\textwidth}
  \includegraphics[scale=0.12]{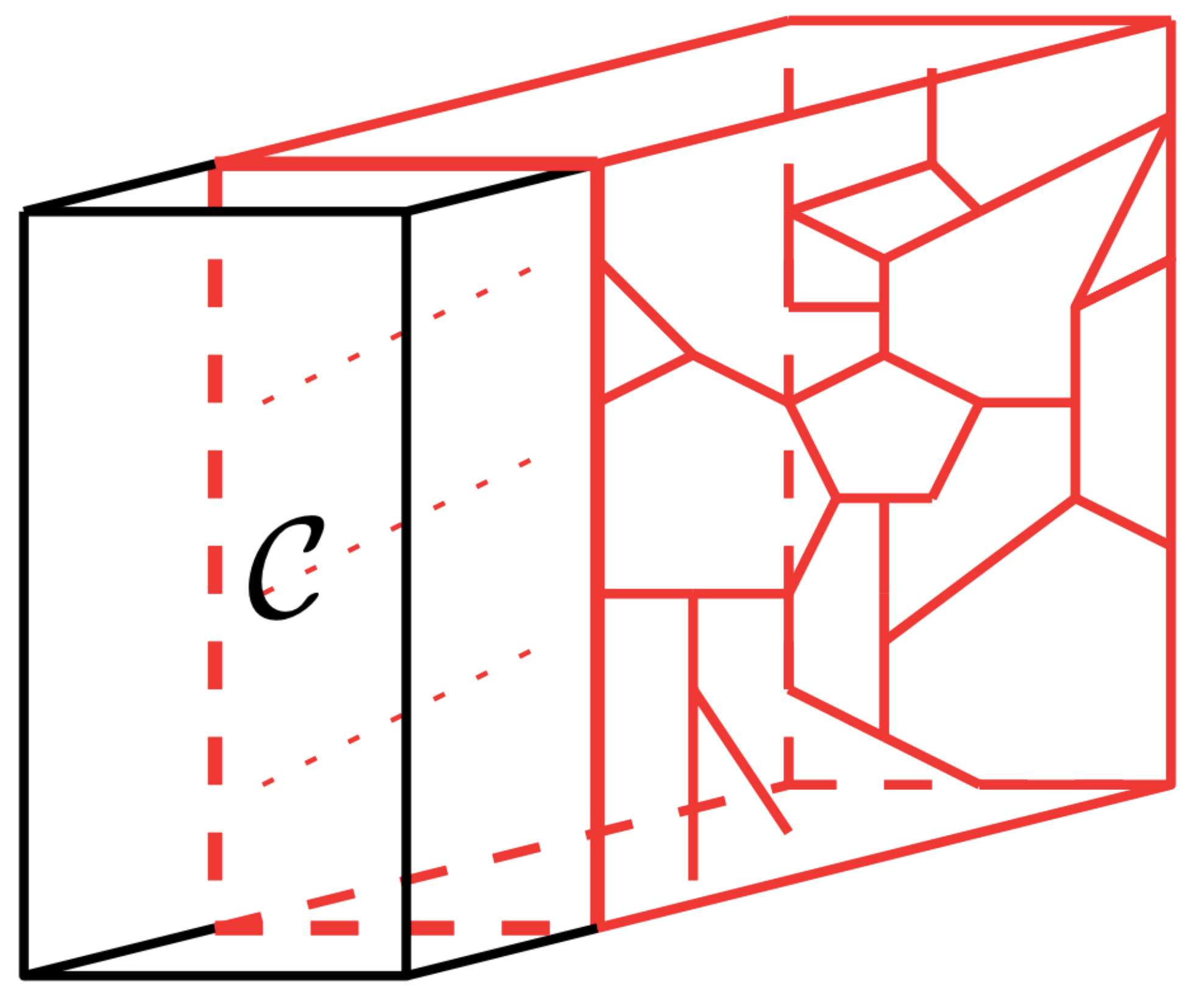}
\end{subfigure}%
\hspace{-0.95cm}
\begin{subfigure}{.45\textwidth}
  \includegraphics[scale=0.075]{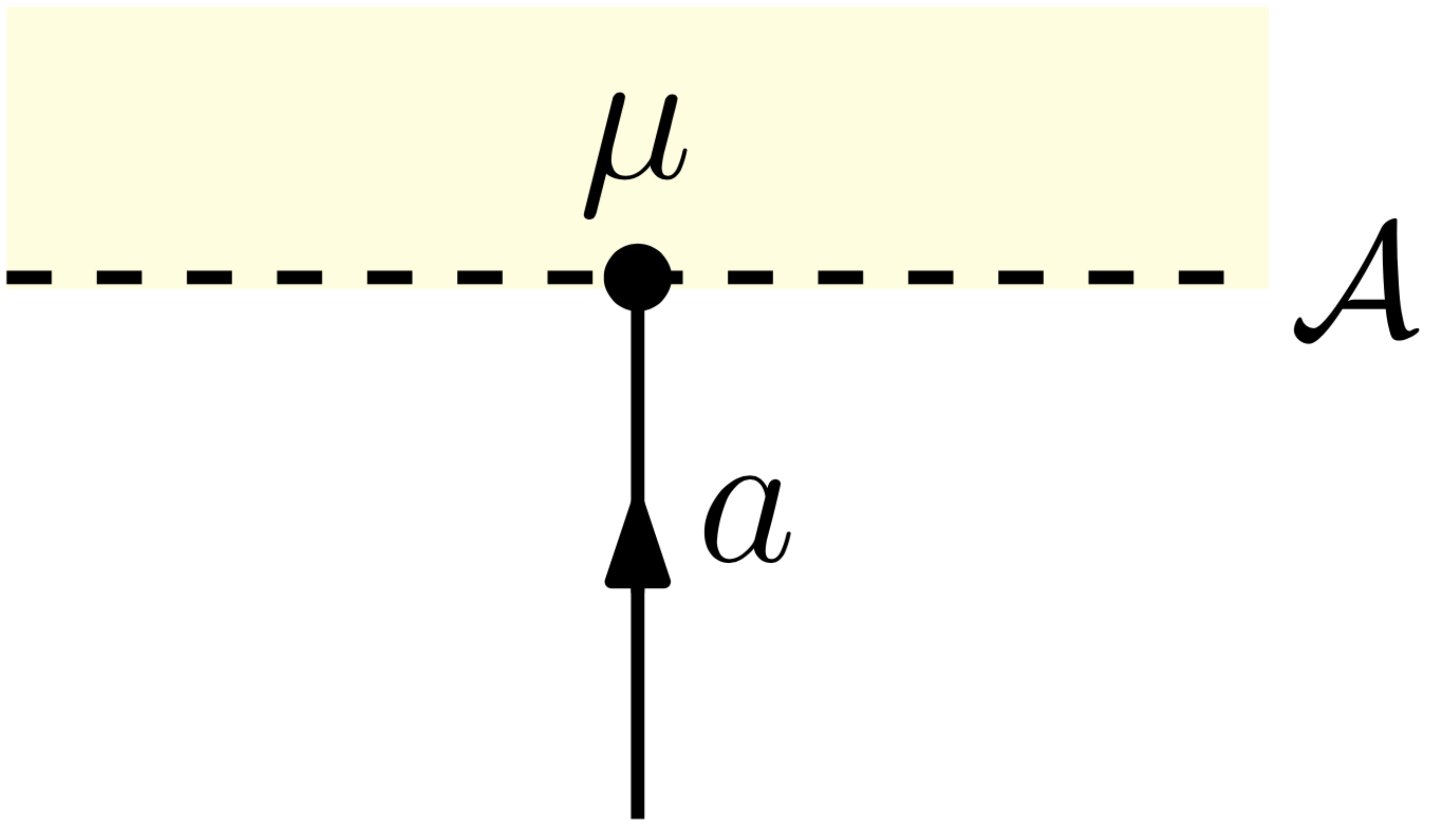}
\end{subfigure}
\caption{On the left: The 3D TQFT described by the MTC $\mathcal{C}$ (in black) exhibits a topological boundary generated by the process of gauging a Lagrangian algebra $\mathcal{A}$ on one half of the spacetime (in red). In the abelian case, the Lagrangian algebra reduces to the notion of a Lagrangian subgroup, and we have $L = \bigoplus_{h \in H} h$ for $H$ a Lagrangian subgroup of $G$ (see main text). On the right: Simple anyons $a$ in the Lagrangian algebra $\mathcal{A}$ are allowed to end perpendicularly at the topological boundary (dashed line) on a set of topological (quasi-)local junctions labeled by $\mu$, a channel in which $a$ embeds into $\mathcal{A}$.}
\label{TopologicalBoundary} 
\end{figure}

In this setting, we must determine then which objects are gaugable in a bulk region of the 3D spacetime so that the result after gauging is a trivial theory. In the context of abelian MTCs the answer is well-known, and it corresponds to a Lagrangian subgroup $L$ of the given abelian MTC \cite{Levin:2013gaa,Barkeshli_2013,Kapustin:2010hk,Fuchs:2012dt}. This is a subset of the anyons that fulfills $H \subset G$ fusion rules, and such that all anyons in $L$ are bosons.\footnote{A boson is a simple anyon $a$ with topological spin $\theta_{a} = 1$. See Appendix \ref{MTCsection} for a brief review.} In particular, notice that this subset of anyons is closed under fusion. Additionally, the following two conditions have to be satisfied: \\
\begin{itemize}
    \item Any two simple anyons in $L$ have trivial braiding phase \eqref{BraidingPhase} with each other. This condition can be interpreted as follows: Recall that the braiding phase \eqref{BraidingPhase} measures the charge of a simple anyon $a$ under the symmetry implemented by an anyon $b$ encircling $a$. Then, this condition states that the lines generating the gauge symmetry must be gauge-invariant amongst themselves. When we gauge these lines we sum over all possible insertions of them and they become indistinguishable from the identity line in the gauged theory. \\

    \item Any simple anyon not in $L$ has non-trivial braiding phase \eqref{BraidingPhase} with at least one simple anyon belonging to $L$. This is the statement that the remaining lines are charged under $L$ and thus not gauge-invariant. They are left out of the spectrum of the gauged theory. Overall, after gauging, the result is a completely trivial theory. \\
\end{itemize}
Performing this gauging operation on half of spacetime leads then to a gapped boundary, as depicted in the left of Figure \ref{TopologicalBoundary}.\footnote{In practical calculations the gauging of abelian anyons may be implemented along the lines of the three-step gauging rule of \cite{Moore:1989yh, Hsin:2018vcg} (for a more recent review on this topic, see Appendix A.1. in \cite{Delmastro:2021xox}). A gapped boundary is then identified if the outcome of the three-step gauging rule returns the trivial MTC.} The order of $H$ is always $|H| = \sqrt{|G|}$. We require anyons in the Lagrangian subgroup to be bosons since otherwise twisting lines into loops gives phases (see \eqref{TopologicalTwist}) that makes the definition of gauging ambiguous \cite{Gaiotto:2014kfa, Gomis:2017ixy, Hsin:2018vcg}. Thus, only bosons can participate in a gaugable symmetry.

\begin{figure}[!b]
\centering
        \includegraphics[scale=0.9]{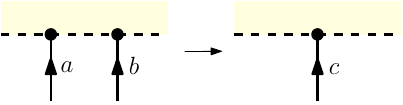} 
        \caption{Two simple lines ending perpendicularly at a topological boundary always fuse to a single line when the topological boundary is described by a Lagrangian subgroup. In particular, the junctions follow the same fusion rules as those of the corresponding Lagrangian subgroup.} \label{AbelianBoundaryFusion}
\end{figure}

As previously stated, lines in the Lagrangian subgroup become indistinguishable from the identity line on the side of the trivial theory. More precisely, the Lagrangian subgroup --seen as a non-simple line of the original theory-- becomes the identity line on the side of the trivial theory after gauging:
\begin{equation}
    \bigoplus_{h \in H} h \to 0_{\mathrm{Trivial}}.
\end{equation}
A simple line in the Lagrangian subgroup is thus mapped by the gapped boundary to the identity line on the side of the trivial theory. When the simple line ends perpendicularly at the gapped boundary, this defines a topological junction for such a simple line at the boundary. See Figure \ref{TopologicalBoundary}. Fusion of these junctions follows in the obvious way. Since a Lagrangian subgroup is closed under fusion and fusion of abelian anyons results in a single simple anyon, the fusion of these junctions at the boundary follow the same fusion rules as those of the corresponding Lagrangian subgroup. See Figure \ref{AbelianBoundaryFusion}. The special case of Lagrangian subgroups in abelian MTCs will be important in Section \ref{chiral2DQCDSection} below.

\subsubsection{Non-Abelian Case}

The general situation for non-abelian MTCs is more delicate, but the general characterization of gapped boundaries is known. In general, any elementary gapped boundary condition of a 3D TQFT is described in terms of a generalization of the concept of a Lagrangian subgroup known as \textit{Lagrangian algebra} (see e.g., \cite{davydov2013witt, kong2014anyon, Cong:2017ffh, Kaidi:2021gbs}). In parallel to the situation of gauging invertible one-form symmetries reviewed above, a gapped boundary is generated by inserting a suitable fine mesh of anyons on half of spacetime and performing a weighted summation over such insertions so that the resulting theory is trivial. (See  Figure \ref{TopologicalBoundary})). A Lagrangian algebra is, essentially, the precise description of such a mesh of anyons and their insertions.

In the rest of this subsection, our goal will be to provide a more accurate definition of the concept of Lagrangian algebra. In the subsequent discussion we mainly follow the algebraic definitions of \cite{Fuchs:2002cm}. We follow their interpretation in terms of anyon condensation from \cite{kong2014anyon}.

An \textit{algebra} $\mathcal{A}$ in a MTC $\mathcal{C}$ consists, first, of some (non-simple) anyon in $\mathcal{C}$:
\begin{equation} \label{LagrangianAlgebra}
    \mathcal{A} = \bigoplus_{a \in \mathcal{C}} n_{a} a, \quad n_{a} \in \mathbb{N}.
\end{equation}
This is the anyon that we will gauge on half of spacetime, so in particular we demand that it becomes indistinguishable from the identity line on the side of the trivial theory. The fact that the anyon $\mathcal{A}$ is (generically) non-simple implies that there exist topological junctions describing the embedding of simple anyons $a$ into $\mathcal{A}$. Generally, there are $n_{a}$ such junctions, so we label them by Greek letters and draw them as gray arrows:
\begin{equation} \label{embeddingintoalgebra}
    \includegraphics[scale=1.0, valign=c]{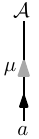} \in \mathrm{Hom}(a, \mathcal{A}) \, .
\end{equation}
An analogous diagram holds where an algebra is ``decomposed'' by a topological junction into its simples.

An algebra is also defined by a ``multiplication map'' that we denote by $m$.\footnote{Technically, an algebra also requires a unit morphism $u \in \mathrm{Hom}(0,\mathcal{A})$ satisfying a number of properties. In the following we will actually not make use of the unit, and instead refer the reader to \cite{Fuchs:2002cm} for further details.} The multiplication in the algebra corresponds to a trivalent vertex of the algebra object, and may be expressed in terms of the simple lines of the MTC as a collection of complex numbers encoding how the simple lines embed into the algebra object $\mathcal{A}$. Diagrammatically:
\begin{equation} \label{Product}
    \includegraphics[scale=0.75, valign=c]{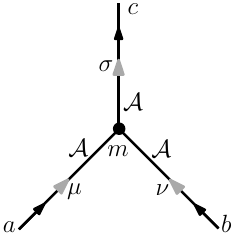} \ \ = \sum_{j=1}^{N_{ab}^{c}} m_{(a,\mu)(b,\nu)}^{(c,\sigma),j} \includegraphics[scale=0.105, valign=c]{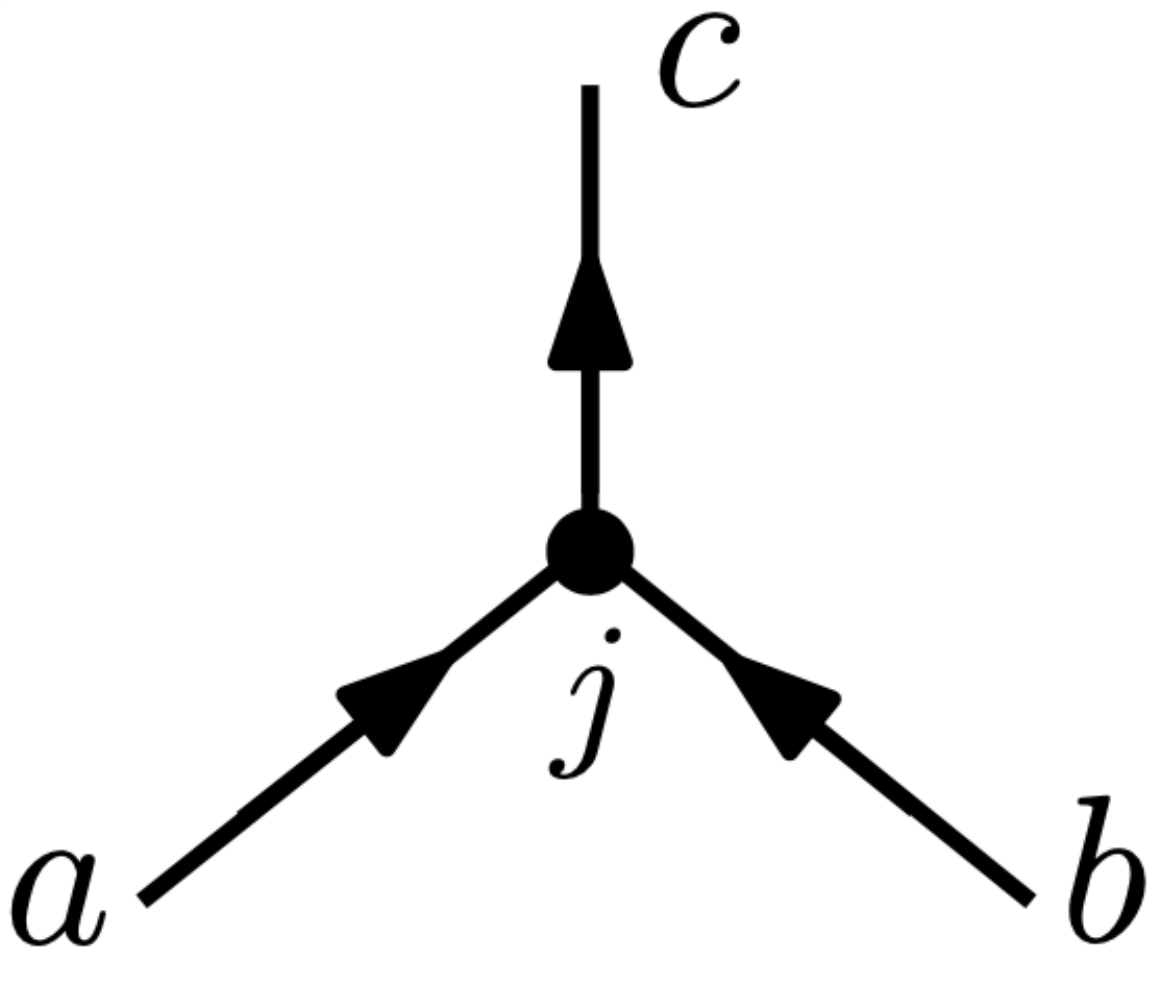},
\end{equation}
It is important to note that two different algebras may have the same underlying object \eqref{LagrangianAlgebra} and differ only in their multiplication, so strictly speaking we should write a pair $(\mathcal{A},m)$ to denote an algebra appropriately. However, as there should be no misunderstanding given context, we will abuse language and use the same notation $\mathcal{A}$ to refer to both the abstract notion of algebra with multiplication, or just the underlying (non-simple) anyon.

A technical point is that a Lagrangian algebra is both an algebra and a \textit{coalgebra}. A coalgebra in a MTC $\mathcal{C}$ is defined similarly by a pair $(\mathcal{A}, m^{\vee})$,\footnote{More precisely, by a triple $(\mathcal{A}, m^{\vee},u^{\vee})$, where $u^{\vee}$ is a counit morphism $ u^{\vee} \in \mathrm{Hom}(\mathcal{A},0)$. As with the unit morphism, the counit will not be needed in the following, so we do not refer to it from now on.} where now $m^{\vee}$ is a ``comultiplication map'', whose diagrammatic expression is similar to \eqref{Product}, but where we replace $m \to m^{\vee}$ and the diagrams now involve splitting spaces instead of fusing spaces. In a unitary theory, there exists a basis where the comultiplication is determined by the multiplication as $m^{\vee} = m^{\dagger}$.\footnote{Sometimes it is useful to consider the general basis, e.g. when one does not know the $F$-symbols in unitary basis.}

Recall that for higher-form symmetries, a gauge transformation corresponds to a rearrangement of the trivalent vertices defining the mesh of higher-form symmetry \cite{Gaiotto:2014kfa}. Similarly, we demand that performing a topological manipulation of our mesh of non-invertible anyons leaves the result invariant. This more general form of gauge invariance implies that the multiplication and comultiplication must satisfy a number of constraints -- often expressed diagrammatically -- in order to properly trivialize the aforementioned topological manipulations. The first of these are the (co)associativity conditions:\footnote{We could also write the associativity and Frobenius conditions, but since we assume we are working with unitary theories where a basis exist where $m^{\vee} = m^{\dagger}$, these are not independent conditions to consider.}
\begin{align} \label{diagrammatic(co)associativity}
    \includegraphics[scale=0.09, valign=c]{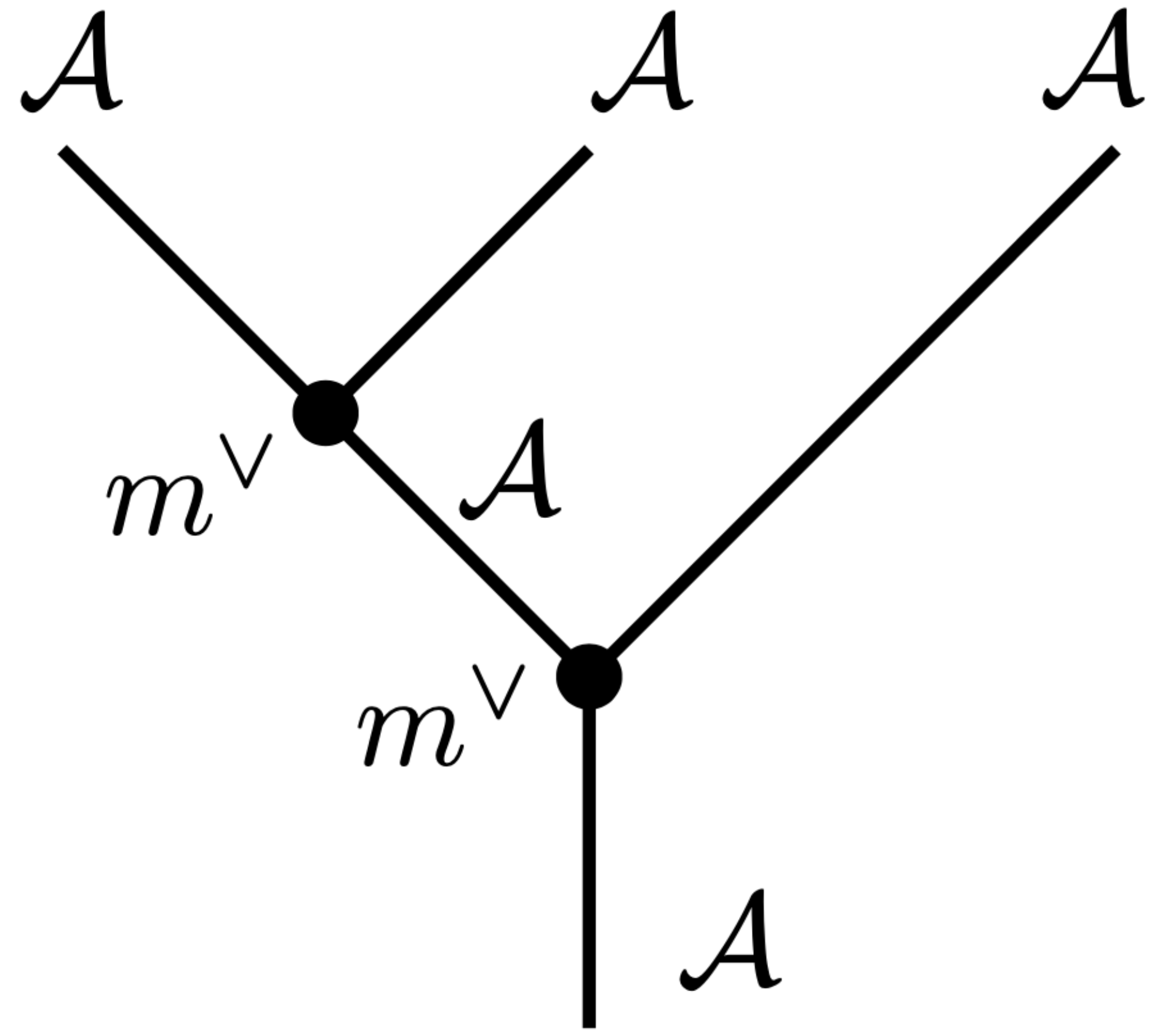} = \includegraphics[scale=0.09, valign=c]{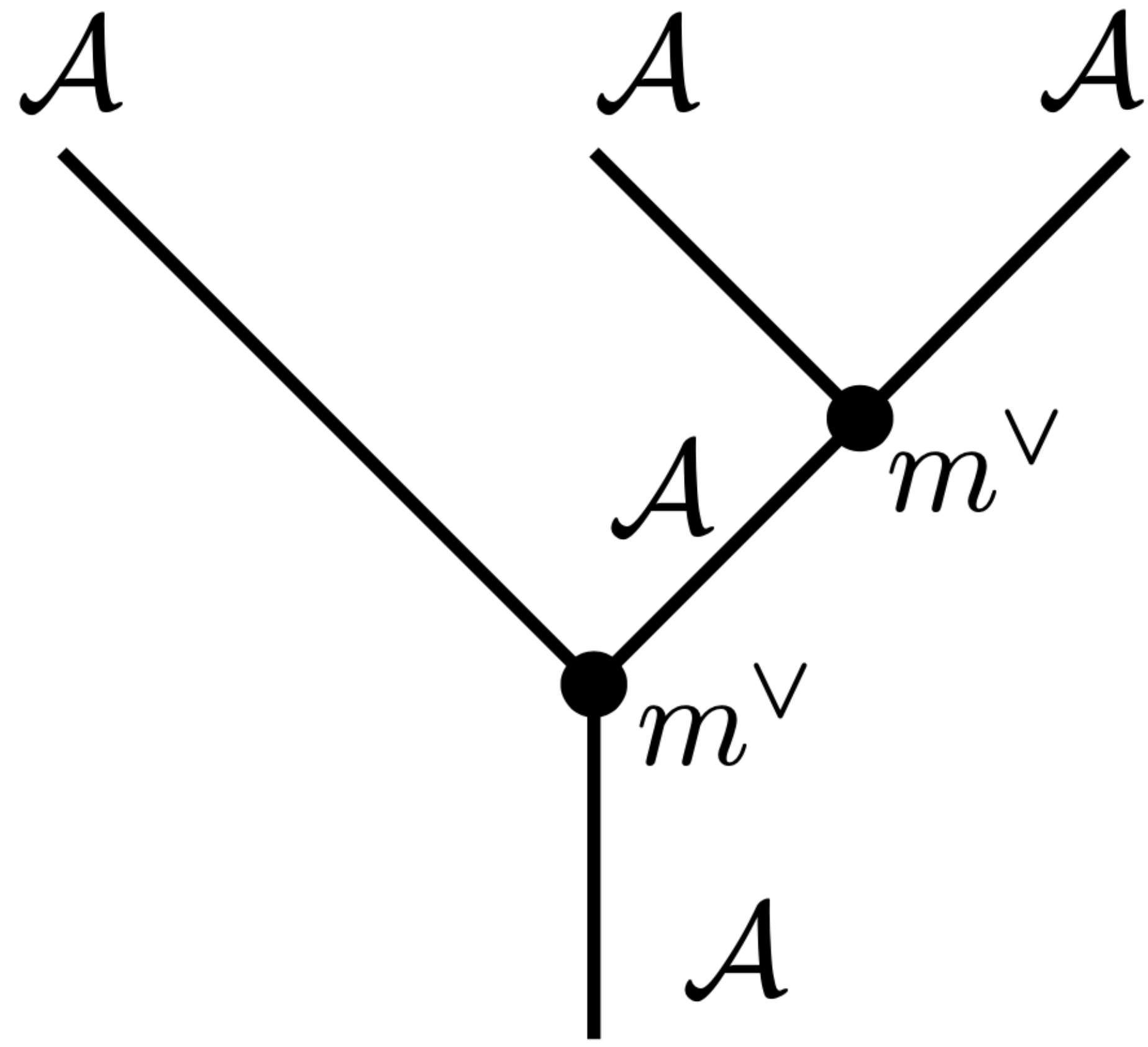} \ ,
\end{align}

The previous diagrams can be understood as a series of polynomial (quadratic) equations that the multiplication and comultiplication must fulfill, analogous to the pentagon and hexagon equations constraining the $F$ and $R$-symbols. To be explicit, we can express the coassociativity condition in components:
\begin{equation} \label{coassociativity}
    \sum_{\lambda} m^{\vee (d,\phi),k}_{\ \, (a,\mu)(f,\lambda)} m^{\vee (f, \lambda), j}_{\ \, (b, \nu)(c, \sigma)} = \sum_{(e,\rho),\ell,p} m^{\vee (e, \rho), \ell}_{\ \, (a, \mu)(b, \nu)} m^{\vee (d, \phi), p}_{\ \, (e, \rho)(c, \sigma)} [F^{abc}_{d}]_{(e,\ell,p),(f,j,k)}
\end{equation}

Another topological manipulation that must trivialize on the side of the gauged theory corresponds to the braiding of one algebra object around each other. Diagrammatically, one finds the condition:
\begin{equation} \label{AlgebraCommutativity}
    \includegraphics[scale=0.075, valign=c]{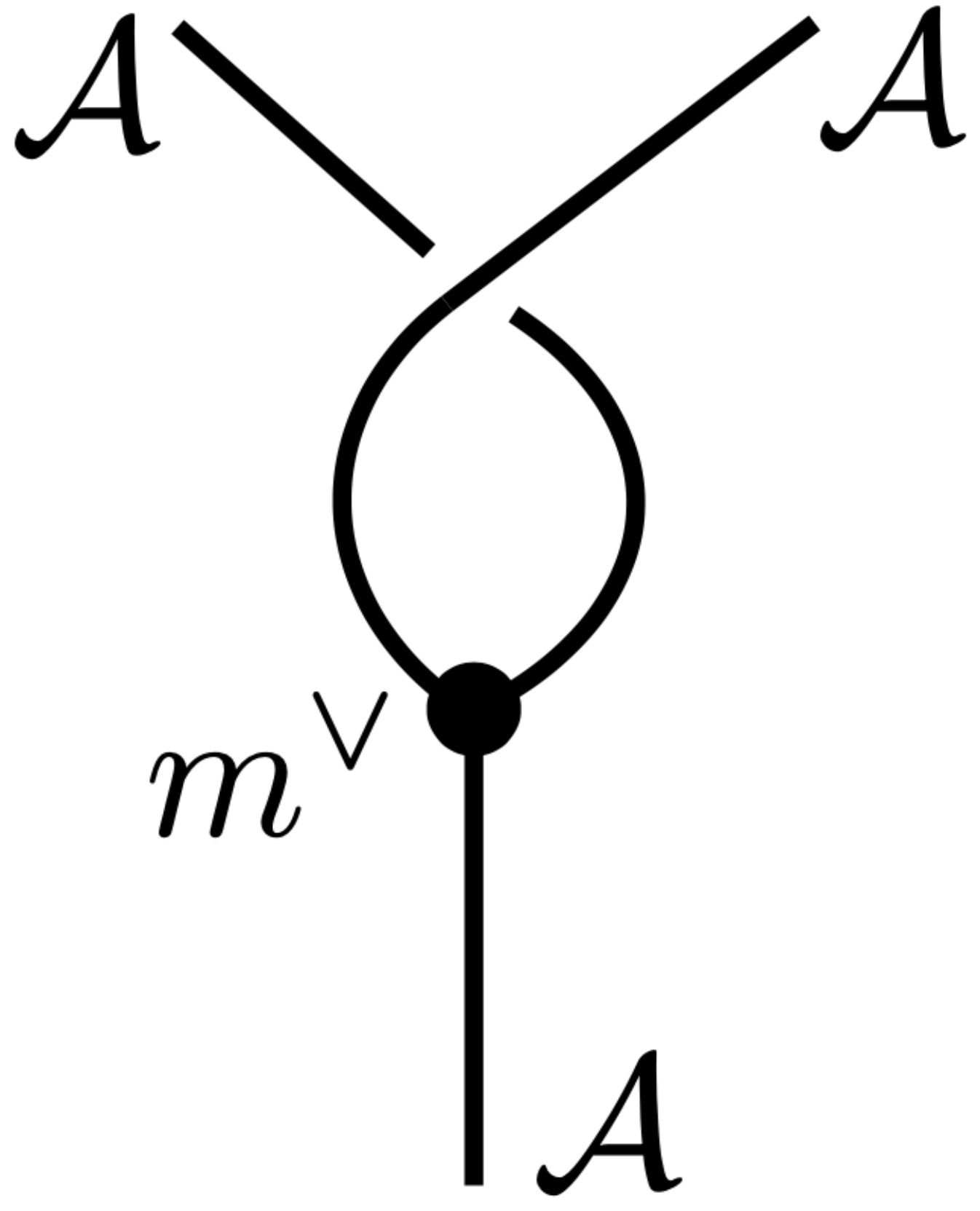} = \includegraphics[scale=0.075, valign=c]{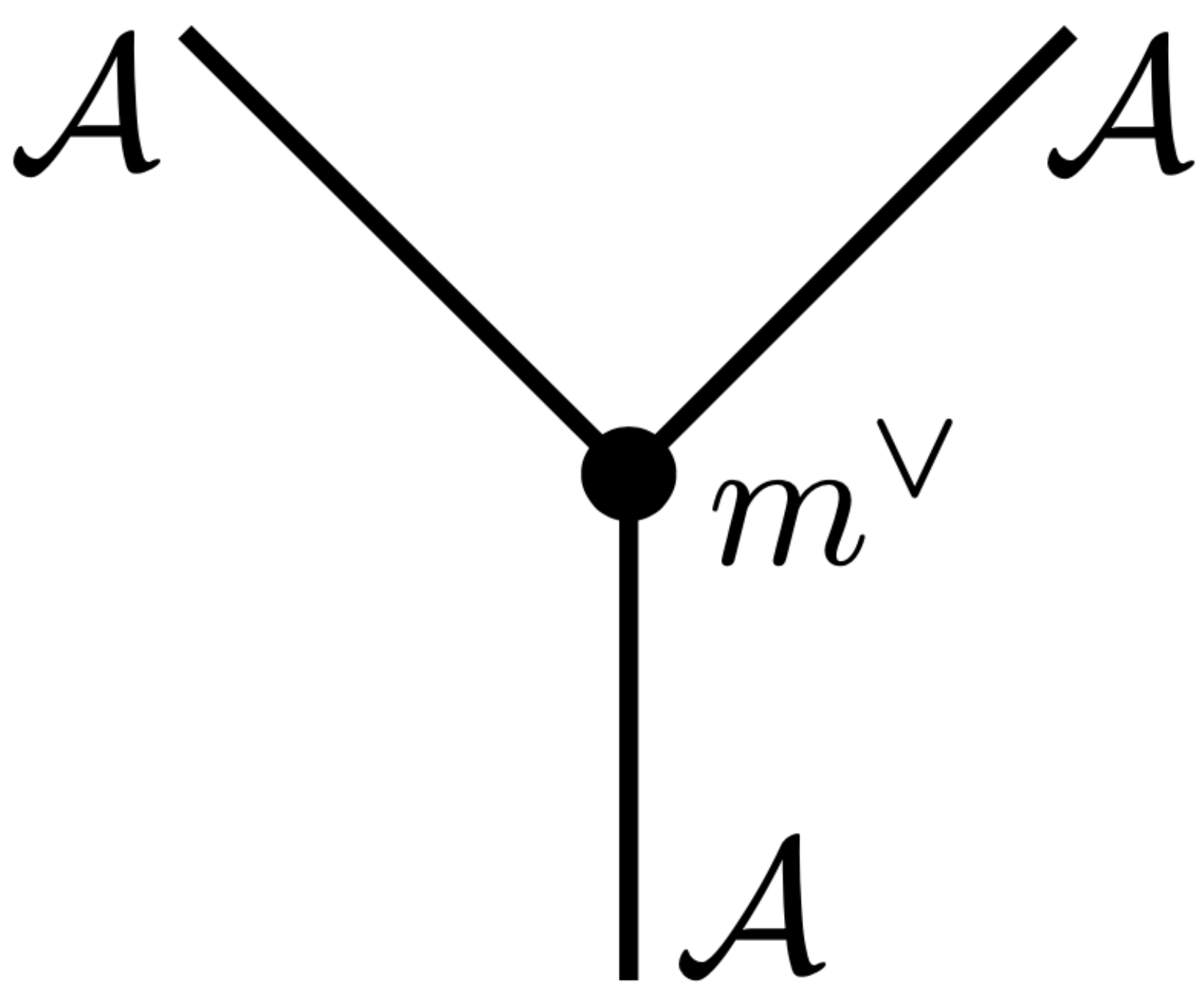}.
\end{equation}
Or in components:
\begin{equation} \label{commutativityLagrangianAlgebra}
    m^{\vee (c, \sigma), j}_{\ \, (a, \mu)(b, \nu)} = \sum_{k} m^{\vee (c, \sigma), k}_{\ \, (b, \nu)(a, \mu)} [R^{ab}_{c}]_{kj}.
\end{equation}

We also require that a bubble of the vacuum line on the side of the gauged theory can always be attached or removed to the vacuum line. In terms of the algebra object in the original theory, this leads to the  separability condition:\footnote{In two dimensions, the analogous rearrangement of lines relating different triangulations of a Riemann surface is known as the ``bubble move'' \cite{Fuchs:2002cm}.}\footnote{Notice that the constant on the right-hand side of \eqref{NormSpecial} is often set to one. As described in \cite{Fuchs:2002cm, Frohlich:2003hm}, this amounts to a normalization choice of the unit and counit morphisms that we have made to obtain simple-looking expressions in our context. Changing this normalization merely amounts to renormalizing the local operators defined below by a global constant.}
\begin{equation} \label{NormSpecial} 
    \includegraphics[scale=0.10, valign=c]{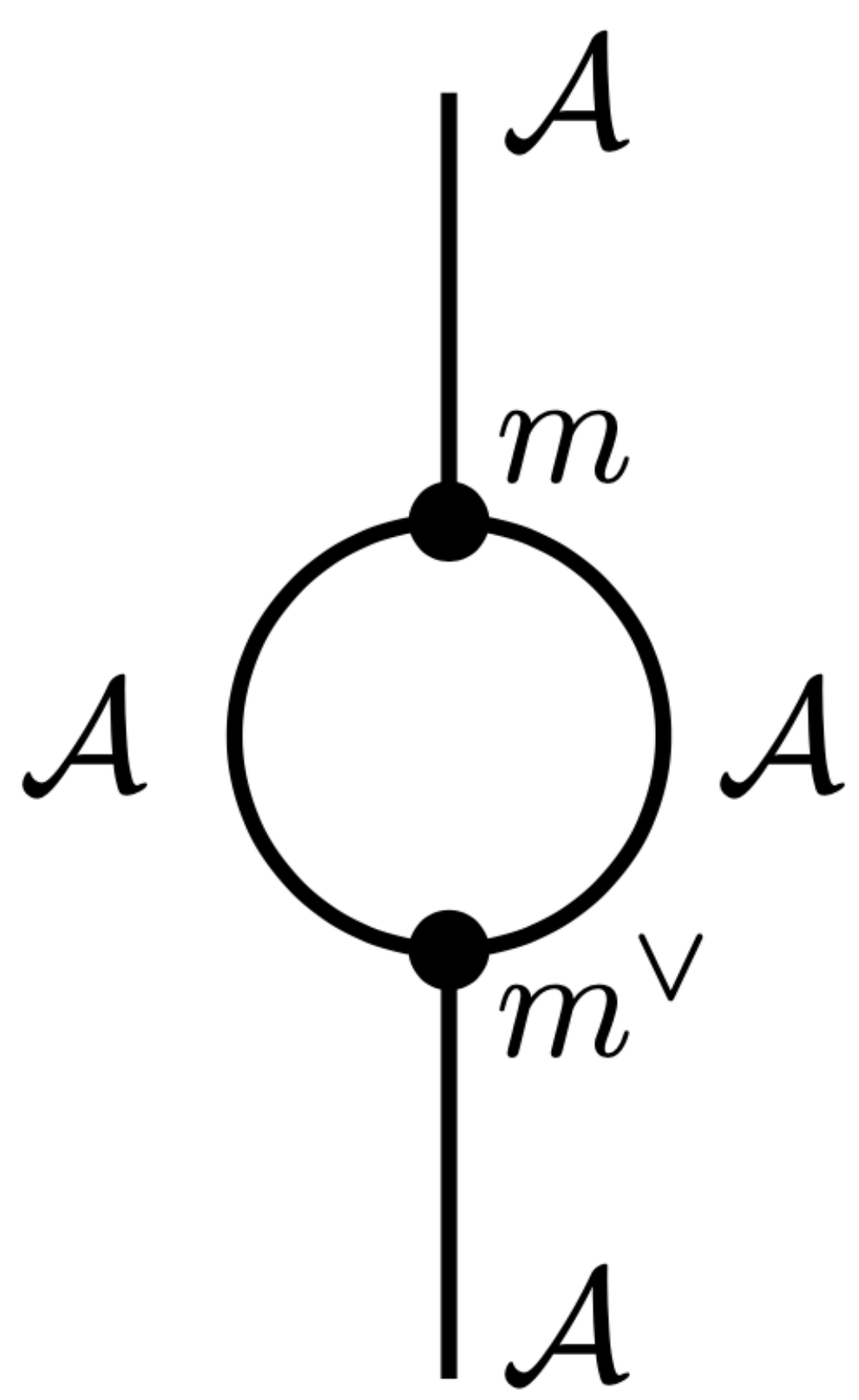} \ = \mathrm{dim}(\mathcal{A})\ \includegraphics[scale=0.10, valign=c]{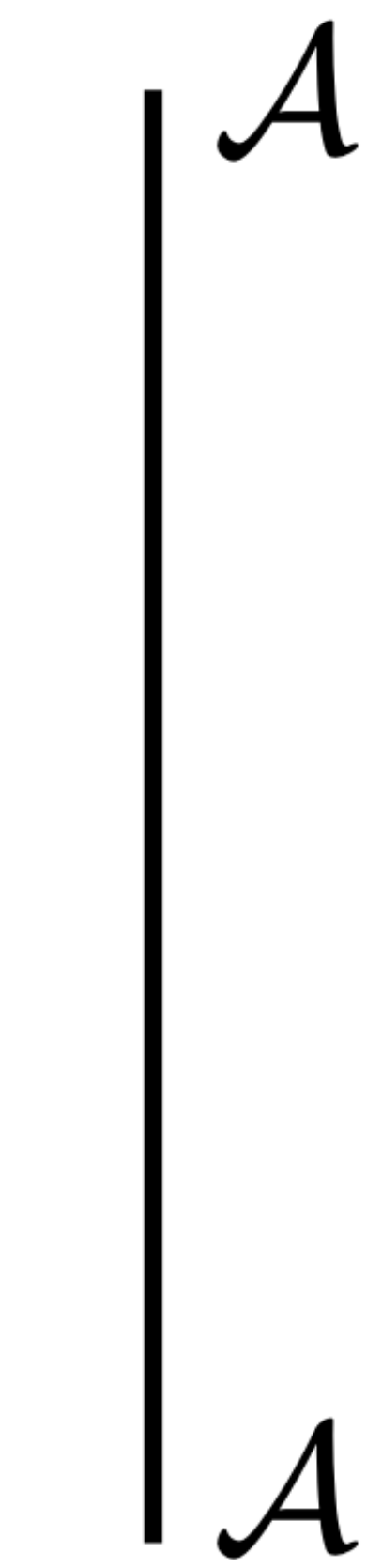} \quad .
\end{equation}

Notice that a choice of multiplication in the Frobenius algebra is unique only up to a ``gauge transformation'' that replaces $m^{(c, \sigma), j}_{(a, \mu)(b, \nu)} \to \Gamma^{a}_{\mu \mu'} \Gamma^{b}_{\nu \nu'} m^{(c, \sigma'), j}_{(a, \mu')(b, \nu')} (\Gamma^{c}_{\sigma \sigma'})^{-1}$, for a choice of unitary transformations $\Gamma^{a}_{\mu \mu'}$.

Finally, we demand that a gaugable algebra consists only of bosonic simple anyons. Similarly as with Lagrangian subgroups, this condition arises to avoid ambiguous phases coming from twisting loops (recall \eqref{TopologicalTwist}).

Thus far, the conditions written down define a gaugable algebra in 3D TQFTs along a three-dimensional region. This is, the gauged theory has not been demanded to be trivial. A Lagrangian algebra is furthermore defined by the condition\footnote{From the abstract definitions of a gaugable algebra one can deduce that $\mathrm{dim}(\mathcal{D}) = \mathrm{dim}(\mathcal{C})/\big( \mathrm{dim}(\mathcal{A}) \big)^{2}$, where $\mathrm{dim}(\mathcal{D})$ corresponds to the dimension of the MTC $\mathcal{D}$ obtained after gauging the algebra $\mathcal{A}$ in $\mathcal{C}$ (see e.g. \cite{Frohlich:2003hm}). When the result is the trivial theory $\mathrm{dim}(\mathcal{D}) = 1$, and \eqref{quantumdimensioncondition} follows. Alternatively, one can deduce \eqref{quantumdimensioncondition} must be fulfilled by a gapped boundary by analyzing the Hopf link between $\mathcal{A}$ and a simple line $a$ as in \cite{Kaidi:2021gbs}.}
\begin{equation} \label{quantumdimensioncondition}
    \mathrm{dim}(\mathcal{C}) = \big( \mathrm{dim}(\mathcal{A}) \big)^{2}.
\end{equation}
This finishes the definition of a gapped boundary in 3D TQFTs in terms of Lagrangian algebras.

To give a general example, note that if a MTC is of the form $\mathcal{C} \times \bar{\mathcal{C}}$, then a gapped boundary with Lagrangian algebra object 
\begin{equation} \label{DiagonalLagrangian}
    \mathcal{A} = \bigoplus_{a \in \mathcal{I}} (a, a)
\end{equation}
always exists. (Recall that $\mathcal{I}$ labels the simple anyons in $\mathcal{C}$, see Appendix \ref{MTCsection}.) This is the so-called diagonal Lagrangian algebra. Physically, the existence of this algebra follows from the trivial fact that the identity interface in $\mathcal{C}$ always exists. After folding, the identity interface becomes the topological boundary in $\mathcal{C} \times \bar{\mathcal{C}}$ described by the diagonal Lagrangian algebra. For a mathematical proof that the object \eqref{DiagonalLagrangian} can always be extended to satisfy the definitions of a Lagrangian algebra above, see e.g.\ Lemma 6.19 in \cite{Frohlich:2003hm}.

As a quick example of a gapped boundary given by the diagonal Lagrangian consider $\mathcal{C} = (G_{2})_{1}$. Then, we have a gapped boundary:
\begin{equation}
    \mathcal{A} = (0,0) \oplus (\tau, \tau),
\end{equation}
where $\tau$ is the unique non-trivial anyon in $(G_{2})_{1}$ and the second entry corresponds to the associated anyon in $(G_{2})_{-1}$ (For additional details on notation, see the end of Appendix \ref{MTCsection}). Using the data of the Fibonacci MTC in Appendix \ref{MTCdataAppendix}, it is straightforward to check that the associativity, separability, and commutativity conditions just discussed are easily solved by
\begin{equation}
    m^{(0,0)}_{(0,0)(0,0)} = m^{(\tau,\tau)}_{(\tau,\tau) (0,0)} = m^{(\tau,\tau)}_{(0,0) (\tau,\tau)} = m^{(0,0)}_{(\tau,\tau)(\tau,\tau)} = m^{(\tau,\tau)}_{(\tau,\tau) (\tau,\tau)} = 1,
\end{equation}
which gives the full algebra for this gapped boundary.

As in the case of abelian gauging discussed above, one can study the behavior of the topological endpoints of anyons at the topological boundary. Here, we highlight the key differences arising for our more general situation. First, notice that since the Lagrangian algebra behaves as the identity line in the resulting (trivial) gauged theory, a simple object in the Lagrangian algebra is allowed to have as many topological junctions at the topological boundary as the multiplicity with which it appears in the decomposition of the algebra object \eqref{LagrangianAlgebra} into simple anyons (see  Figure \ref{TopologicalBoundary}). Secondly, fusion of anyons ending at the boundary is more delicate in the general case of non-invertible anyon condensation since fusion of simple anyons in the Lagrangian algebra is generically not closed. To capture this phenomenon, one introduces the $M$-symbols (originally introduced in \cite{Cong:2016ayp}):
\begin{equation}
    \includegraphics[scale=0.1, valign=c]{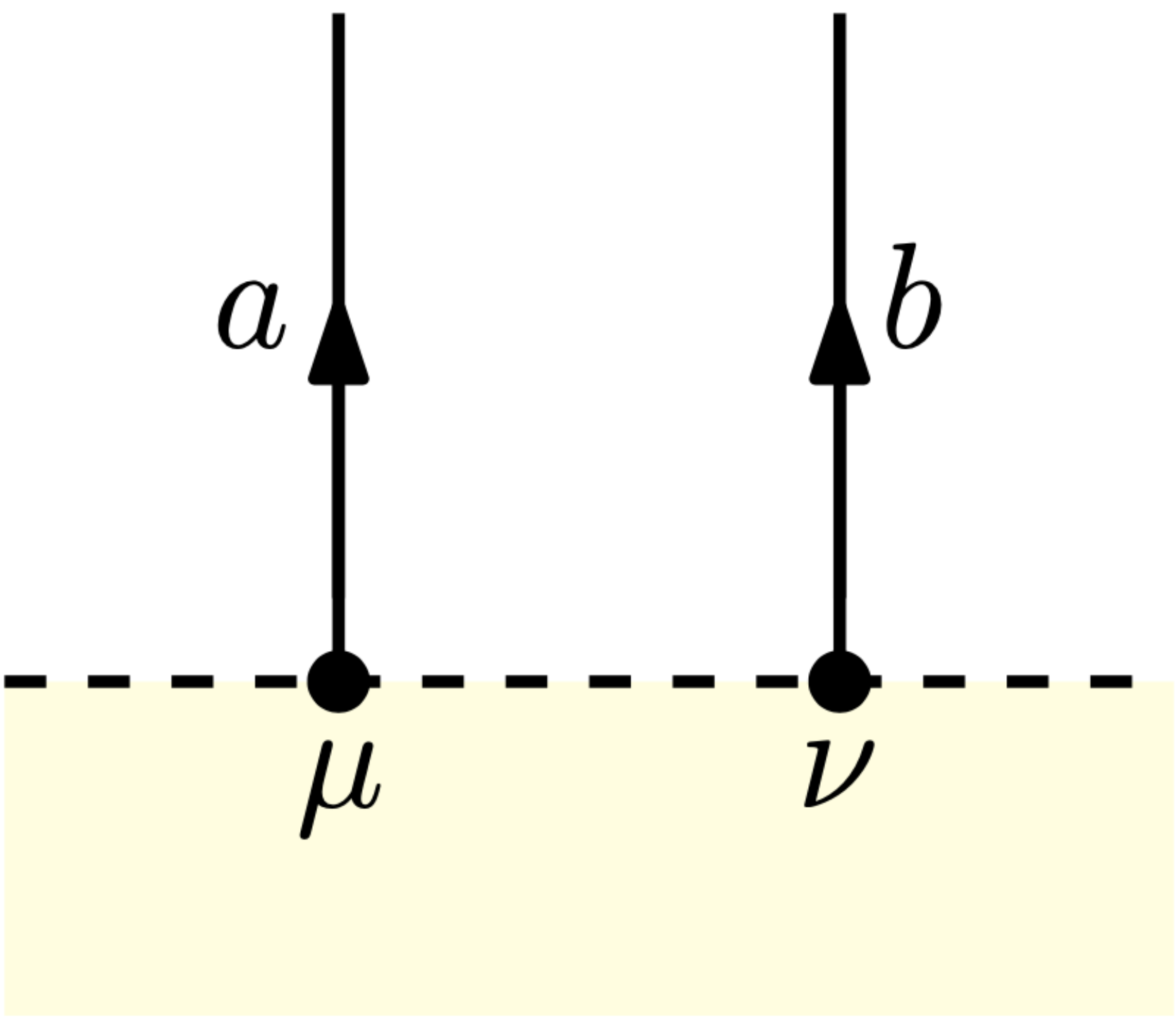} = \sum_{j, (c,\lambda)} M_{(a,\mu)(b,\nu)}^{(c,\lambda),j} \ \includegraphics[scale=0.1, valign=c]{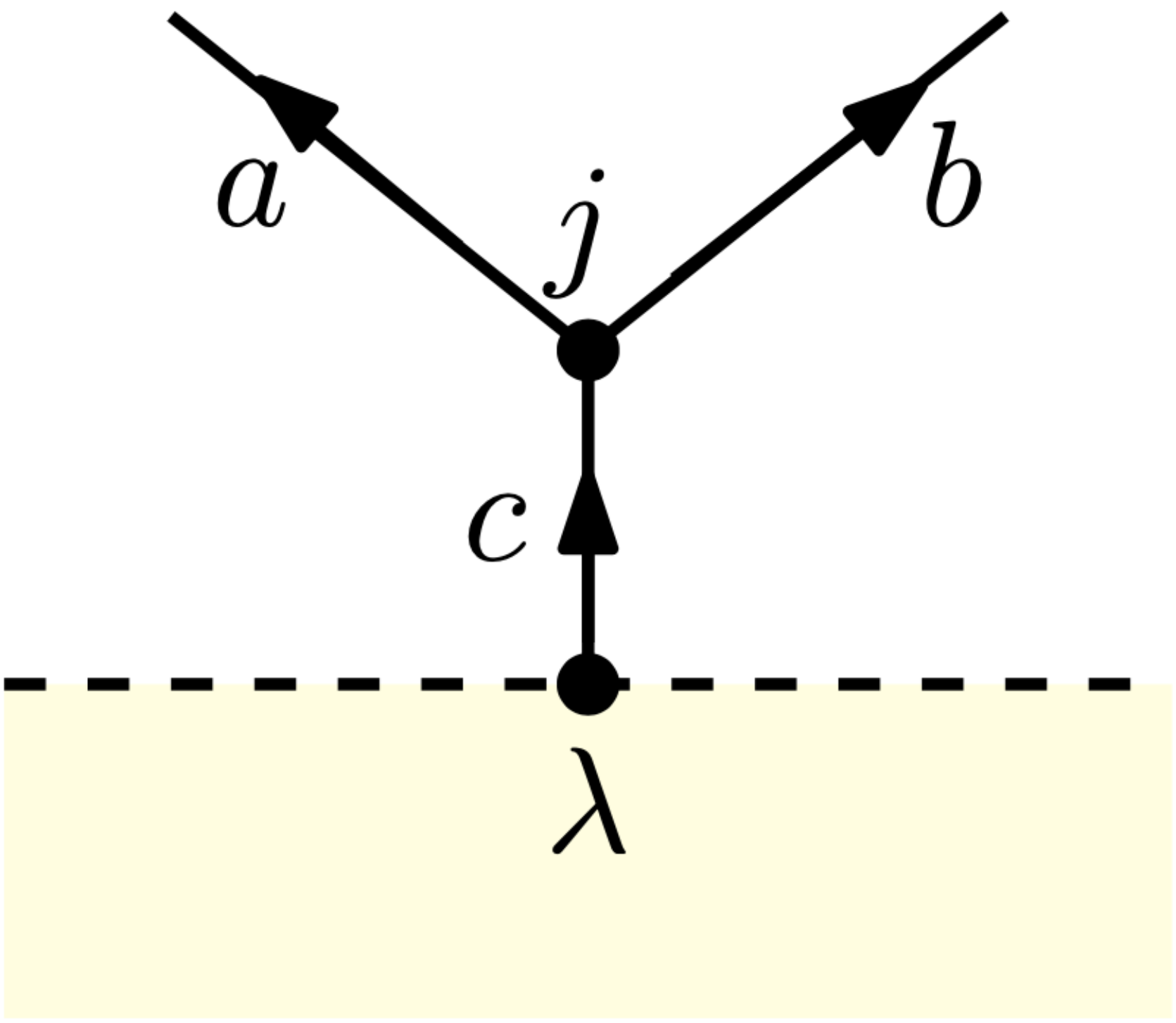}.
\end{equation}

\begin{figure}[t] \hspace{1cm}
        \includegraphics[scale=0.25]{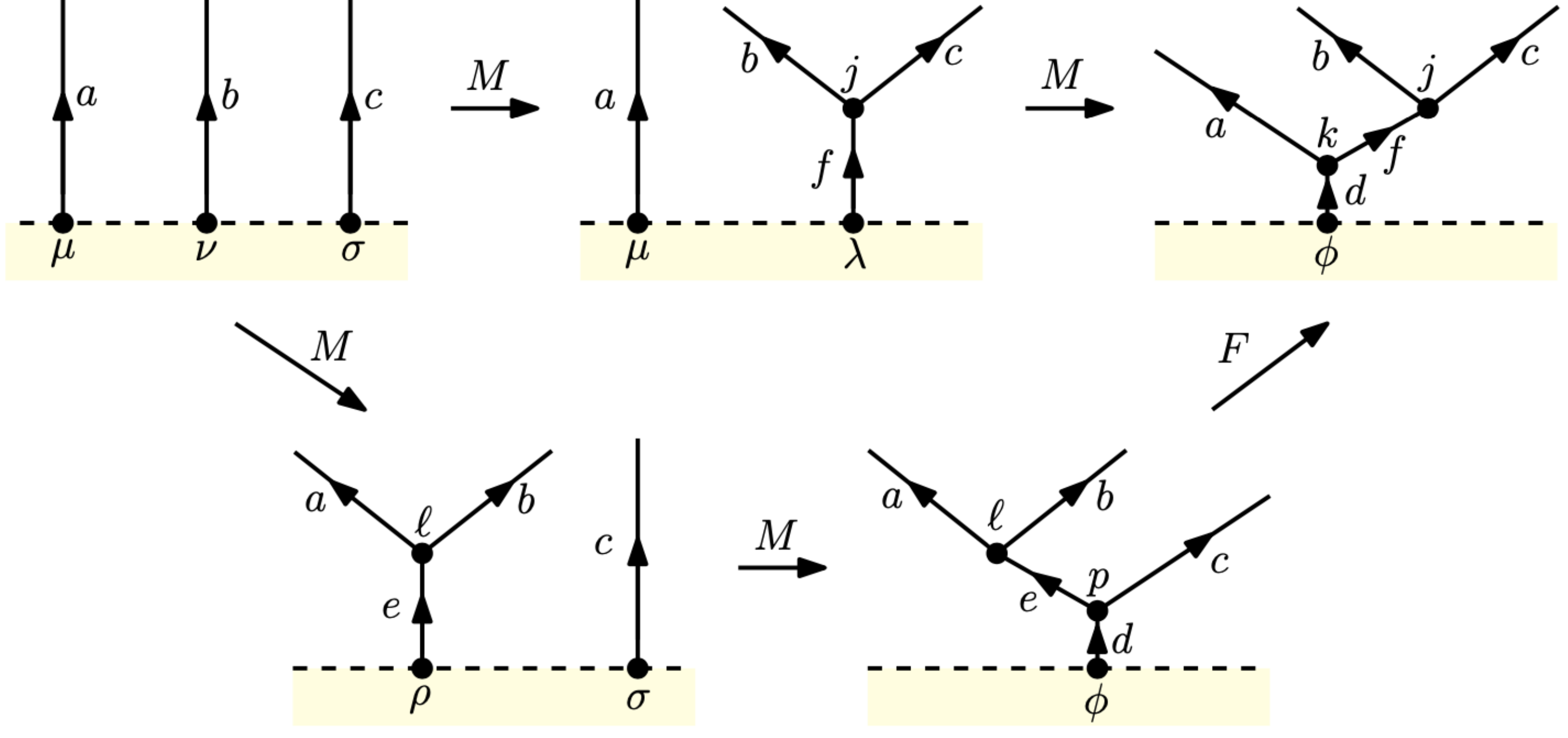} 
        \caption{An associativity condition for the $M$-symbols is established by examining the fusion of boundary junctions in varying orders.} \label{M-Symbol-Associativity}
\end{figure}

One can easily derive constraints that the $M$-symbols must satisfy. For instance, since the boundary is topological one must be able to freely move the endpoints of the anyons around each other so we find the commutativity condition:
\begin{equation} \label{MSymbolCommutativity}
    \includegraphics[scale=0.08, valign=c]{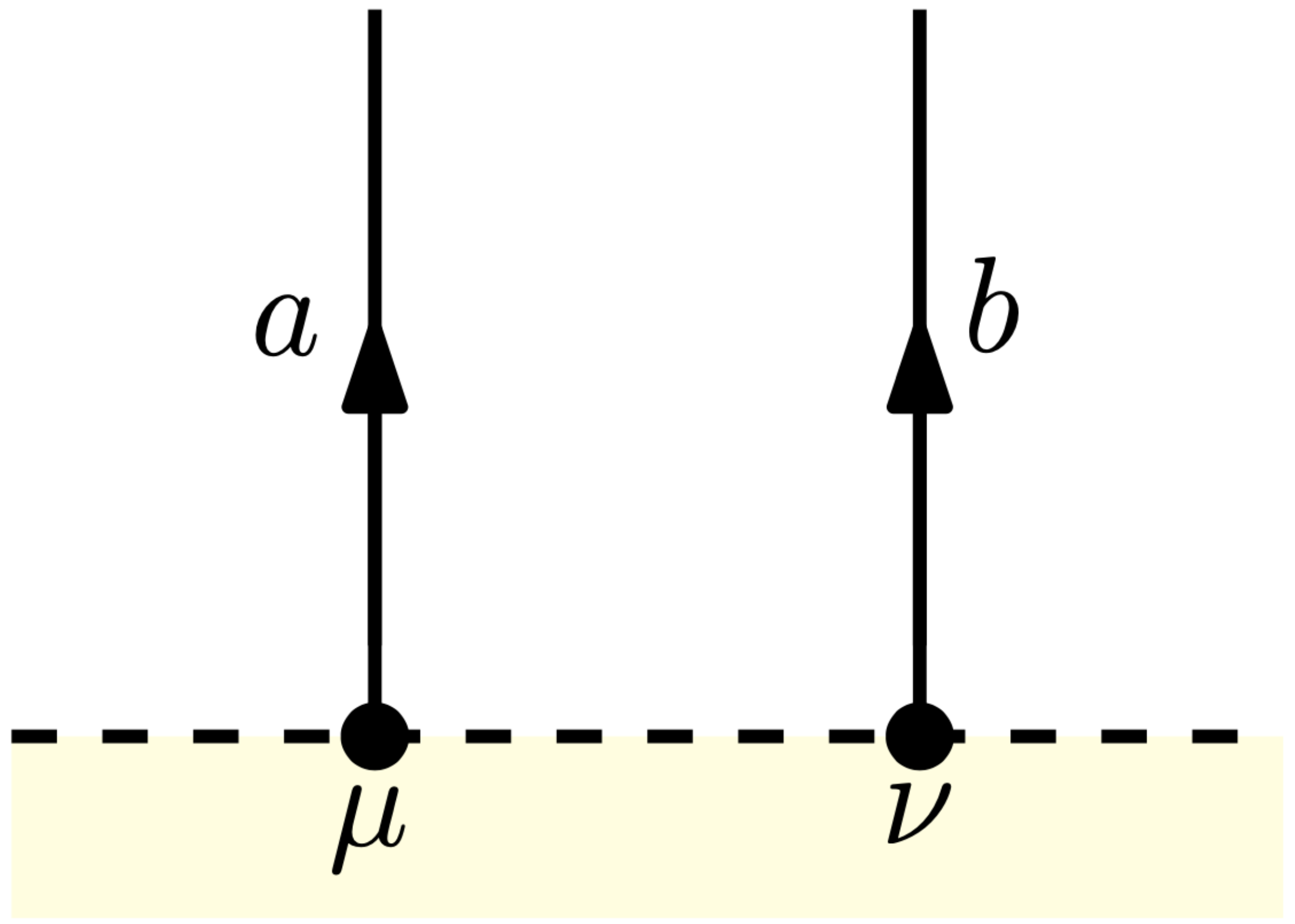} =  \includegraphics[scale=0.085, valign=c]{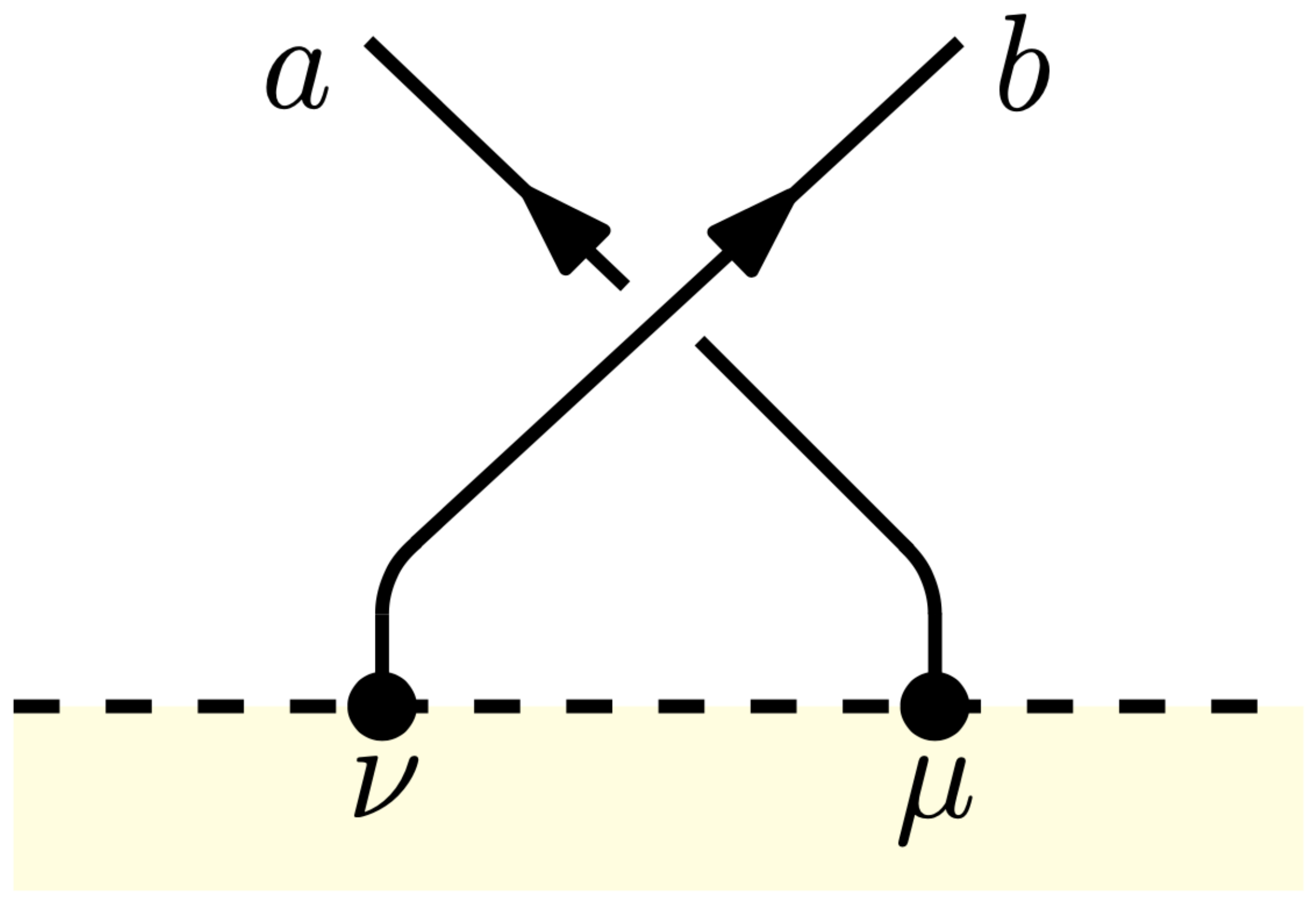}
\end{equation}
Similarly, one can derive an associativity constraint by manipulating three endpoints at the topological boundary as shown in Figure \ref{M-Symbol-Associativity}.

It is now straightforward to see that the constraints thus obtained on the $M$-symbol are equivalent to those satisfied by the (co)multiplication of the Lagrangian algebra characterizing the gapped boundary. See Eqns. \eqref{coassociativity} and \eqref{commutativityLagrangianAlgebra}. Therefore, the corresponding Lagrangian algebra necessarily yields a solution for the $M$-symbols. That is, identifying:
\begin{equation}
    M_{(a,\mu)(b,\nu)}^{(c,\lambda),j} \longleftrightarrow m^{\vee (c, \lambda), j}_{\ \, (a, \mu)(b, \nu)},
\end{equation}
the associativity constraints in Fig. \ref{M-Symbol-Associativity} and the commutativity constraints \eqref{MSymbolCommutativity} are automatically fulfilled. Following this identification, below we will refer to the Frobenius algebra multiplication and the $M$-symbols interchangeably.

\section{Topological Cosets} \label{TopologicalCosets}

Before discussing topological cosets more precisely, let us summarize here the data we seek. Specifically, for a topological coset $G_{1}/H_{\tilde{k}}$ we look for \\
\begin{itemize}
    \item The fusion category of topological line operators living at the boundary of the Chern-Simons theory $G_{1} \times H_{-\tilde{k}}$ for the boundary condition describing the embedding of $H_{\tilde{k}}$ into $G_{1}$. (We place the same boundary condition in both ends of the interval.) \\

    \item The set of local operators, seen as anyons in the 3D theory belonging to the Lagrangian algebra stretching in between both ends of the interval, possibly considering multiplicity, and the OPE between these local operators. \\

    \item The linking action of the line operators on the local operators. \\
\end{itemize}

Let us clarify the relationship between the OPE and the action of lines on local operators.  In any 2D TQFT the set of local operators always admits an idempotent complete basis \cite{Durhuus:1993cq, Sawin:1995rh, Moore:2006dw, Huang:2021zvu}. Up to normalization, this is the statement that there always exists a basis of topological local operators where the following operator product holds:
\begin{equation} \label{IdempotentCompleteBasisEqn}
    \mathcal{O}_{m} \mathcal{O}_{n} = \frac{\delta_{m,n}}{d_{m}} \mathcal{O}_{m}.
\end{equation}
This idempotent complete basis can be obtained by shrinking topological boundary conditions satisfying clustering to a point. See e.g., \cite{Moore:2006dw, Cordova:2024vsq}.

Thus, if we do not consider the action of lines on the local operators, the OPE does not contain any information independent of simply the count of the number of local operators.  By contrast, when we consider the action of lines, we can find an alternative basis of local operators, which instead diagonalize the action of the lines.  It is this latter basis that is obtained, e.g. in the context of 2D QCD, by keeping track of the flow of operators from the UV. In the following, when we present the OPE of local operators, we will therefore work in the basis that diagonalizes the fusion category of line operators, and thus has the interpretation of the leading contribution to the OPE of local operators as they approach the IR fixed point. 

In the context of gapped 2D QCD, an important fact about the 3D construction is that it makes manifest that the topological coset zero-form symmetry is present along the whole RG flow.\footnote{Here, we are deliberately not keeping track of possible accidental symmetries which may appear in the strict IR limit but do not extend to the entire RG trajectory.} See Appendix \ref{CircleInterval2DQCD}. Then, the assumption that the IR is given by the topological coset in Figure \ref{GoverKCoset} with the same topological coset boundary conditions on the left and on the right means that the topological coset symmetry acts regularly on the vacua of the theory \cite{Huang:2021zvu, Zhang:2023wlu}. In physical terms, this means that the fusion category symmetry arising from the coset construction is fully spontaneously broken at long distances. Note that this is not, in general, the full symmetry of the flow, but rather a subset that is intrinsically defined by the 3D construction.\footnote{Alternatively, as pointed out in \cite{Delmastro:2022prj}, one can also see directly in 2D that the coset chiral algebra is preserved along the flow by analysis of the 2D Hamiltonian \cite{Kutasov:1994xq, Delmastro:2021otj}.} Hence, the set of vacua, labeled by topological point operators, is in one to one correspondence with the simple lines in coset zero-form symmetry.\footnote{Mathematically, if $\mathcal{F}_{G_{1}/H_{\tilde{k}}}$ is the fusion category describing the topological coset symmetry, then the statement is that the IR is described by a regular $\mathcal{F}_{G_{1}/H_{\tilde{k}}}$-symmetric 2D TQFT.} In practice, this means that the topological lines of the topological coset act over the idempotent complete basis as
\begin{equation} \label{nondiagonalactionoflines}
    m (\mathcal{O}_{n}) = \sum_{p} \hat{N}^{p}_{mn} \mathcal{O}_{p},
\end{equation}
where $m,n,p,\ldots$ and $\hat{N}^{p}_{mn}$ above stands for the topological lines and their fusion coefficients in the 2D theory respectively, and $d_{m}$ stands for the quantum dimension of the $m$-th line. \footnote{Here we have specialized the discussion to the case where the 2D theory is obtained upon interval compactification of a 3D TQFT with the same left and right topological boundary conditions, as in our current interest of topological cosets depicted in Figure \ref{GoverKCoset}, and the expressions are more general when the two topological boundary conditions are different. See \cite{Huang:2021zvu} for more details.} The notation $m(\mathcal{O}_{n})$ stands for the action of the $m$-th line over the local operators $\mathcal{O}_{n}$ in the idempotent complete basis. Diagrammatically:
\begin{equation}
    \includegraphics[scale=1.2, valign=c]{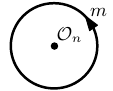} = \sum_{p} \hat{N}^{p}_{mn} \ \mathcal{O}_{p} \hspace{-0.7cm} \includegraphics[scale=1., valign=c]{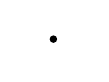}  \ .
\end{equation}
We will see many concrete examples of these equations below.

We now move to a detailed discussion of topological cosets. The coset construction of 2D CFT \cite{Goddard:1984vk, Goddard:1986ee} instructs us that a 2D CFT denoted $G_{k}/H_{\tilde{k}}$ can be obtained by decomposing the characters of the $G_{k}$ affine Lie algebra into those of the $H_{\tilde{k}}$ affine Lie algebra:
\begin{equation}
    \chi^{G_{k}}_{\Lambda}(q) = \sum_{\lambda} b_{(\Lambda, \lambda)}(q) \, \chi^{H_{\tilde{k}}}_{\lambda}(q),
\end{equation}
where $\Lambda$ and $\lambda$ stand for the integrable representations of $G_{k}$ and $H_{\tilde{k}}$ respectively, and $q = e^{2 \pi i \tau}$ with $\tau$ the modular parameter as usual. The quantities $b_{(\Lambda, \lambda)}(q)$ are known as branching functions and describe the spectrum of local operators in the coset 2D CFT. More precisely, the torus partition function for the coset is given by \cite{Goddard:1986ee, Gawedzki:1988nj, Gawedzki:1988hq, Witten:1991mm, hori1994global}:
\begin{equation}
    Z_{G_{k}/H_{\tilde{k}}} \, (T^{2}) = \sum_{\Lambda,\lambda} |b_{\Lambda}^{\lambda}(q)|^{2}, \label{cosetpartitionfunction}
\end{equation}
where we assume that the coset theory we are considering corresponds to that of a diagonal modular invariant. The central charge of the coset theory is
\begin{equation}
    c_{G_{k}/H_{\tilde{k}}} = c_{G_{k}} - c_{H_{\tilde{k}}}.
\end{equation}

The class of cosets we will be interested in arise when the branching rules consist only of non-negative integers, with no non-trivial $q$-expansion:
\begin{equation} \label{ConformalEmbeddingsBranching}
    \chi^{G_{k}}_{\Lambda}(q) = \sum_{\lambda} b_{(\Lambda, \lambda)} \, \chi^{H_{\tilde{k}}}_{\lambda}(q), \quad b_{(\Lambda, \lambda)}  \in \mathbb{N}.
\end{equation}
Then, the coset theory has no excited states and all the states are vacua, or topological local operators of the theory. Following \cite{Delmastro:2021otj}, therefore, we refer to this type of coset as \textit{topological coset}. It also follows that the central charge of the coset vanishes $c_{G_{k}/H_{\tilde{k}}} = 0$, in which case the embedding of affine Lie algebras is known as a conformal embedding. As reviewed in \cite{DiFrancesco:1997nk}, unless $H_{\tilde{k}} = G_{k}$, conformal embeddings can only occur when the numerator has level $k=1$. Our interest in the following will be in the latter case, so we set $k=1$ from now on.

It is well-known that a conformal embedding $H_{\tilde{k}} \hookrightarrow G_{1}$ gives rise to an extension of the $H_{\tilde{k}}$ chiral algebra into $G_{1}$ \cite{BOUWKNEGT1987359, Moore:1988ss, DiFrancesco:1997nk}, and the branching rules express how the integrable representations of $G_{1}$ are given from those of $H_{\tilde{k}}$ upon extension. In turn, we know from \cite{Huang:2014ixa} that the notions of extension of a chiral algebra $V$ and that of gauging an algebra in $V$\footnote{More precisely, gauging an algebra in the category of $V$-modules.} are equivalent. In our context, this means that a gaugable algebra $\mathcal{B}$ (in the sense explained in the previous subsection) exists in $H_{\tilde{k}}$ such that
\begin{equation}
    H_{\tilde{k}} / \mathcal{B} = G_{1},
\end{equation}
and the integrable representations of $G_{1}$ are constructed from $H_{\tilde{k}}$ from \eqref{ConformalEmbeddingsBranching}. Interpreting this from the 3D TQFT point of view, we see that this observation leads to a topological interface separating $H_{\tilde{k}}$ and $G_{1}$ with the line operators connected by a junction(s) at the interface according to \eqref{ConformalEmbeddingsBranching}. Upon folding, we conclude that the branching rules induce a Lagrangian algebra object in $G_{1} \times H_{-\tilde{k}}$ given by:
\begin{equation} \label{TopologicalCosetLagrangian}
    \mathcal{A} = \bigoplus_{(\Lambda, \lambda)} b_{(\Lambda, \lambda)} (\Lambda, \lambda_{\mathrm{rev}}),
\end{equation}
where $\lambda_{\mathrm{rev}}$ stands for the representation $\lambda$ in \eqref{ConformalEmbeddingsBranching}, but in the theory $H_{-\tilde{k}}$ after orientation-reversal of $H_{\tilde{k}}$. We will see many examples of this Lagrangian algebra below. As a quick check, notice that the anyons $(\Lambda, \lambda_{\mathrm{rev}})$ on the right-hand side are always bosons, since in order for \eqref{ConformalEmbeddingsBranching} to hold, the conformal weights of the primaries on the left and right-hand sides must be equal mod 1. The conclusion then follows from the standard relationship between the topological spins and the conformal weights (see Eqn. \eqref{TopologicalTwist}).

In passing, notice that the partition function \eqref{cosetpartitionfunction} for a topological coset arises naturally in terms of the Lagrangian algebra \eqref{TopologicalCosetLagrangian}. Indeed, as recalled in the Introduction, each contribution to the partition function arises from an anyon stretching in between a left and a right boundary. In our construction, each simple anyon $(\Lambda, \lambda_{\mathrm{rev}})$ allows for $b_{(\Lambda, \lambda)}$ topological endpoints at the gapped boundary given by \eqref{TopologicalCosetLagrangian}. Setting the same boundary condition at the left and at the right, we see that the anyon $(\Lambda, \lambda_{\mathrm{rev}})$ contributes $|b_{(\Lambda, \lambda)}|^{2}$ units to the partition function, thus reproducing the expected result \eqref{cosetpartitionfunction} overall.

Once we collect all the previous facts, we see that the problem of describing topological cosets is that of determining the boundary theory for the coset 3D topological order $G_{1} \times H_{-\tilde{k}}$, which always has a gapped boundary determined by the branching rules of the conformal embedding. Now that we know that topological cosets imply the existence of a Lagrangian algebra, we must move on to describing the spectrum of line and point operators in the 2D theory.

\subsection*{Point Operators}

A 2D topological local operator is constructed from the 3D bulk by stretching a boson belonging to the Lagrangian algebra from one boundary to the other, possibly with different junctions at both endpoints, as follows:
\begin{align} \label{TopologicalLocalOperator-Boundary}
    \phi_{(a; \mu_{1}, \mu_{2})} \coloneqq \ \includegraphics[scale=0.125, valign=c]{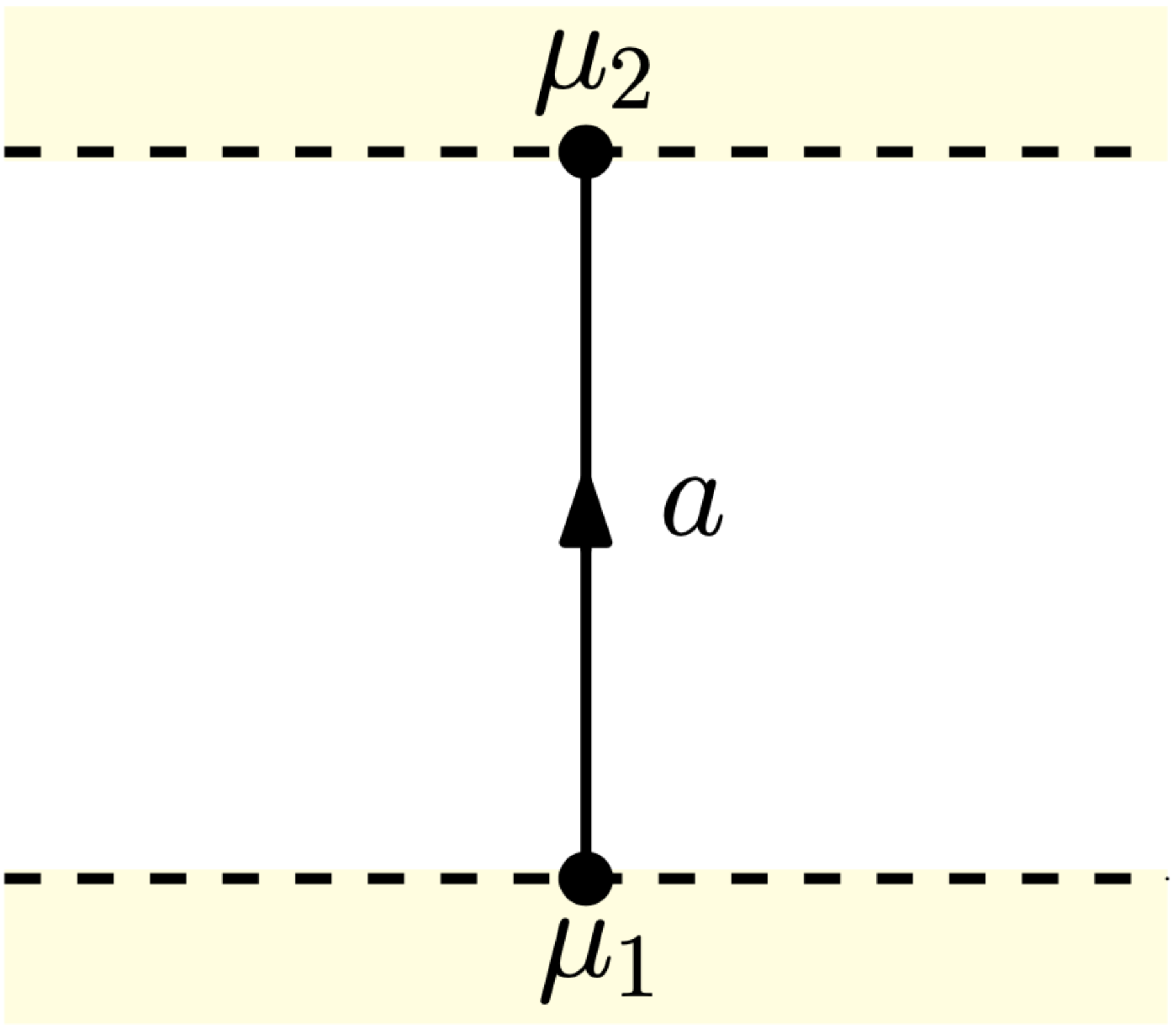},
\end{align}
where in a topological coset, the space of anyons that can end on the boundary are given by the $(\Lambda, \lambda_{\mathrm{rev}})$ in \eqref{TopologicalCosetLagrangian}, with $b_{(\Lambda, \lambda)}$ possible junctions at the boundary. The number of local operators is then $\sum_{\Lambda, \lambda} b^{2}_{\Lambda, \lambda}$. The OPE coefficients for the topological local operators in the 2D TQFT must be supported by the Lagrangian algebra (co)multiplication/M-symbols, since one has the following manipulations:
\begin{align} 
    \includegraphics[scale=0.15, valign=c]{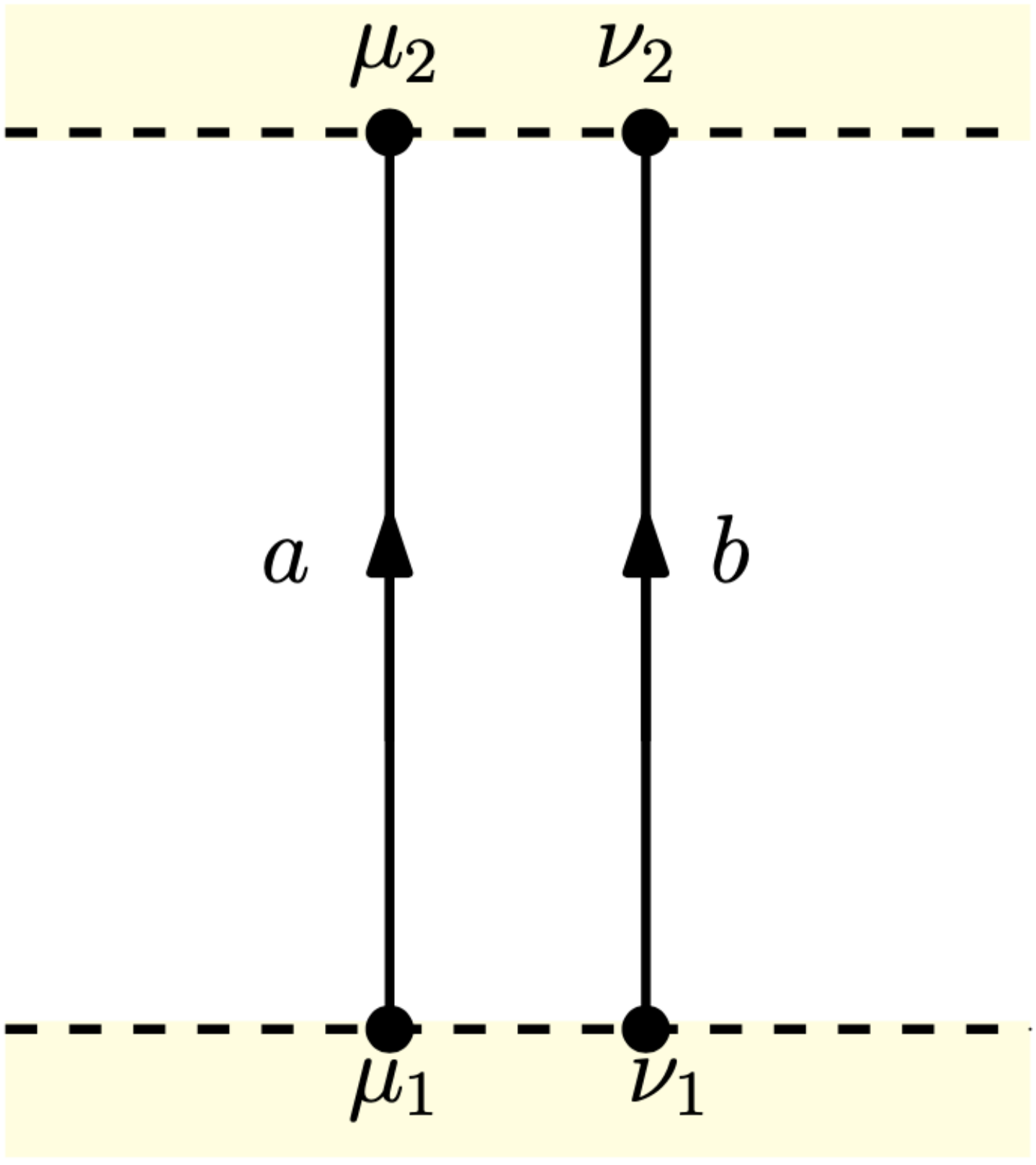} &= \sum_{j_{1}, j_{2}, (c,\sigma_{1}), (d,\sigma_{2})} m^{\vee (c,\sigma_{1}), j_{1}}_{\ \, (a,\mu_{1})(b, \nu_{1})} \, m^{(d,\sigma_{2}), j_{2}}_{(a,\mu_{2})(b, \nu_{2})}  \ \includegraphics[scale=0.15, valign=c]{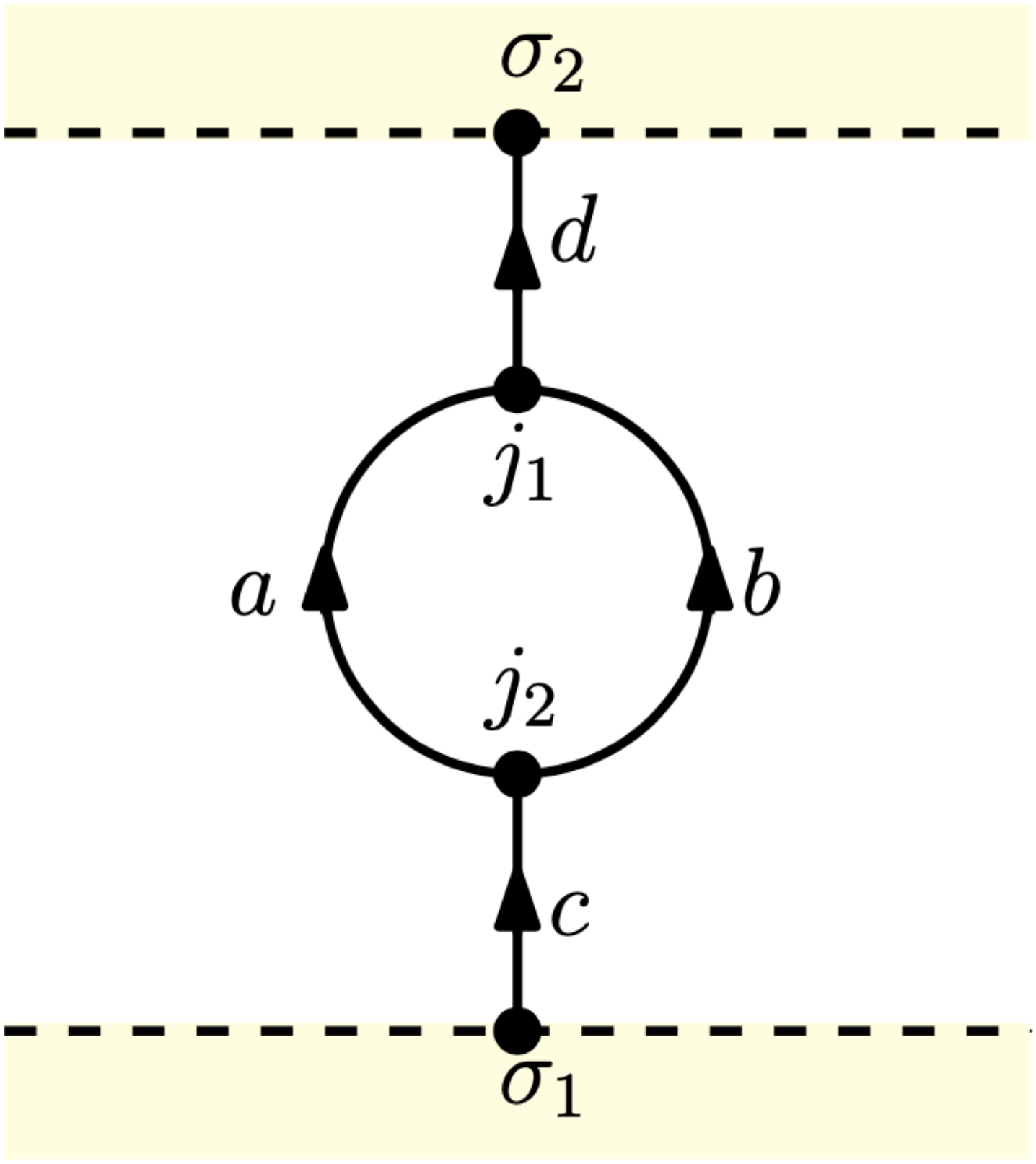} \nonumber \\[0.3cm]  & \hspace{-0.2cm} = \sum_{j_{1}, (c,\sigma_{1},\sigma_{2})} \sqrt{\frac{d_{a} d_{b}}{d_{c}}} \, m^{\vee (c,\sigma_{1}), j_{1}}_{\ \, (a,\mu_{1})(b, \nu_{1})} \, m^{(c,\sigma_{2}), j_{1}}_{(a,\mu_{2})(b, \nu_{2})}  \ 
    \includegraphics[scale=0.15, valign=c]{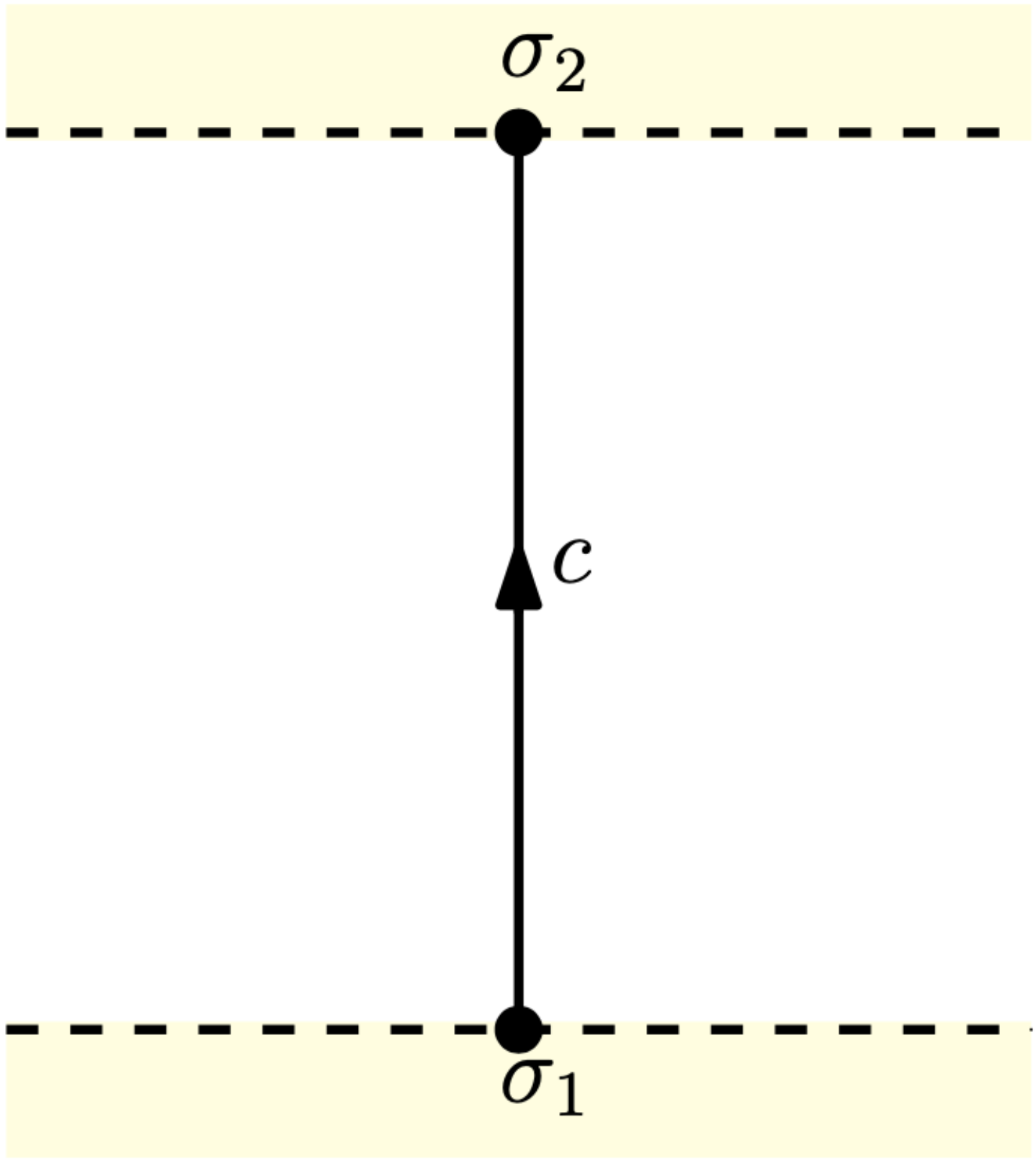}.
\end{align}
In particular, notice that the multiplication of the Lagrangian algebra instructs one to consider in the fusion of bulk lines only junctions of the simple anyons belonging to the Lagrangian algebra, as one would have intuitively expected. Thus, one finds:
\begin{equation} \label{LocalOpsOPE}
    \phi_{(a; \mu_{1}, \mu_{2})} \boldsymbol{\cdot} \phi_{(b; \nu_{1}, \nu_{2})} = \sum_{(c; \sigma_{1}, \sigma_{2})} \mathcal{N}_{(a; \mu_{1}, \mu_{2}), (b; \nu_{1}, \nu_{2})}^{(c, \sigma_{1}, \sigma_{2})} \, \phi_{(c; \sigma_{1}, \sigma_{2})},
\end{equation}
where
\begin{equation} \label{OPEcoeffs}
    \mathcal{N}_{(a; \mu_{1}, \mu_{2}), (b; \nu_{1}, \nu_{2})}^{(c, \sigma_{1}, \sigma_{2})} \coloneqq \sqrt{\frac{d_{a} d_{b}}{d_{c}}} \sum_{j_{1}} \  m^{\vee (c,\sigma_{1}), j_{1}}_{\ \, (a,\mu_{1})(b, \nu_{1})} \, m^{(c,\sigma_{2}), j_{1}}_{(a,\mu_{2})(b, \nu_{2})}.
\end{equation}
In principle, this expression solves the general problem of determining the OPE coefficients $\mathcal{N}$ of the topological local operators (in the basis that diagonalizes the action of the lines) in terms of the data that characterizes the topological boundary.\footnote{Notice the resemblance with the fact that the multiplication in a symmetric commutative special Frobenius algebra $\mathcal{E}$ in a RCFT determines the boundary OPE coefficients of the boundary operators living in the conformal boundary conditions associated with the $\mathcal{E}$-modular invariant \cite{Fuchs:2002cm}.} As we saw above, for a topological coset the Lagrangian algebra object and the spectrum of endpoints is moreover determined by the branching rules of the conformal embedding. In particular, notice that the condition \eqref{NormSpecial} translates to the fact that the coefficients satisfy
\begin{equation} \label{SpecialCondition}
\sum_{(a; \mu_{1}, \mu_{2}), \, (b; \nu_{1}, \nu_{2})} \mathcal{N}_{(a; \mu_{1}, \mu_{2}), (b; \nu_{1}, \nu_{2})}^{(c, \sigma_{1}, \sigma_{2})} = \mathrm{dim}(\mathcal{A}).
\end{equation}

In order to illustrate the general solution, below we will study a couple of examples where the coefficients can be directly obtained from Lagrangian algebra multiplications. Of course, in practice, we may use as many allowed constraints to solve for the OPE coefficients as we wish. Below, we will also see examples where the precise coefficients will be determined from the condition \eqref{SpecialCondition} along with the constraints of commutativity and associativity of the OPE of local operators.

\subsection*{Line Operators}

\begin{figure}[t] \hspace{2.5cm}
\centering
\begin{subfigure}{.45\textwidth}
  \includegraphics[scale=1.24]{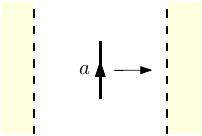}
\end{subfigure}%
\hspace{-2cm}
\begin{subfigure}{.45\textwidth}
  \includegraphics[scale=1.24]{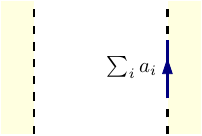}
\end{subfigure}
\caption{A simple line $a$ in bulk generically becomes non-simple when pushed to a topological boundary: $a \to \sum_{i} a_{i}$.}
\label{Lines-to-Boundary} 
\end{figure}

We move on now to discuss the topological line operators of the topological coset. The fusion category of line operators of the 2D theory can be constructed starting from the line operators in bulk and pushing them to the boundary. See Figure \ref{Lines-to-Boundary}. Mathematically, the precise way to calculate the result is to construct the category of $\mathcal{A}$-modules, or the Karoubi envelope of the quotient category $\mathcal{C}/\mathcal{A}$ for the bulk MTC $\mathcal{C}$. This construction is rather abstract, so we present the rigorous definitions in Appendix \ref{KaroubiEnvelope} and here we content ourselves with outlining the more heuristic approach of \cite{Bais:2008ni}. In the following, we denote by $\otimes$ and $\oplus$ the fusion and direct sum of lines in the bulk MTC respectively, while we use $\times$ and + for the fusion and direct sum of topological lines at the boundary respectively.

As stated previously, to find the line operators of the 2D theory we take a simple anyon $a$ in bulk and push it to the boundary. In general, when we do this a simple anyon in bulk becomes non-simple at the boundary. This is, the bulk simple anyon $a$ ``splits'' in terms of many boundary components:
\begin{equation}
    a \longrightarrow \sum_{i} z_{i}^{a}a_{i}, \quad z_{i}^{a} \in \mathbb{N}, \label{restriction}
\end{equation}
where the subindex $i$ labels the distinct line operators $a_{i}$ that arise when $a$ is pushed to the boundary. The integers $z_{i}^{a}$ label multiplicities with which the $a_{i}$ may appear in the decomposition. On the right-hand side of \eqref{restriction}, it is important to note that the labels $a_{i}$ that correspond to different simple anyons $a$ in the bulk MTC do not always correspond to different simple line operators of the boundary theory. It generically happens that $a_{i} = b_{j}$ for bulk simple anyons $a \neq b$. In other words, not only does there exist a splitting procedure for the bulk anyons at the boundary, but there are also identifications between the boundary labels $a_{i}$.

In practice, one can often determine the splittings and identifications by imposing a set of consistency conditions. Firstly, it is important to observe that if one evaluates a loop of a simple anyon $a$ before and after pushing it to the boundary (as depicted in Figure \ref{Lines-to-Boundary}), one discovers that the quantum dimensions of the bulk lines must be conserved when they are pushed to the boundary:
\begin{equation}
    a \longrightarrow \sum_{i} z_{i}^{a} a_{i} \Longrightarrow d_{a} = \sum_{b} z_{i}^{a} d_{a_{i}}.
\end{equation} 
Thus, splitting is constrained to preserve quantum dimension. 

Secondly, we require that if 
\begin{equation}
    a \longrightarrow \sum_{i} z_{i}^{a}a_{i}, \Longrightarrow  \bar{a} \longrightarrow \sum_{i} z_{i}^{a} \bar{a_{i}}.
\end{equation}
Thus, the conjugate of $a$ splits into the conjugates of the split of $a$.

Thirdly, simple anyons in the Lagrangian algebra $\mathcal{A} = \oplus_{a} n_{a} a$ always split into components that contain the identity line of the boundary theory:
\begin{equation} \label{LagrangianCondensation}
    a \rightarrow n_{a} \, 0 + \cdots.
\end{equation}
In the anyon condensation jargon, these anyons are said to \textit{condense}. As per the discussion in Section \ref{GappedBoundariesandTopologicalCosets}, only bosons can condense. 

Finally, one requires consistency between bulk fusion and splitting. This is, when one gauges non-invertible anyons, the resulting boundary theory has a fusion ring satisfying the standard conditions of associativity, existence of a unique identity line, and existence of unique conjugate representations with a unique way to fuse to the identity line. Consistency between bulk fusion and splitting then means that one has that:
\begin{equation} \label{fusionsplittingconsistency}
    a \otimes b = \bigoplus_{c} N_{ab}^{c} c \Longrightarrow \Big( \sum_{i}z_{i}^{\ a}a_{i} \Big) \times \Big( \sum_{j}z_{j}^{\ b}b_{j} \Big) = \sum_{c,k}N_{ab}^{c}z_{k}^{\ c}c_{k} \, .
\end{equation}
Thus, for instance, one can deduce that an anyon $a$ splits if in the fusion $a \otimes \bar{a}$ more than one identity line appears on the right-hand side after splitting and using \eqref{LagrangianCondensation}. Similarly, identifications can be found by studying $a \otimes b$ and studying the identity lines that appear after splitting. Once one has found the splitting and identifications of the bulk lines into boundary lines, one can furthermore exploit the previous equation and tightly constrain the fusion ring of the boundary lines. In many concrete cases, like the ones studied later in this work, this allows one to find the fusion ring exactly.

Once we have obtained the spectrum of line and point operators following the previous discussion we can discuss the interplay of topological line operators with topological local operators on the topological coset. This is straightforward to do by considering a configuration where a loop wraps a line that ends at the boundary. Requiring that we obtain the same result either by calculating the braiding phase \eqref{BraidingPhase} in bulk first, or by pushing the loop to the boundary first and calculating the action of the boundary lines with the point operators, as in:
\begin{equation}
    \includegraphics[scale=1.2, valign=c]{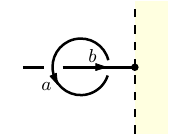} = \includegraphics[scale=1.2, valign=c]{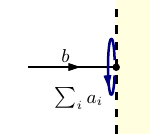}  
\end{equation}
gives a set of equations from which we can readily find the action of the line operators on the point operators in the 2D theory.

\section{Examples} \label{examples}

In previous sections we have outlined the rules to describe a topological coset. In this section we provide concrete examples. One of them is essentially the simplest instance of a topological coset, and can be completely worked out in practice. The other examples are important in that they are (conjecturally) the IR fixed point of a massless 2D QCD theory. Other examples of topological cosets can be found from the table of conformal embeddings in \cite{davydov2013witt}.

In the following, we denote by $\otimes$ and $\oplus$ the fusion and direct sum of lines in the bulk MTC respectively, $\times$ and + for the fusion and direct sum of topological lines at the boundary respectively, and by $\boldsymbol{\cdot}$ the product of topological local operators in the diagonal basis (also called bootstrap basis \cite{Huang:2021zvu}) inherited by the endpoints of the simple anyons in the bulk Lagrangian algebra $\mathcal{A}$. The spectrum and fusion rules of the MTCs used in this section can be obtained from the KAC software program \cite{KAC}.

\subsection{A Pedagogical Example}

\subsubsection*{$SU(3)_{1}/SU(2)_{4}$} \label{PedagogicalExample}

In this subsection we study the topological coset $SU(3)_{1}/SU(2)_{4}$. This is arguably the simplest topological coset, so for the sake of illustration we describe it here completely. The branching rules of this conformal embedding are well-known (see \cite{DiFrancesco:1997nk}) and are given by:
\begin{align}
    & \chi^{SU(3)_{1}}_{\mathbf{1}}(q) = \chi^{SU(2)_{4}}_{0}(q) + \chi^{SU(2)_{4}}_{4}(q) \, , \\[0.3cm]
    & \chi^{SU(3)_{1}}_{\mathbf{3}}(q) = \chi^{SU(2)_{4}}_{2}(q) \, , \\[0.3cm] 
    & \chi^{SU(3)_{1}}_{\mathbf{\bar{3}}}(q) = \chi^{SU(2)_{4}}_{2}(q) \, ,
\end{align}
where we label an integrable representation in $SU(3)_{1}$ by its dimension in boldface and we use the standard indexing $i=0,1,\ldots,k$ to label the $i$-th integrable representation in $SU(2)_{k}$ (see Appendix \ref{MTCdataAppendix} for a more detailed description of the MTC data of $SU(2)_{k}$). 

Recall from the general discussion above that, as expected from a conformal embedding, all the branching functions are non-negative integers, independent of $q$. These non-negative integers are interpreted as a contribution to the spectrum of local operators in the coset theory (see e.g., \cite{Delmastro:2021otj}). On the other hand, from the perspective of the bulk-boundary correspondence, these contributions arise from the allowed endpoints of bosons in $SU(3)_{1} \times SU(2)_{-4}$ at the topological boundary. That is, the branching rules yield a Lagrangian algebra $\mathcal{A}$ for the MTC $ \mathcal{C}= SU(3)_{1} \times SU(2)_{-4}$. Explicitly, the algebra object is given by:
\begin{equation} \label{SU(3)1su2lv4object}
    \mathcal{A} = (\mathbf{1}, 0) \oplus (\mathbf{1}, 4) \oplus (\mathbf{3}, 2) \oplus (\mathbf{\bar{3}}, 2).
\end{equation}
Indeed, it is straightforward to compute the quantum dimensions of the algebra object and the total quantum dimension of $\mathcal{C} = SU(3)_{1} \times SU(2)_{-4}$ (recall Eqn. \eqref{totalquantumdimension}) to check that $$\mathrm{dim}\big(SU(3)_{1} \times SU(2)_{-4} \big) = \mathrm{dim}(\mathcal{A})^{2},$$
as required by the definition of a Lagrangian algebra in Eqn. \eqref{quantumdimensioncondition}.

Although we do not reproduce the calculation here, it is straightforward to use the setup described in Section \ref{GappedBoundariesandTopologicalCosets} to check that gauging $\mathcal{A}$ in $SU(3)_{1} \times SU(2)_{-4}$ in a bulk region of 3D spacetime results in the trivial MTC, thus generating a topological boundary when gauging on half of spacetime.

We move on now to discuss local operators. In the following, as in Section \ref{GappedBoundariesandTopologicalCosets}, we denote a topological local operator at the end of a bulk line $a$ as $\phi_{a}$.\footnote{Notice that in Section \ref{GappedBoundariesandTopologicalCosets}, for generality, we kept track of multiple possible junctions in the subindices. In the following, we will suppress such indices since the junctions we will consider below will have single multiplicity.} The allowed OPE channels for the topological endpoints of bosons in the Lagrangian algebra may be obtained from the corresponding fusion rules in $SU(3)_{1} \times SU(2)_{-4}$ and projecting into the elements of the algebra. Specifically:
\begin{align}
    &(\mathbf{1},4) \otimes (\mathbf{1},4) \longrightarrow (\mathbf{1},0), \quad (\mathbf{1},4) \otimes (\mathbf{3},2) \longrightarrow (\mathbf{3},2), \quad (\mathbf{1},4) \otimes (\mathbf{\bar{3}},2) \longrightarrow (\mathbf{\bar{3}},2), \nonumber \\[0.3cm]
    &(\mathbf{3},2) \otimes (\mathbf{3},2) \longrightarrow (\mathbf{\bar{3}},2), \quad (\mathbf{\bar{3}},2) \otimes (\mathbf{\bar{3}},2) \longrightarrow (\mathbf{3},2), \\[0.3cm]
    &(\mathbf{3},2) \otimes (\mathbf{\bar{3}},2) \longrightarrow (\mathbf{1},0) + (\mathbf{1},4). \nonumber
\end{align}

In order to obtain a well-defined OPE in between the local operators $\phi_{a}$ in the diagonal basis, one must find an associative and commutative product consistent with the bulk Lagrangian algebra and the allowed OPE channels just presented. Indeed, it is straightforward to check that the fusion product of topological local operators given by
\begin{align} \label{CoefsSU31SU24COSET}
    & \phi_{(\mathbf{1},0)} \boldsymbol{\cdot} \phi_{(a,b)} = \phi_{(a,b)}, \ \mathrm{for} \ (a,b) \in A, \nonumber \\[0.3cm]
    & \phi_{(\mathbf{1},4)} \boldsymbol{\cdot} \phi_{(\mathbf{1},4)} = \phi_{(\mathbf{1},0)}, \quad \phi_{(\mathbf{1},4)} \boldsymbol{\cdot} \phi_{(\mathbf{3},2)} = \phi_{(\mathbf{3},2)}, \quad \phi_{(\mathbf{1},4)} \boldsymbol{\cdot} \phi_{(\mathbf{\bar{3}},2)} = \phi_{(\mathbf{\bar{3}},2)},  \\[0.3cm]
    & \phi_{(\mathbf{3},2)} \boldsymbol{\cdot} \phi_{(\mathbf{3},2)} = 
2 \, \phi_{(\bar{\mathbf{3}}, 2)}, \quad \phi_{(\bar{\mathbf{3}}, 2)} \boldsymbol{\cdot} \phi_{(\bar{\mathbf{3}}, 2)} = 2 \, \phi_{(\mathbf{3},2)}, \quad \phi_{(\mathbf{3},2)} \boldsymbol{\cdot} \phi_{(\bar{\mathbf{3}}, 2)} = 2 \, \phi_{(\mathbf{1},0)} + 
 \, 2\phi_{(\mathbf{1},4)}. \nonumber
\end{align}
is commutative, associative, and that \eqref{SpecialCondition} holds. As stressed previously, this topological coset can be worked out in detail, and in particular one can see explicitly that the previous OPE coefficients are supported by appropriate Lagrangian algebra multiplications. Specifically:
\begin{flalign}
    \hspace{1cm}  m_{(\mathbf{1}, 0) (\mathbf{1}, 0) }^{(\mathbf{1}, 0)} & = m_{(\mathbf{1}, 0) (\mathbf{1}, 4) }^{(\mathbf{1}, 4)} = m_{(\mathbf{1}, 0) (\mathbf{3}, 2) }^{(\mathbf{3}, 2)} = m_{(\mathbf{1}, 0) (\bar{\mathbf{3}}, 2) }^{( \bar{\mathbf{3}}, 2)} = 1, && \\[0.3cm]
    \hspace{1cm}  m_{(\mathbf{1}, 4) (\mathbf{1}, 4) }^{(\mathbf{1}, 0)} & = m_{(\mathbf{3}, 2) (\bar{\mathbf{3}}, 2) }^{(\mathbf{1}, 0)} = m_{( \bar{\mathbf{3}},2) (\mathbf{3},2) }^{(\mathbf{1},0)} = 1, && \\[0.3cm]
    \hspace{1cm}  m_{(\mathbf{3}, 2) (\bar{\mathbf{3}}, 2) }^{(\mathbf{1},4)} & = - m_{( \bar{\mathbf{3}}, 2) (\mathbf{3}, 2) }^{(\mathbf{1}, 4)} = 1, && \\[0.3cm]
    \hspace{1cm}  m_{(\mathbf{1}, 4) (\mathbf{3}, 2) }^{(\mathbf{3}, 2)} & = - m_{(\mathbf{3}, 2) (\mathbf{1}, 4) }^{(\mathbf{3}, 2)} = 1, && \\[0.3cm]
    \hspace{1cm}  m_{(\mathbf{1}, 4) (\bar{\mathbf{3}}, 2) }^{(\bar{\mathbf{3}},2)} & = - m_{(\bar{\mathbf{3}},2) (\mathbf{1},4) }^{(\bar{\mathbf{3}},2)} = - 1, && \\[0.3cm]
    \hspace{1cm}  m_{(\mathbf{3}, 2) (\mathbf{3}, 2) }^{(\bar{\mathbf{3}}, 2)} & = 2^{1/4}, && \\[0.3cm]
    \hspace{1cm}  m_{(\bar{\mathbf{3}}, 2) (\bar{\mathbf{3}}, 2) }^{(\mathbf{3}, 2)} & = -2^{1/4}. && 
\end{flalign}
This can be checked directly by plugging the multiplications into the (co)associativity conditions \eqref{diagrammatic(co)associativity}, or Eqn. \eqref{coassociativity} in components. Using \eqref{LocalOpsOPE} and \eqref{OPEcoeffs}, it is then straightforward to reproduce the coefficients in \eqref{CoefsSU31SU24COSET}.

Having discussed the point operators in the theory, we may now proceed to examine the spectrum of line operators. Given the limited number of objects in this example, it is possible to be particularly explicit about the computation based on Karoubian envelopes discussed in Appendix \ref{KaroubiEnvelope}. Calculating a few fusions with the Lagrangian algebra object \eqref{SU(3)1su2lv4object} will illustrate the procedure. Consider:
\begin{align}
    \mathcal{A} \otimes (\mathbf{1},0) & = (\mathbf{1},0) \oplus (\mathbf{1},4) \oplus (\mathbf{3},2) \oplus (\bar{\mathbf{3}},2),  \label{examplefusion1} \\[0.3cm]
    \mathcal{A} \otimes (\mathbf{1},1) & = (\mathbf{1},1) \oplus (\mathbf{1},3) \oplus (\mathbf{3},1) \oplus (\mathbf{3},3) \oplus (\bar{\mathbf{3}},1) \oplus (\bar{\mathbf{3}},3), \label{examplefusion2} \\[0.3cm]
    \mathcal{A} \otimes (\mathbf{1},2) & = 2(\mathbf{1},2) \oplus (\mathbf{3},0) \oplus (\mathbf{3},2) \oplus (\mathbf{3},4) \oplus (\bar{\mathbf{3}},0) \oplus (\bar{\mathbf{3}},2) \oplus (\bar{\mathbf{3}},4), \label{examplefusion3} \\[0.3cm]
    \mathcal{A} \otimes (\mathbf{3},0) & = (\mathbf{3},0) \oplus (\mathbf{3},4) \oplus (\bar{\mathbf{3}},2) \oplus (\mathbf{1},2), \label{examplefusion4} \\[0.3cm]
    \mathcal{A} \otimes (\bar{\mathbf{3}},0) & = (\bar{\mathbf{3}},0) \oplus (\bar{\mathbf{3}},4) \oplus (\mathbf{3},2) \oplus (\mathbf{1},2). \label{examplefusion5}
\end{align}
From this we deduce, for example:
\begin{align}
    \mathrm{Hom}((\mathbf{1},2), \mathcal{A} \otimes (\mathbf{1},2)) = \mathbb{C}^{2}, \\[0.3cm]
    \mathrm{Hom}((\mathbf{1},2), \mathcal{A} \otimes (\mathbf{3},0)) = \mathbb{C}, \\[0.3cm]
    \mathrm{Hom}((\mathbf{1},2), \mathcal{A} \otimes (\bar{\mathbf{3}},0)) = \mathbb{C}, \\[0.3cm]
     \mathrm{Hom}((\mathbf{3},0), \mathcal{A} \otimes (\mathbf{3},0)) = \mathbb{C}, \\[0.3cm]
    \mathrm{Hom}((\bar{\mathbf{3}},0), \mathcal{A} \otimes (\bar{\mathbf{3}},0)) = \mathbb{C}.
\end{align}
A few consequences follow from this calculation. For example, $\mathrm{Hom}((\mathbf{1},2), \mathcal{A} \otimes (\mathbf{1},2)) = \mathbb{C}^{2}$ implies that $(\mathbf{1},2)$ splits into two lines at the boundary. Similarly, $(\mathbf{3},0)$ and $(\bar{\mathbf{3}},0)$ do not split. The fact that $\mathrm{Hom}((\mathbf{1},2), \mathcal{A} \otimes (\mathbf{3},0)) = \mathbb{C}$ means that one of the splitting channels of $(\mathbf{1},2)$ coincides with the line to which $(\mathbf{3},0)$ descends at the boundary. We denote this common line at the boundary $\mathbf{3}$. A similar statement holds for $(\bar{\mathbf{3}},0)$, whose boundary line is denoted as $\bar{\mathbf{3}}$. However, from \eqref{examplefusion4} and \eqref{examplefusion5}, we deduce that $(\mathbf{3},0)$ and $(\bar{\mathbf{3}},0)$ share no common channel at the boundary, so they descend into different boundary lines: $\bar{\mathbf{3}} \neq \mathbf{3}$. A similar procedure can be performed for the rest of the lines, revealing that there is only one further additional line at the boundary, denoted $\mathcal{N}$. The overall result for all the lines and what they descend to at the boundary is presented in Table \ref{SplittingSU(2)LV4}.

\begin{table}
\centering
\begin{tabular}{ |c| }
 \hline
\hspace{-1.5cm} $(\mathbf{1},0)$, $(\mathbf{1},4) \to \mathbf{1}$ \quad \quad  $(\mathbf{1},2) \to \mathbf{3} + \bar{\mathbf{3}}$  \\ 
\hspace{-1.5cm} $(\mathbf{3},0)$, $(\mathbf{3},4) \to \mathbf{3}$ \quad \quad  $(\mathbf{3},2) \to \mathbf{1} + \bar{\mathbf{3}}$  \\ 
\hspace{-1.5cm} $(\bar{\mathbf{3}},0)$, $(\bar{\mathbf{3}},4) \to \bar{\mathbf{3}}$ \quad \quad $(\bar{\mathbf{3}},2) \to \mathbf{1} + \mathbf{3}$ \\ 
$(\mathbf{1},1)$, $(\mathbf{3},1)$, $(\bar{\mathbf{3}},1)$, $(\mathbf{1},3)$, $(\mathbf{3},3)$, $(\bar{\mathbf{3}},3) \to \mathcal{N}$  \\ 
 \hline
\end{tabular}
\caption{Descent of the bulk topological lines of $SU(3)_{1} \times SU(2)_{-4}$ (left of the arrows) to the topological boundary. The resulting lines (right of the arrows) describe the topological line operators in the $SU(3)_{1}/SU(2)_{4}$ topological coset.}
\label{SplittingSU(2)LV4}
\end{table}

The fusion rules of the lines at the boundary can be found from the discussion around Eqn. \eqref{fusionsplittingconsistency}. Indeed, from the spectrum already found, we see that the only consistent fusion ring is given by
\begin{equation}
    \mathbf{3} \times \mathbf{3} = \bar{\mathbf{3}}, \quad \mathbf{3} \times \bar{\mathbf{3}} = \bar{\mathbf{3}} \times \mathbf{3} = \mathbf{1}, \quad \bar{\mathbf{3}} \times \bar{\mathbf{3}} = \mathbf{3},
\end{equation}
\begin{equation}
    \mathbf{3} \times \mathcal{N} = \mathcal{N} \times \mathbf{3} = \mathcal{N}, \quad \bar{\mathbf{3}} \times  
 \mathcal{N} =  \mathcal{N} \times \bar{\mathbf{3}} = \mathcal{N}, \quad \mathcal{N} \times \mathcal{N} = \mathbf{1} + \mathbf{3} + \bar{\mathbf{3}},
\end{equation}
which we recognize as the fusion ring of $\mathbb{Z}_{3}$ Tambara-Yamagami. Actually, from the simplicity of the topological coset we can further identify the full fusion category as the $\mathbb{Z}_{3}$ Tambara-Yamagami fusion category with non-trivial bicharacter and non-trivial Frobenius-Schur indicator. See, for instance \cite{Lin:2020bak}, where it is easy to identify that the Drinfeld center of the fusion category just mentioned indeed coincides with $SU(3)_{1} \times SU(2)_{-4}$.

Finally, to check that everything is in order we can check the existence of an idempotent complete basis where the line operators act according to \eqref{nondiagonalactionoflines}. This can be straightforwardly be checked to be correct, with:
\begin{align}
    \mathcal{O}_{0} &= \frac{1}{6}(\phi_{(\mathbf{1},0)} + \phi_{(\mathbf{1},4)} + \phi_{(\mathbf{3},2)} + \phi_{(\bar{\mathbf{3}},2)}), \\[0.3cm]
    \mathcal{O}_{\mathbf{3}} &= \frac{1}{6}( \phi_{(\mathbf{1},0)} + \phi_{(\mathbf{1},4)} + \omega \phi_{(\mathbf{3},2)} + \omega^{2}\phi_{(\bar{\mathbf{3}},2)}), \\[0.3cm]
    \mathcal{O}_{\bar{\mathbf{3}}} &= \frac{1}{6}(\phi_{(\mathbf{1},0)} + \phi_{(\mathbf{1},4)} + \omega^{2}\phi_{(\mathbf{3},2)} + \omega \phi_{(\bar{\mathbf{3}},2)}), \\[0.3cm]
    \mathcal{O}_{\mathcal{N}} &= \frac{1}{6}(\sqrt{3}\phi_{(\mathbf{1},0)} - \sqrt{3} \phi_{(\mathbf{1},4)}).
\end{align}

In upcoming examples the calculations are essentially equivalent to those displayed in this example, but more complicated to showcase explicitly because of the higher number of objects involved. Thus, we will present results without presenting detailed calculations, leaving implicit that they proceed in the same way as presented in Section \ref{GappedBoundariesandTopologicalCosets} and this example.

\subsection{QCD-like Examples}

\subsubsection{$Spin(5)_{1}/SU(2)_{10}$} 

We now move on to describe a topological coset describing the IR fixed point of a 2D QCD theory. Specifically, we study the $Spin(5)_{1}/SU(2)_{10}$ topological coset, describing the IR fixed point of (bosonized) $SU(2)$ Yang-Mills theory with fermions in the spin 2 representation. The branching rules of the conformal embedding are given by:
\begin{align}
    & \chi^{Spin(5)_{1}}_{0}(q) = \chi^{SU(2)_{10}}_{0}(q) + \chi^{SU(2)_{10}}_{6}(q) \, , \\[0.3cm]
    & \chi^{Spin(5)_{1}}_{v}(q) = \chi^{SU(2)_{10}}_{4}(q) + \chi^{SU(2)_{10}}_{10}(q) \, , \\[0.3cm] 
    & \chi^{Spin(5)_{1}}_{\sigma}(q) = \chi^{SU(2)_{10}}_{3}(q) + \chi^{SU(2)_{10}}_{7}(q) \, .
\end{align}
As before, interpreting the non-negative integers in the branching functions as the allowed topological endpoints of bosons at the gapped boundaries leads us to a Lagrangian algebra for the MTC $ \mathcal{C}= Spin(5)_{1} \times SU(2)_{-10}$. Explicitly, the algebra object is:
\begin{equation} \label{SU(2)LV4overSU(3)LV1Object}
    \mathcal{A} = (0, 0) \oplus (v, 10) \oplus (0, 6) \oplus (v, 4) \oplus (\sigma, 3) \oplus (\sigma, 7),
\end{equation}
where clearly all the entries are bosons, and using the MTC data summarized in Appendix \ref{MTCdataAppendix}, it is straightforward to check that Eqn. \eqref{quantumdimensioncondition} defining a Lagrangian algebra is fulfilled.

\begin{table}[!b]
\centering
\begin{tabular}{|c|c|c|c|} 
\hline
\parbox{3cm}{\begin{align*}
(0,0) &\to 0 \\ 
(0,1) &\to 1 \\ 
(0,2) &\to 2 \\ 
(0,3) &\to 9 + \sigma \\ 
(0,4) &\to 2 + v \\
(0,5) &\to 1 + 9 \\
(0,6) &\to 0 + 2 \\
(0,7) &\to 1 + \sigma \\
(0,8) &\to 2 \\
(0,9) &\to 9 \\
(0,10) &\to v
\end{align*}} &
\parbox{3cm}{\begin{align*}
(v,0) &\to v \\ 
(v,1) &\to 9 \\ 
(v,2) &\to 2 \\ 
(v,3) &\to 1 + \sigma \\ 
(v,4) &\to 0 + 2 \\
(v,5) &\to 1 + 9 \\
(v,6) &\to 2 + v \\
(v,7) &\to 9 + \sigma \\
(v,8) &\to 2 \\
(v,9) &\to 1 \\
(v,10) &\to 0
\end{align*}} &
\parbox{3cm}{\begin{align*}
(\sigma, 0) &\to \sigma \\ 
(\sigma, 1) &\to 2 \\ 
(\sigma, 2) &\to 1 + 9 \\ 
(\sigma, 3) &\to 0 + 2 + v \\ 
(\sigma, 4) &\to 1 + 9 + \sigma \\
(\sigma, 5) &\to 2(2) \\
(\sigma, 6) &\to 1 + 9 + \sigma \\
(\sigma, 7) &\to 0 + 2 + v \\
(\sigma, 8) &\to 1 + 9 \\
(\sigma, 9) &\to 2 \\
(\sigma, 10) &\to \sigma
\end{align*}} \\ \hline
\end{tabular}
\caption{Splitting of the bulk topological lines of $Spin(5)_{1} \times SU(2)_{-10}$ (left of the arrows) to the topological boundary. The resulting lines (right of the arrows) describe the topological line operators in the $Spin(5)_{1}/SU(2)_{10}$ topological coset.}
\label{SplittingSU(2)LV10}
\end{table}

\begin{table}[t]
\centering
\begin{tabular}{ |c|c|c|c|c|c| } 
\hline
$\times$  & $v$ & $1$ & $2$ & $9$ & $\sigma$ \\
\hline
$v$ & $0$ & $9$  & $2$  & $1$  & $\sigma$ \\ 
\hline
$1$ & $9$ & $0 + 2$  & $1 + 9 + \sigma$  & $2 + v$  & $2$ \\ 
\hline
$2$ & $2$ & $1 + 9 + \sigma$  & $0 + v + 2(2) $  & $1 + 9 + \sigma$  & $1 + 9$ \\ 
\hline
$9$ & $1$ & $2 + v$  & $1 + 9 + \sigma$  & $0 + 2$  & $2$ \\ 
\hline
$\sigma$ & $\sigma$ & $2$  & $1 + 9$  & $2$  & $0 + v$ \\ 
\hline
\end{tabular}
\caption{Fusion ring of the topological defect lines in the $Spin(5)_{1}/SU(2)_{10}$ topological coset. The line labeled as $0$ corresponds to the identity line.}
\label{SU(2)LV10FusionRing}
\end{table}

From the MTC data we can also obtain the allowed OPE channels between topological endpoints in the Lagrangian algebra, from which following the discussion in Section \ref{GappedBoundariesandTopologicalCosets} we can derive the following set of OPE coefficients:
\begin{align}
    & \phi_{(0,0)} \boldsymbol{\cdot} \phi_{(a,b)} = \phi_{(a,b)}, \ \mathrm{for} \ (a,b) \in \mathcal{A}, \label{coef1} \\[0.3cm] 
    & \phi_{(v, 10)} \boldsymbol{\cdot} \phi_{(v, 10)} = \phi_{(0,0)}, \quad \phi_{(v,10)} \boldsymbol{\cdot} \phi_{(0, 6)} = \phi_{(v, 4)}, \quad \phi_{(v,10)} \boldsymbol{\cdot} \phi_{(\sigma, 3)} = \phi_{(\sigma, 7)}, \\[0.3cm] 
    &\phi_{(0,6)} \boldsymbol{\cdot} \phi_{(0,6)} = \phi_{(v, 4)} \boldsymbol{\cdot} \phi_{(v, 4)} = (2 + \sqrt{3})\phi_{(0,0)} + (1 + \sqrt{3}) \, \phi_{(0,6)},
    \\[0.3cm] &\phi_{(0,6)} \boldsymbol{\cdot} \phi_{(v, 4)} = (2 + \sqrt{3}) \phi_{(v, 10)} + (1 + \sqrt{3}) \, \phi_{(v, 4)},
    \\[0.3cm] &\phi_{(0,6)} \boldsymbol{\cdot} \phi_{(\sigma, 3)} = \phi_{(v, 4)} \boldsymbol{\cdot} \phi_{(\sigma, 7)} = \frac{1}{2} (1 + \sqrt{3}) \phi_{(\sigma, 3)} + \frac{1}{2} (3 + \sqrt{3})\phi_{(\sigma, 7)},
    \\[0.3cm] &\phi_{(0,6)} \boldsymbol{\cdot}\phi_{(\sigma, 7)} = \phi_{(v, 4)} \boldsymbol{\cdot} \phi_{(\sigma, 3)} =  \frac{1}{2} (3 + \sqrt{3})\phi_{(\sigma, 3)} + \frac{1}{2} (1 + \sqrt{3})\phi_{(\sigma, 7)},
    \\[0.3cm] &\phi_{(\sigma, 3)} \boldsymbol{\cdot} \phi_{(\sigma, 3)} = (3+\sqrt{3})\phi_{(0,0)} + \sqrt{3} \phi_{(0, 6)} + 3 \phi_{(v, 4)},
    \\[0.3cm] &\phi_{(\sigma, 3)} \boldsymbol{\cdot} \phi_{(\sigma, 7)} = (3+\sqrt{3}) \phi_{(v, 10)} + 3 \phi_{(0, 6)} + \sqrt{3} \phi_{(v, 4)},
    \\[0.3cm] &\phi_{(\sigma, 7)} \boldsymbol{\cdot} \phi_{(\sigma, 7)} = (3+\sqrt{3}) \phi_{(0,0)} + \sqrt{3} \phi_{(0,6)} + 3 \phi_{(v, 4)}. \label{coef2}
\end{align}
This result is exact in the extreme IR limit, and the OPE coefficients along the RG flow must tend towards these expressions as one approaches the IR fixed point. Notice that associativity alone does not completely fix the coefficients above, but the special Frobenius condition \eqref{SpecialCondition} derived from consistency with the Lagrangian algebra multiplication adds an additional equation that readily fixes their values.

Similar to the previous example, it is possible to verify that the previous coefficients are supported by suitable Lagrangian algebra multiplications (see Eqn. \eqref{OPEcoeffs}). The specific values are presented in Appendix \ref{Spin5multiplications}, however, since there are too many entries to present here in a streamlined manner. As explained in Section \ref{GappedBoundariesandTopologicalCosets}, the multiplications are derived using the (co)associativity conditions \eqref{diagrammatic(co)associativity} and the specific input obtained from the MTC data in Appendix \ref{MTCdataAppendix}.

Now that we have discussed the point operators in the theory, we move on to obtain the spectrum of line operators and their fusion ring. To find the topological lines operators of the topological coset, we have to compute the spectrum of topological lines at the boundary. Either if we use Karoubi completions, or using the consistency conditions of splitting discussed in Section \ref{GappedBoundariesandTopologicalCosets} directly, we find the spectrum of boundary lines presented in Table \ref{SplittingSU(2)LV10} with the fusion ring presented in Table \ref{SU(2)LV10FusionRing}.

To check the consistency of the data obtained one may check the existence of the idempotent complete basis \eqref{IdempotentCompleteBasisEqn}. The specific transformation in this case is given by:
\begin{align}
    4 (3 + \sqrt{3}) \mathcal{O}_{0} &= \phi_{(0,0)} + \phi_{(0,6)} + \phi_{(v,10)} + \phi_{(v,4)} + \phi_{(\sigma,3)} + \phi_{(\sigma,7)}, \nonumber \\[0.3cm]
     4 (3 + \sqrt{3})\mathcal{O}_{1} &= \sqrt{2 + \sqrt{3}}\phi_{(0,0)} - \sqrt{2 - \sqrt{3}} \phi_{(0,6)} - \sqrt{2 + \sqrt{3}} \phi_{(v,10)} + \sqrt{2 - \sqrt{3}}\phi_{(v,4)} + \phi_{(\sigma,3)} - \phi_{(\sigma,7)}, \nonumber \\[0.3cm]
     4 (3 + \sqrt{3}) \mathcal{O}_{2} &= (1+\sqrt{3})\phi_{(0,0)} + (1-\sqrt{3})\phi_{(0,6)} + (1+\sqrt{3})\phi_{(v,10)} + (1-\sqrt{3})\phi_{(v,4)}, \nonumber \\[0.3cm]
    4 (3 + \sqrt{3}) \mathcal{O}_{v} &= \phi_{(0,0)} + \phi_{(0,6)} + \phi_{(v,10)} + \phi_{(v,4)} - \phi_{(\sigma,3)} - \phi_{(\sigma,7)}, \nonumber \\[0.3cm]
     4 (3 + \sqrt{3})\mathcal{O}_{9} &= \sqrt{2 + \sqrt{3}}\phi_{(0,0)} - \sqrt{2 - \sqrt{3}} \phi_{(0,6)} - \sqrt{2 + \sqrt{3}}\phi_{(v,10)} + \sqrt{2 - \sqrt{3}} \phi_{(v,4)} - \phi_{(\sigma,3)} + \phi_{(\sigma,7)}, \nonumber \\[0.3cm]
    4 (3 + \sqrt{3}) \mathcal{O}_{\sigma} &= \sqrt{2}\phi_{(0,0)} + \sqrt{2}\phi_{(0,6)} - \sqrt{2}\phi_{(v,10)} - \sqrt{2} \phi_{(v,4)}, 
\end{align}

\subsubsection{$Spin(8)_{1}/SU(3)_{3}$}

In this subsection we work out the topological coset $Spin(8)_{1}/SU(3)_{3}$, which describes the IR fixed point of (bosonized) 2D QCD with gauge group $SU(3)$ and fermions in the adjoint representation. The branching rules for this coset are well-known and are given by:
\begin{align}
    & \chi^{Spin(8)_{1}}_{0}(q) = \chi^{SU(3)_{3}}_{\mathbf{1}}(q) + \chi^{SU(3)_{3}}_{\mathbf{10}}(q) + \chi^{SU(3)_{3}}_{\overline{\mathbf{10}}}(q), \\[0.3cm]
    & \chi^{Spin(8)_{1}}_{\mathbf{v}}(q) = \chi^{SU(3)_{3}}_{\mathbf{8}}(q), \\[0.3cm] 
    & \chi^{Spin(8)_{1}}_{\mathbf{s}}(q) = \chi^{SU(3)_{3}}_{\mathbf{8}}(q), \\[0.3cm]
    & \chi^{Spin(8)_{1}}_{\mathbf{c}}(q) = \chi^{SU(3)_{3}}_{\mathbf{8}}(q).
\end{align}
Correspondingly, we can construct the following Lagrangian algebra object:
\begin{equation}
    \mathcal{A} = (0, \mathbf{1}) \oplus (0, \mathbf{10}) \oplus (0, \bar{\mathbf{10}}) \oplus (\mathbf{v}, \mathbf{8}) \oplus (\mathbf{s}, \mathbf{8}) \oplus (\mathbf{c}, \mathbf{8}),
\end{equation}
where it is easy to check, for instance, the constraint on the quantum dimensions \eqref{quantumdimensioncondition} demanded by the definition of a Lagrangian algebra. Working out the spectrum of topological local operators in the diagonal basis and their OPE along the lines of Section \ref{GappedBoundariesandTopologicalCosets} (see also previous examples), we find the OPE coefficients displayed in the following table:
\begin{table}[h] 
\hspace{-2.1cm} 
\begin{tabular}{|c|c|c|c|c|c|c|} 
\hline
$\boldsymbol{\cdot}$  & $\phi_{(0, \mathbf{10})}$ & $\phi_{(0, \overline{\mathbf{10}})}$ & $\phi_{(\mathbf{v}, \mathbf{8})}$ & $\phi_{(\mathbf{s}, \mathbf{8})}$ & $\phi_{(\mathbf{c}, \mathbf{8})}$ \\
\hline
$\phi_{(0, \mathbf{10})}$  & $\phi_{(0, \overline{\mathbf{10}})}$  & $\phi_{(0, \mathbf{1})}$  & $\phi_{(\mathbf{v}, \mathbf{8})}$ & $\phi_{(\mathbf{s}, \mathbf{8})}$  & $\phi_{(\mathbf{c}, \mathbf{8})}$  \\ 
\hline
$\phi_{(0, \overline{\mathbf{10}})}$ & $\phi_{(0, \mathbf{1})}$  & $\phi_{(0, \mathbf{10})}$  & $\phi_{(\mathbf{v}, \mathbf{8})}$  & $\phi_{(\mathbf{s}, \mathbf{8})}$ & $\phi_{(\mathbf{c}, \mathbf{8})}$ \\ 
\hline
$\phi_{(\mathbf{v}, \mathbf{8})}$ & $\phi_{(\mathbf{v}, \mathbf{8})}$  & $\phi_{(\mathbf{v}, \mathbf{8})}$  & $3\phi_{(0, \mathbf{1})} + 3\phi_{(0, \mathbf{10})} + 3\phi_{(0, \bar{\mathbf{10}})}$   & $3\phi_{(\mathbf{c}, \mathbf{8})}$ & $3\phi_{(\mathbf{s}, \mathbf{8})}$ \\ 
\hline
$\phi_{(\mathbf{s}, \mathbf{8})}$  & $\phi_{(\mathbf{s}, \mathbf{8})}$  & $\phi_{(\mathbf{s}, \mathbf{8})}$  & $3\phi_{(\mathbf{c}, \mathbf{8})}$  & $3\phi_{(0, \mathbf{1})} + 3\phi_{(0, \mathbf{10})} + 3 \phi_{(0, \bar{\mathbf{10}})}$ & $3\phi_{(\mathbf{v}, \mathbf{8})}$ \\ 
\hline
$\phi_{(\mathbf{c}, \mathbf{8})}$ & $\phi_{(\mathbf{c}, \mathbf{8})}$  & $\phi_{(\mathbf{c}, \mathbf{8})}$  & $3\phi_{(\mathbf{s}, \mathbf{8})}$   & $3\phi_{(\mathbf{v}, \mathbf{8})}$ & $3\phi_{(0, \mathbf{1})} + 3\phi_{(0, \mathbf{10})} +3\phi_{(0, \bar{\mathbf{10}})}$ \\ 
\hline
\end{tabular} 
\label{OPECoeffsSU(3)LV3}
\end{table}

\noindent and where obviously $\phi_{(0,\mathbf{1})} \boldsymbol{\cdot} \phi_{a} = \phi_{a}$ for any $a \in \mathcal{A}$. It is straightforward to check that the product above is commutative, associative, and that the special Frobenius condition \eqref{NormSpecial} is fulfilled. This result is exact in the extreme IR limit, and of course approximate along the RG flow slightly above the IR fixed point. Notice that associativity alone does not completely fix the coefficients above, but the special Frobenius condition \eqref{SpecialCondition} adds an additional equation that readily fixes their values.

\begin{table}[t] 
\centering
\begin{tabular}{|c|c|c|c|}
\hline
\parbox{3cm}{\begin{align*}
&(0,0) && \hspace{-0.4cm} \to 0 \\ 
&(0,\mathbf{10}) && \hspace{-0.4cm} \to 0 \\ 
&(0,\bar{\mathbf{10}}) && \hspace{-0.4cm} \to 0 \\ 
&(0,\mathbf{3}) && \hspace{-0.4cm} \to \mathcal{N} \\ 
&(0,\mathbf{15}) && \hspace{-0.4cm} \to \mathcal{N} \\
&(0,\bar{\mathbf{6}}) && \hspace{-0.4cm} \to \mathcal{N} \\
&(0,\mathbf{6})  && \hspace{-0.4cm} \to \bar{\mathcal{N}} \\
&(0,\bar{\mathbf{15}}) && \hspace{-0.4cm} \to \bar{\mathcal{N}} \\
&(0,\bar{\mathbf{3}}) && \hspace{-0.4cm} \to \bar{\mathcal{N}} \\
&(0,\mathbf{8}) && \hspace{-0.4cm} \to \mathbf{v} + \mathbf{s} + \mathbf{c}
\end{align*}} &
\parbox{3cm}{\begin{align*}
&(v,0) && \hspace{-0.4cm} \to \mathbf{v} \\ 
&(v,\mathbf{10}) && \hspace{-0.4cm} \to \mathbf{v} \\
&(v,\bar{\mathbf{10}}) && \hspace{-0.4cm} \to \mathbf{v} \\ 
&(v,\mathbf{3}) && \hspace{-0.4cm} \to \mathcal{N} \\ 
&(v,\mathbf{15}) && \hspace{-0.4cm} \to \mathcal{N} \\
&(v,\bar{\mathbf{6}}) && \hspace{-0.4cm} \to \mathcal{N} \\
&(v,\mathbf{6})  && \hspace{-0.4cm} \to \bar{\mathcal{N}} \\
&(v,\bar{\mathbf{15}}) && \hspace{-0.4cm} \to \bar{\mathcal{N}} \\
&(v,\bar{\mathbf{3}}) && \hspace{-0.4cm} \to \bar{\mathcal{N}} \\
&(v,\mathbf{8}) && \hspace{-0.4cm} \to 0 + \mathbf{s} + \mathbf{c}
\end{align*}} &
\parbox{3cm}{\begin{align*}
&(s,0) && \hspace{-0.4cm} \to \mathbf{s}  \\ 
&(s,\mathbf{10}) && \hspace{-0.4cm} \to \mathbf{s}  \\ 
&(s,\bar{\mathbf{10}}) && \hspace{-0.4cm} \to \mathbf{s}  \\ 
&(s,\mathbf{3}) && \hspace{-0.4cm} \to \mathcal{N} \\ 
&(s,\mathbf{15}) && \hspace{-0.4cm} \to \mathcal{N} \\
&(s,\bar{\mathbf{6}}) && \hspace{-0.4cm} \to \mathcal{N} \\
&(s,\mathbf{6})  && \hspace{-0.4cm} \to \bar{\mathcal{N}} \\
&(s,\bar{\mathbf{15}}) && \hspace{-0.4cm} \to \bar{\mathcal{N}} \\
&(s,\bar{\mathbf{3}}) && \hspace{-0.4cm} \to \bar{\mathcal{N}} \\
&(s,\mathbf{8}) && \hspace{-0.4cm} \to 0 + \mathbf{v} + \mathbf{c}
\end{align*}} &
\parbox{3cm}{\begin{align*}
&(c,0) && \hspace{-0.4cm} \to \mathbf{c} \\ 
&(c,\mathbf{10}) && \hspace{-0.4cm} \to \mathbf{c} \\ 
&(c,\bar{\mathbf{10}}) && \hspace{-0.4cm} \to \mathbf{c} \\ 
&(c,\mathbf{3}) && \hspace{-0.4cm} \to \mathcal{N} \\ 
&(c,\mathbf{15}) && \hspace{-0.4cm} \to \mathcal{N} \\
&(c,\bar{\mathbf{6}}) && \hspace{-0.4cm} \to \mathcal{N} \\
&(c,\mathbf{6})  && \hspace{-0.4cm} \to \bar{\mathcal{N}} \\
&(c,\bar{\mathbf{15}}) && \hspace{-0.4cm} \to \bar{\mathcal{N}} \\
&(c,\bar{\mathbf{3}}) && \hspace{-0.4cm} \to \bar{\mathcal{N}} \\
&(c,\mathbf{8}) && \hspace{-0.4cm} \to 0 + \mathbf{v} + \mathbf{s} 
\end{align*}} \\ \hline
\end{tabular}
\caption{Descent of the bulk topological lines of $Spin(8)_{1} \times SU(3)_{-3}$ (left of the arrows) to the topological boundary. The resulting lines (right of the arrows) describe the topological line operators in the $Spin(8)_{1}/SU(3)_{3}$  topological coset.}
\label{SplittingSU(3)LV3}
\end{table}

Similarly, we can study the topological line operators of the topological coset. The bulk line operators descend to line operators at the boundary according to the splitting and identifications shown in Table \ref{SplittingSU(3)LV3}. It is straightforward to calculate the fusion ring of the line defects, and the result is shown in Table \ref{SU(3)LV3FusionRing}.
\begin{table}[!b]
\centering
\begin{tabular}{ |c|c|c|c|c|c| } 
\hline
$\times$  & $\mathcal{N}$ & $\bar{\mathcal{N}}$ & $\mathbf{v}$ & $\mathbf{s}$ & $\mathbf{c}$ \\
\hline
$\mathcal{N}$ & $2 \, \bar{\mathcal{N}}$ & $0 + \mathbf{v} + \mathbf{s} + \mathbf{c}$  & $\mathcal{N}$  & $\mathcal{N}$  & $\mathcal{N}$ \\ 
\hline
$\bar{\mathcal{N}}$ & $0 + \mathbf{v} + \mathbf{s} + \mathbf{c}$ & $2 \, \mathcal{N}$  & $\bar{\mathcal{N}}$  & $\bar{\mathcal{N}}$  & $\bar{\mathcal{N}}$ \\ 
\hline
$\mathbf{v}$ & $\mathcal{N}$ & $\bar{\mathcal{N}}$  & $0$  & $\mathbf{c}$  & $\mathbf{s}$ \\ 
\hline
$\mathbf{s}$ & $\mathcal{N}$ & $\bar{\mathcal{N}}$  & $\mathbf{c}$  & $0$  & $\mathbf{v}$ \\ 
\hline
$\mathbf{c}$ & $\mathcal{N}$ & $\bar{\mathcal{N}}$  & $\mathbf{s}$  & $\mathbf{v}$  & $0$ \\ 
\hline
\end{tabular}
\caption{Fusion ring of the topological defect lines in the $Spin(8)_{1}/SU(3)_{3}$ topological coset. The line labeled as $0$ corresponds to the identity line.}
\label{SU(3)LV3FusionRing}
\end{table}

As a check of the formalism, we can transform the topological local operators to the idempotent complete basis. The transformation is given by: 
\begin{align}
    \mathcal{O}_{0} &= \phi_{(0,0)} + \phi_{(0,\mathbf{10})} + \phi_{(0, \bar{\mathbf{10}})} + \phi_{(\mathbf{v},\mathbf{8})} + \phi_{(\mathbf{s},\mathbf{8})} + \phi_{(\mathbf{c},\mathbf{8})},  \\[0.3cm]
    \mathcal{O}_{\mathcal{N}} &= 2 \phi_{(0,0)} + 2 \omega \phi_{(0,\mathbf{10})} + 2 \omega^{2} \phi_{(0, \bar{\mathbf{10}})},  \\[0.3cm]
    \mathcal{O}_{\bar{\mathcal{N}}} &= 2 \phi_{(0,0)} + 2 \omega^{2} \phi_{(0,\mathbf{10})} + 2 \omega \phi_{(0, \bar{\mathbf{10}})},  \\[0.3cm]
    \mathcal{O}_{\mathbf{v}} &= \phi_{(0,0)} + \phi_{(0,\mathbf{10})} + \phi_{(0, \bar{\mathbf{10}})} + \phi_{(\mathbf{v},\mathbf{8})} - \phi_{(\mathbf{s},\mathbf{8})} - \phi_{(\mathbf{c},\mathbf{8})}, \\[0.3cm]
    \mathcal{O}_{\mathbf{s}} &= \phi_{(0,0)} +\phi_{(0,\mathbf{10})} + \phi_{(0, \bar{\mathbf{10}})} - \phi_{(\mathbf{v},\mathbf{8})} + \phi_{(\mathbf{s},\mathbf{8})} - \phi_{(\mathbf{c},\mathbf{8})}, \\[0.3cm]
    \mathcal{O}_{\mathbf{c}} &= \phi_{(0,0)} + \phi_{(0,\mathbf{10})} + \phi_{(0, \bar{\mathbf{10}})} - \phi_{(\mathbf{v},\mathbf{8})} - \phi_{(\mathbf{s},\mathbf{8})} + \phi_{(\mathbf{c},\mathbf{8})}.
\end{align}
Indeed, it is straightforward to check that the lines in the 2D theory act according to \eqref{nondiagonalactionoflines} for any of the line operators in Table \ref{SplittingSU(3)LV3} and the fusion rules in the fusion ring displayed in Table \ref{SU(3)LV3FusionRing}.

\subsubsection{$Spin(16)_{1}/Spin(9)_{2}$}

In this subsection we work out the topological coset $Spin(16)_{1}/Spin(9)_{2}$, which describes the IR fixed point of (bosonized) 2D QCD with gauge group $Spin(9)$ with fermions in the spinorial representation. We borrow the branching rules from \cite{Delmastro:2021otj}:
\begin{align}
    & \chi^{Spin(16)_{1}}_{0}(q) = \chi^{Spin(9)_{2}}_{\mathbf{1}}(q) + \chi^{Spin(9)_{2}}_{\mathbf{84}}(q), \\[0.3cm]
    & \chi^{Spin(16)_{1}}_{\mathbf{v}}(q) = \chi^{Spin(9)_{2}}_{\mathbf{16}}(q), \\[0.3cm] 
    & \chi^{Spin(16)_{1}}_{\mathbf{s}}(q) = \chi^{Spin(9)_{2}}_{\mathbf{44}}(q) + \chi^{Spin(9)_{2}}_{\mathbf{84}}(q), \\[0.3cm]
    & \chi^{Spin(16)_{1}}_{\mathbf{c}}(q) = \chi^{Spin(9)_{2}}_{\mathbf{128}}(q).
\end{align}
The associated Lagrangian algebra object is:
\begin{equation}
    \mathcal{A} = (0, \mathbf{1}) \oplus (0, \mathbf{84}) \oplus (\mathbf{v}, \mathbf{16}) \oplus (\mathbf{s}, \mathbf{44}) \oplus (\mathbf{s}, \mathbf{84}) \oplus (\mathbf{c}, \mathbf{128}).
\end{equation}
Again, it is straightforward to check the constraint on the quantum dimensions: $\mathrm{dim}(\mathcal{A})^{2} = \mathrm{dim}(Spin(16)_{1} \times Spin(9)_{-2})$. Working out the spectrum of topological local operators and their OPE along the lines of Section \ref{GappedBoundariesandTopologicalCosets}, we find the OPE coefficients displayed in the following table:
\begin{table}[h]
\hspace{-1.5cm}
\begin{tabular}{|c|c|c|c|c|c|c|} 
\hline
$\boldsymbol{\cdot}$  & $\phi_{(0, \mathbf{1})}$ & $\phi_{(\mathbf{s}, \mathbf{44})}$ & $\phi_{(\mathbf{v}, \mathbf{16})}$ & $\phi_{(\mathbf{c}, \mathbf{128})}$ & $\phi_{(0, \mathbf{84})}$ & $\phi_{(\mathbf{s}, \mathbf{84})}$ \\
\hline
$\phi_{(0, \mathbf{1})}$ & $\phi_{(0, \mathbf{1})}$ & $\phi_{(\mathbf{s}, \mathbf{44})}$  & $\phi_{(\mathbf{v}, \mathbf{16})}$  & $\phi_{(\mathbf{c}, \mathbf{128})}$  & $\phi_{(0, \mathbf{84})}$ & $\phi_{(\mathbf{s}, \mathbf{84})}$ \\ 
\hline
$\phi_{(\mathbf{s}, \mathbf{44})}$ & $\phi_{(\mathbf{s}, \mathbf{44})}$  & $\phi_{(0, \mathbf{1})}$  & $\phi_{(\mathbf{c}, \mathbf{128})}$  & $\phi_{(\mathbf{v}, \mathbf{16})}$ & $\phi_{(\mathbf{s}, \mathbf{84})}$  & $\phi_{(0, \mathbf{84})}$  \\ 
\hline
$\phi_{(\mathbf{v}, \mathbf{16})}$ & $\phi_{(\mathbf{v}, \mathbf{16})}$ & $\phi_{(\mathbf{c}, \mathbf{128})}$  & $3 \phi_{(0, \mathbf{1})} + 3 \phi_{(0, \mathbf{84})}$  & $3 \phi_{(\mathbf{s}, \mathbf{44})} + 3 \phi_{(\mathbf{s}, \mathbf{84})}$  & $2 \phi_{(\mathbf{v}, \mathbf{16})}$ & $2 \phi_{(\mathbf{c}, \mathbf{128})}$ \\ 
\hline
$\phi_{(\mathbf{c}, \mathbf{128})}$ & $\phi_{(\mathbf{c}, \mathbf{128})}$ & $\phi_{(\mathbf{v}, \mathbf{16})}$  & $3 \phi_{(\mathbf{s}, \mathbf{44})} + 3 \phi_{(\mathbf{s}, \mathbf{84})}$  & $3 \phi_{(0, \mathbf{1})} + 3 \phi_{(0, \mathbf{84})}$  & $2 \phi_{(\mathbf{c}, \mathbf{128})}$ & $2 \phi_{(\mathbf{v}, \mathbf{16})}$ \\ 
\hline
$\phi_{(0, \mathbf{84})}$ & $\phi_{(0, \mathbf{84})}$ & $\phi_{(\mathbf{s}, \mathbf{84})}$  & $2 \phi_{(\mathbf{v}, \mathbf{16})}$  & $2 \phi_{(\mathbf{c}, \mathbf{128})}$  & $2 \phi_{(0, \mathbf{1})} + \phi_{(0, \mathbf{84})}$ & $2 \phi_{(\mathbf{s}, \mathbf{44})} + \phi_{(\mathbf{s}, \mathbf{84})}$ \\ 
\hline
$\phi_{(\mathbf{s}, \mathbf{84})}$ & $\phi_{(\mathbf{s}, \mathbf{84})}$ & $\phi_{(0, \mathbf{84})}$  & $2 \phi_{(\mathbf{c}, \mathbf{128})}$  & $2 \phi_{(\mathbf{v}, \mathbf{16})}$   & $2 \phi_{(\mathbf{s}, \mathbf{44})} + \phi_{(\mathbf{s}, \mathbf{84})}$ & $2 \phi_{(0, \mathbf{1})} + \phi_{(0, \mathbf{84})}$ \\ 
\hline
\end{tabular} 
\end{table}

\noindent Once again, it is straightforward to check that the product of local operators above is commutative, associative, and that the special Frobenius condition \eqref{NormSpecial} is fulfilled.

Meanwhile, the spectrum of line operators of the topological coset can be obtained from the splitting and identifications shown in Table \ref{Spin(9)lv2Splittings}. It is straightforward to calculate the fusion ring of the resulting line defects, and the result is shown in Table \ref{Spin(9)2cosetfusionring}. The obtained fusion ring can be recognized as that of $\mathbb{Z}_{2} \times \mathrm{Rep}(S_{3})$. As before, we can transform to the idempotent complete basis to check that everything is in order. The transformation is given by: 
\begin{align}
    \mathcal{O}_{0} &= \phi_{(0,\mathbf{1})} + \phi_{(0,\mathbf{84})} + \phi_{(\mathbf{s},\mathbf{44})} + \phi_{(\mathbf{v},\mathbf{16})} + \phi_{(\mathbf{c},\mathbf{128})} + \phi_{(\mathbf{s}, \mathbf{84})},  \\[0.3cm]
    \mathcal{O}_{\mathbf{s}} &= \phi_{(0,\mathbf{1})} + \phi_{(0,\mathbf{84})} + \phi_{(\mathbf{s},\mathbf{44})} - \phi_{(\mathbf{v},\mathbf{16})} - \phi_{(\mathbf{c},\mathbf{128})} + \phi_{(\mathbf{s}, \mathbf{84})},  \\[0.3cm]
    \mathcal{O}_{\mathbf{v}} &= \phi_{(0,\mathbf{1})} + \phi_{(0,\mathbf{84})} - \phi_{(\mathbf{s},\mathbf{44})} + \phi_{(\mathbf{v},\mathbf{16})} - \phi_{(\mathbf{c},\mathbf{128})} - \phi_{(\mathbf{s}, \mathbf{84})}, \\[0.3cm]
    \mathcal{O}_{\mathbf{c}} &= \phi_{(0,\mathbf{1})} + \phi_{(0,\mathbf{84})} - \phi_{(\mathbf{s},\mathbf{44})} - \phi_{(\mathbf{v},\mathbf{16})} + \phi_{(\mathbf{c},\mathbf{128})} - \phi_{(\mathbf{s}, \mathbf{84})}, \\[0.3cm]
    \mathcal{O}_{A} &= 2 \phi_{(0,\mathbf{1})} - \phi_{(0,\mathbf{84})} + 2 \phi_{(\mathbf{s},\mathbf{44})} - \phi_{(\mathbf{s}, \mathbf{84})},  \\[0.3cm]
    \mathcal{O}_{B} &= 2 \phi_{(0,\mathbf{1})} - \phi_{(0,\mathbf{84})} - 2 \phi_{(\mathbf{s},\mathbf{44})} + \phi_{(\mathbf{s}, \mathbf{84})}, 
\end{align}
and it is straightforward to check that the operators so defined are acted on by the lines according to \eqref{nondiagonalactionoflines}.

\begin{table}[t]
\centering
\begin{tabular}{|c|c|c|c|} 
\hline
\parbox{3cm}{\begin{align*}
(0,\mathbf{1}) &\to 0 \\ 
(0,\mathbf{44}) &\to \mathbf{s} \\ 
(0,\mathbf{16}) &\to \mathbf{v} + B \\ 
(0,\mathbf{128}) &\to \mathbf{c} + B \\ 
(0,\mathbf{126}) &\to A \\
(0,\mathbf{84}) &\to 0 + \mathbf{s} \\
(0,\mathbf{36}) &\to A \\
(0,\mathbf{9}) &\to A 
\end{align*}} &
\parbox{3cm}{\begin{align*}
(\mathbf{v},\mathbf{1}) &\to \mathbf{v} \\ 
(\mathbf{v},\mathbf{44}) &\to \mathbf{c} \\ 
(\mathbf{v},\mathbf{16}) &\to 0 + A \\ 
(\mathbf{v},\mathbf{128}) &\to \mathbf{s} + A \\ 
(\mathbf{v},\mathbf{126}) &\to B \\
(\mathbf{v},\mathbf{84}) &\to \mathbf{v} + \mathbf{c} \\
(\mathbf{v},\mathbf{36}) &\to B \\
(\mathbf{v},\mathbf{9}) &\to B 
\end{align*}} &
\parbox{3cm}{\begin{align*}
(\mathbf{s},\mathbf{1}) &\to \mathbf{s} \\ 
(\mathbf{s},\mathbf{44}) &\to 0 \\ 
(\mathbf{s},\mathbf{16}) &\to \mathbf{c} + B \\ 
(\mathbf{s},\mathbf{128}) &\to \mathbf{v} + B \\ 
(\mathbf{s},\mathbf{126}) &\to A \\
(\mathbf{s},\mathbf{84}) &\to 0 + \mathbf{s} \\
(\mathbf{s},\mathbf{36}) &\to A \\
(\mathbf{s},\mathbf{9}) &\to A 
\end{align*}} &
\parbox{3cm}{\begin{align*}
(\mathbf{c},\mathbf{1}) &\to \mathbf{c} \\ 
(\mathbf{c},\mathbf{44}) &\to \mathbf{v} \\ 
(\mathbf{c},\mathbf{16}) &\to \mathbf{s} + A \\ 
(\mathbf{c},\mathbf{128}) &\to 0 + A \\ 
(\mathbf{c},\mathbf{126}) &\to B \\
(\mathbf{c},\mathbf{84}) &\to \mathbf{v} + \mathbf{c} \\
(\mathbf{c},\mathbf{36}) &\to B \\
(\mathbf{c},\mathbf{9}) &\to B 
\end{align*}} \\ \hline
\end{tabular}
\caption{Descent of the bulk topological lines of $Spin(16)_{1} \times Spin(9)_{-2}$ (left of the arrows) to the topological boundary. The resulting lines (right of the arrows) describe the topological line operators in the $Spin(16)_{1}/Spin(9)_{2}$ topological coset.} \label{Spin(9)lv2Splittings}
\end{table}

\begin{table}[h]
\centering
\begin{tabular}{ |c|c|c|c|c|c| } 
\hline
$\times$  & $\mathbf{s}$ & $\mathbf{v}$ & $\mathbf{c}$ & $A$ & $B$ \\
\hline
$\mathbf{s}$ & $0$ & $\mathbf{c}$  & $\mathbf{v}$  & $A$  & $B$ \\ 
\hline
$\mathbf{v}$ & $\mathbf{c}$ & $0$  & $\mathbf{s}$  & $B$  & $A$ \\ 
\hline
$\mathbf{c}$ & $\mathbf{v}$ & $\mathbf{s}$  & $0$  & $B$  & $A$ \\ 
\hline
$A$ & $A$ & $B$  & $B$  & $0 + \mathbf{s} + A$  & $\mathbf{v} + \mathbf{c} + B$ \\ 
\hline
$B$ & $B$ & $A$  & $A$  & $\mathbf{v} + \mathbf{c} + B$  & $0 + \mathbf{s} + A$ \\ 
\hline
\end{tabular}
\caption{Fusion ring of topological defect lines in the $Spin(16)_{1}/Spin(9)_{2}$ topological coset. The line labeled as $0$ corresponds to the identity line.}
\label{Spin(9)2cosetfusionring}
\end{table}

\section{A Trivially Gapped Chiral $\mathrm{\bf{QCD_{2}}}$ Theory} \label{chiral2DQCDSection}

As reviewed above, it was shown in \cite{Delmastro:2021otj} that the criterion for a massless 2D QCD theory to be gapped is that the coset CFT
\begin{equation}
    Spin(\mathrm{dim}(R))_{1}/G_{I(R)}
\end{equation}
    is actually a topological theory.\footnote{More precisely, the authors of \cite{Delmastro:2021otj} consider $SO(\mathrm{dim}(R))_{1}/G_{I(R)}$, where $SO(\mathrm{dim}(R))_{1}$ corresponds to the fermionization of $Spin(\mathrm{dim}(R))_{1}$. Since the statement concerns the central charge, we can take the criterion to hold either in the fermionic or bosonic version of the theory.} That is, if the coset is a conformal embedding, the corresponding QCD theory is gapped. Remarkably, \cite{Delmastro:2021otj} also found that there exist \emph{chiral} QCD theories that are gapped. One mechanism to construct such chiral theories is given as follows: Starting from a vector-like theory $(G,R,R)$ which is gapped, the chiral theory $(G,R_{\ell},R_{r})$ is also gapped, where $(R_{\ell},R_{r}) = (\sigma_{\ell} \cdot R, \sigma_{r} \cdot R)$ and $\sigma_{\ell}$ and $\sigma_{r}$ are outer automorphisms of the Lie algebra of $G$. A concrete example is given by $Spin(8)$ gauge theory coupled to massless fermions in the vectorial and spinorial representations, where the automorphism $\sigma$ is given by the triality of $Spin(8)$:
\begin{equation} \label{chiralspin8theory}
    (Spin(8), \mathbf{8}_{\mathbf{v}}, \mathbf{8}_{\mathbf{c}}).
\end{equation}
In the following, we will argue that this theory is in fact trivially gapped. For this, we will make use of some details regarding fermionic 2D CFTs and bosonization, for which we provide a quick summary in Appendix \ref{Fermionization}. To understand how the chiral theory is different from its vector-like counterpart, notice that if the right-moving fermions transform in the spinorial $\mathbf{c}$ representation, we may regard the branching rules of the fermionic characters in the UV theory as given by
\begin{alignat}{3}
    d_{\mathrm{NS},\mathrm{NS}}(q) &= \chi_{\mathbf{1}}(q) + \chi_{\mathbf{v}}(q), \quad \, \tilde{d}_{\mathrm{NS},\mathrm{NS}}(\bar{q}) &&= \tilde{\chi}_{\mathbf{1}}(\bar{q}) + \tilde{\chi}_{\mathbf{c}}(\bar{q}), \\[0.3cm]
    d_{\mathrm{NS},\mathrm{R}}(q) &= \chi_{\mathbf{1}}(q) - \chi_{\mathbf{v}}(q), \quad \ \ \tilde{d}_{\mathrm{NS},\mathrm{R}}(\bar{q}) &&= \tilde{\chi}_{\mathbf{1}}(\bar{q}) - \tilde{\chi}_{\mathbf{c}}(\bar{q}), \\[0.3cm]
    d_{\mathrm{R},\mathrm{NS}}(q) &= \chi_{\mathbf{s}}(q) + \chi_{\mathbf{c}}(q), \quad \ \ \ \tilde{d}_{\mathrm{R},\mathrm{NS}}(\bar{q}) &&= \tilde{\chi}_{\mathbf{s}}(\bar{q}) + \tilde{\chi}_{\mathbf{v}}(\bar{q}), \\[0.3cm]
    d_{\mathrm{R},\mathrm{R}}(q) &= \chi_{\mathbf{s}}(q) - \chi_{\mathbf{c}}(q), \quad \ \ \ \ \tilde{d}_{\mathrm{R},\mathrm{R}}(\bar{q}) &&= \tilde{\chi}_{\mathbf{s}}(\bar{q}) - \tilde{\chi}_{\mathbf{v}}(\bar{q}),
\end{alignat}
where $d_{\mathrm{X},\mathrm{Y}}$ stands for the characters of the fermionic theory with a choice of $\mathrm{X}, \mathrm{Y} = \mathrm{NS},\mathrm{R}$ boundary conditions along the cycles of the torus. Both sides differ by a triality transformation exchanging $\mathbf{v}$ and $\mathbf{c}$, in accordance with \eqref{chiralspin8theory}. Bosonizing, we obtain the following modular invariant capturing the action of triality in the bosonic version of the UV (ungauged) theory:
\begin{equation} \label{trialitymodularinvariant}
    Z_{1} = \tilde{\chi}_{\mathbf{1}} \chi_{\mathbf{1}} + \tilde{\chi}_{\mathbf{c}} \chi_{\mathbf{v}} + \tilde{\chi}_{\mathbf{v}} \chi_{\mathbf{s}} + \tilde{\chi}_{\mathbf{s}} \chi_{\mathbf{c}}.
\end{equation}
It is well-known that bosonization is not unique. Instead, we can redefine the original fermionic chiral QCD theory by stacking with a 2D spin SPT and bosonize (see Appendix \ref{Fermionization} for more details), obtaining
\begin{equation} \label{trialitymodularinvariant2}
    Z_{2} = \tilde{\chi}_{\mathbf{1}} \chi_{\mathbf{1}} + \tilde{\chi}_{\mathbf{c}} \chi_{\mathbf{v}} + \tilde{\chi}_{\mathbf{v}} \chi_{\mathbf{c}} + \tilde{\chi}_{\mathbf{s}} \chi_{\mathbf{s}},
\end{equation}
We will consider both these cases in the following.

Recall that in the bulk-boundary correspondence, as reviewed in the Introduction, for any modular invariant of a chiral algebra, there exists a topological surface operator in the 3D bulk dictating the gluing of holomorphic and antiholomorphic modes \cite{Fuchs:2002cm, Kapustin:2010if, Gaiotto:2020iye, Komargodski:2020mxz}. Recall the left side in Figure \ref{BoundaryGluing}. In more modern terms, the existence of such a topological surface can be understood via higher-gauging of the abelian one-form symmetries generated by the anyons of the bulk \cite{Roumpedakis:2022aik}. More specifically for our purposes, the modular invariant \eqref{trialitymodularinvariant2} can straightforwardly be constructed by the higher-gauging of the $\mathbb{Z}_{2}$ symmetry generated by the $\mathbf{s}$ anyon along a surface, while \eqref{trialitymodularinvariant} arises from the higher-gauging of the $\mathbb{Z}_{2} \times \mathbb{Z}_{2}$ symmetry generated by $\mathbf{s}$ and $\mathbf{c}$ along a surface.\footnote{Higher-gauging with discrete torsion gives rise to the other triality modular invariant $Z_{3} = \tilde{\chi}_{\mathbf{1}} \chi_{\mathbf{1}} + \tilde{\chi}_{\mathbf{s}} \chi_{\mathbf{v}} + \tilde{\chi}_{\mathbf{c}} \chi_{\mathbf{s}} + \tilde{\chi}_{\mathbf{v}} \chi_{\mathbf{c}}$.} We call the latter topological surface $S_{\mathbf{s} \mathbf{c}}$. The specifics of higher-gauging will not be necessary in the following, other than to point out the existence of the $S_{\mathbf{s} \mathbf{c}}$ topological surface and the fact that it acts over the anyons of the theory according to the coupling of antiholomorphic and holomorphic modes in the corresponding modular invariant. For example:
\begin{equation}
    S_{\mathbf{s} \mathbf{c}} [ \mathbf{1} ] = \mathbf{1}, \quad S_{\mathbf{s} \mathbf{c}} [ \mathbf{v} ] = \mathbf{s}, \quad S_{\mathbf{s} \mathbf{c}} [ \mathbf{s} ] = \mathbf{c}, \quad S_{\mathbf{s} \mathbf{c}} [ \mathbf{c} ] = \mathbf{v},
\end{equation}
and thus $S_{\mathbf{s} \mathbf{c}}$ implements the triality action in bulk.

\begin{figure}[t]  
        \hspace{-0.85cm}\includegraphics[scale=1.13]{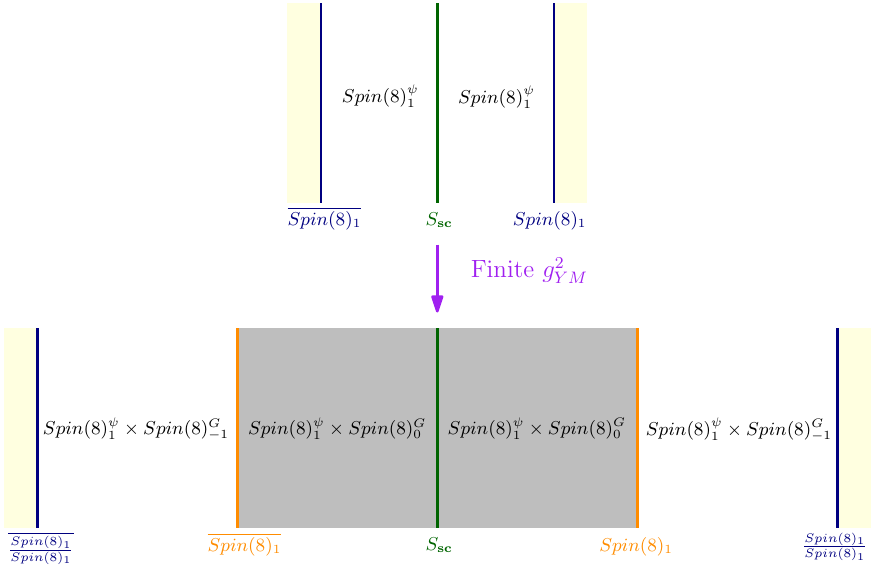} 
        \caption{Three-dimensional construction of chiral $Spin(8)$ 2D QCD coupled to massless fermions in the vectorial and spinorial representations. The top picture represents the UV theory at zero coupling. The chirality of the theory is implied by the surface $S_{\mathbf{s} \mathbf{c}}$ permuting bulk anyons by triality. Notice that at a finite value of $g_{YM}^{2}$ the topological surface $S_{\mathbf{s} \mathbf{c}}$ can be positioned anywhere and not necessarily in the middle region, as the middle interfaces are transparent respect to the $Spin(8)^{\psi}_{1}$ factor, from which we are constructing the surface $S_{\mathbf{s} \mathbf{c}}$.} \label{chiral2DQCD-3DConstruction}
\end{figure}

From the previous discussion, it follows that the 3D realization of the bosonized 2D QCD theory we wish to consider in the UV corresponds to the top picture in Figure \ref{chiral2DQCD-3DConstruction}, as this reproduces the modular invariant \eqref{trialitymodularinvariant}. Along the RG flow, we can run a similar picture as the 3D construction of the non-chiral case reviewed in Appendix \ref{CircleInterval2DQCD}. This gives rise to a similar picture, but with the topological surface $S_{\mathbf{s} \mathbf{c}}$ inserted. Notice that along the flow we have two $Spin(8)$ factors in the bulk TQFT, as shown in the bottom picture in Figure \ref{chiral2DQCD-3DConstruction}. One of these is associated with the UV fermions, which we have written as $Spin(8)^{\psi}$, while the other is associated with the gauge group, which we have written as $Spin(8)^{G}$. Importantly, it is immaterial where the surface $S_{\mathbf{s} \mathbf{c}}$ is inserted in bulk since it has been constructed from the $Spin(8)^{\psi}_{1}$ factor (via higher-gauging), while the three regions differ by $Spin(8)^{G}$ factors.

\begin{figure}[t] \hspace{-0.8cm}
\begin{subfigure}{.45\textwidth}
  \includegraphics[scale=1.24]{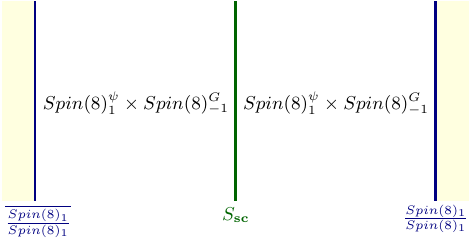}
\end{subfigure}%
\hspace{4cm}
\begin{subfigure}{.45\textwidth}
  \includegraphics[scale=1.24]{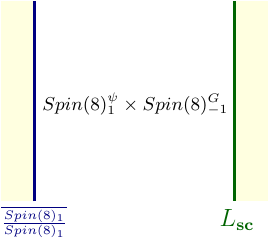}
\end{subfigure}
\caption{On the left: Configuration at the infrared fixed point $g_{YM}^{2} \to \infty$. The collapse of the region with the Yang-Mills kinetic term in Fig.\,\ref{chiral2DQCD-3DConstruction} (grey region) gives the identity interface, and what remains in bulk is the surface $S_{\mathbf{s} \mathbf{c}}$ that implemented the permutation modular invariant in the UV. On the right: Final configuration of the system after we push the topological interface $S_{\mathbf{s} \mathbf{c}}$ to the right boundary. This consists of a bulk 3D TQFT $Spin(8)_{1} \times Spin(8)_{-1}$ with two different topological boundary conditions on the left and on the right, characterized by the Lagrangian subgroups \eqref{DiagonalSpin8lv1Lagrangian} and \eqref{trialityLagrangian} respectively.}
\label{IR-Chiral2DQCD} 
\end{figure}

In the extreme infrared limit, the region containing the pure Yang-Mills kinetic term collapses to an interface. This is analogous to what happens in the non-chiral case that we review in Appendix \ref{CircleInterval2DQCD}, where the assumption that this interface reduces to the trivial interface is equivalent to the assumption that the IR is given by the topological coset in Figure \ref{GoverKCoset}. Notice that the collapse of the bulk region to an interface is a local operation, in the sense that the resulting collapsed interface is not sensitive to the boundary conditions to the left or to the right or to the insertions of any other topological surfaces inserted to its left or right in the bulk. Thus, the assumption in the non-chiral case that the collapsed interface gives the trivial interface can be applied again in the chiral case, and we obtain the configuration shown at the left side of Figure \ref{IR-Chiral2DQCD}.

Recall that the coset boundary conditions appearing at the left and right boundaries in the left side of Figure \ref{IR-Chiral2DQCD} are the same thus far as in the non-chiral theory and are characterized by the diagonal Lagrangian subgroup:
\begin{equation} \label{DiagonalSpin8lv1Lagrangian}
    L_{D} = (0,0) \oplus (\mathbf{v}, \mathbf{v}) \oplus (\mathbf{s}, \mathbf{s}) \oplus (\mathbf{c}, \mathbf{c}),
\end{equation}
which tells us the anyons that end perpendicularly at such topological boundary, as discussed in Section \ref{GappedBoundariesandTopologicalCosets}. The chirality of the full theory is taken into account by the middle interface $S_{\mathbf{s} \mathbf{c}}$, as explained above.

Now, we can push the higher-gauging surface $S_{\mathbf{s} \mathbf{c}}$ to the right boundary. Clearly, this changes the boundary condition, since the anyons that end on the boundary are permuted according to triality. For instance:
\begin{equation}
    \includegraphics[scale=0.2, valign=c]{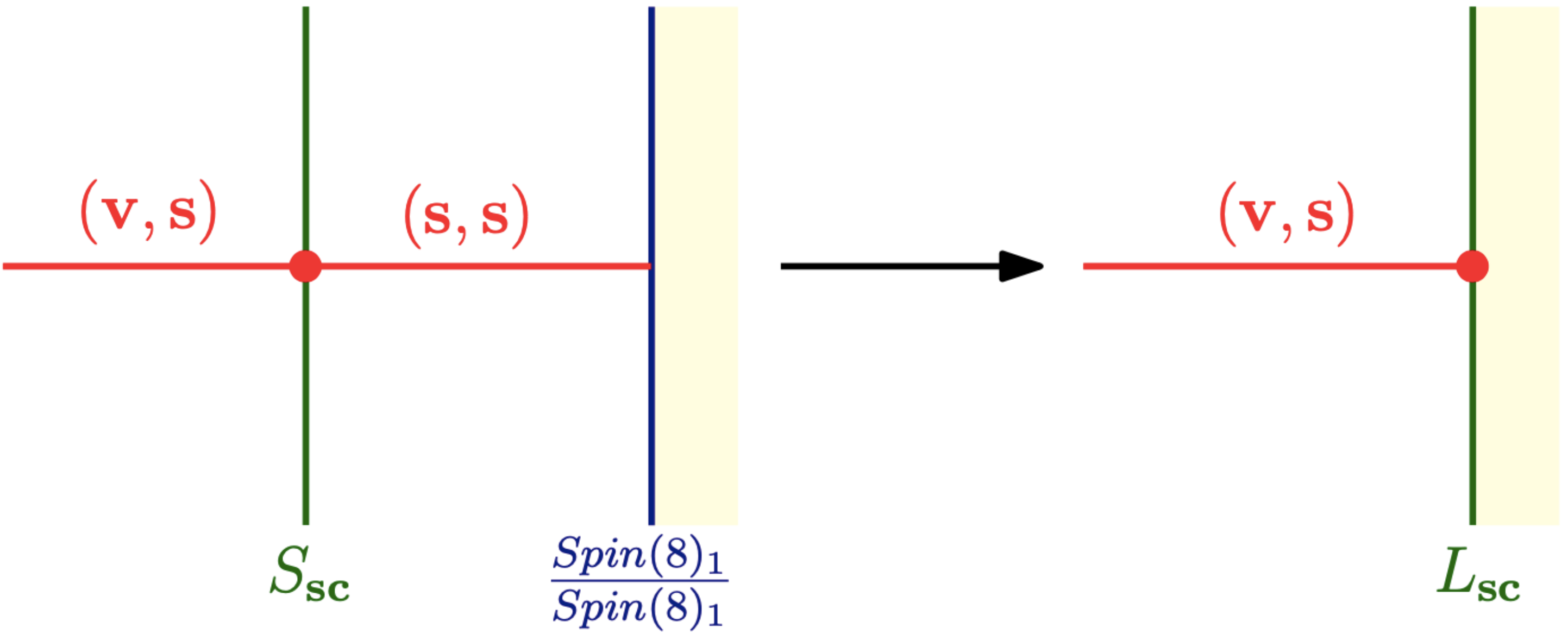} \ , \nonumber
\end{equation}
and similarly for other anyon permutations. The final configuration is shown at the right of Figure \ref{IR-Chiral2DQCD}, where the new right topological boundary is characterized by the Lagrangian subgroup $L_{\mathbf{s} \mathbf{c}}$:
\begin{equation} \label{trialityLagrangian}
    L_{\mathbf{s} \mathbf{c}} = (0,0) \oplus (\mathbf{v}, \mathbf{s}) \oplus (\mathbf{s}, \mathbf{c}) \oplus (\mathbf{c}, \mathbf{v}),
\end{equation}
which is easy to derive from the diagonal Lagrangian subgroup, and the action of triality.

Now that we know the final configuration, all that remains is to study the 2D TQFT arising under compactification to obtain the IR fixed point of $Spin(8)$ chiral 2D QCD. However, this final configuration is rather simple, since the two Lagrangian subgroups \eqref{DiagonalSpin8lv1Lagrangian} and \eqref{trialityLagrangian} characterizing the topological boundaries have no anyons in common other than the identity anyon. As a result, following \cite{Zhang:2023wlu}, no topological local operator (vacua) arises under compactification, and the system must be trivially gapped.

\begin{figure}[t]
        \centering
        \includegraphics[scale=0.80]{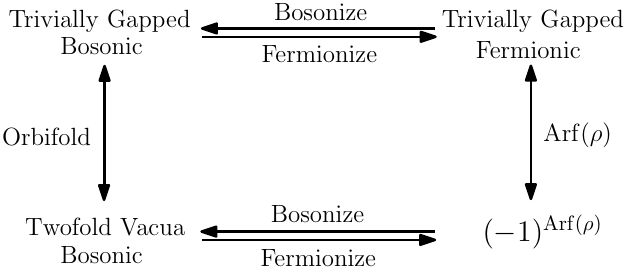} 
        \caption{Commutative diagram showing the interplay between bosonizations and fermionizations of the trivially gapped IR fixed points found in the strongly coupled regime of chiral $Spin(8)$ 2D QCD with massless fermions in the vectorial and spinorial representations.} \label{Bosonization-Fermionization-Map}
\end{figure}

We may now go back to the original fermionic theory by inverting the bosonization map (as summarized in Appendix \ref{Fermionization}), where we find a trivial fermionic theory, with a single trivial state in both the Neveu-Schwarz and Ramond sectors.

It is interesting to ask what happens if instead we consider the modular invariant \eqref{trialitymodularinvariant2}, corresponding to bosonizing the original chiral fermionic theory after we stack it with an Arf invariant. Following the same arguments given in this section, it is straightforward to derive that now the IR has two vacua. Then, inverting the bosonization map we find that, unlike in the previous case, the corresponding fermionic theory has one state in the Neveu-Schwarz sector and one state in the Ramond sector, with the latter charged under $(-1)^{F}$. As expected, both fermionic endpoints just discussed differ by the Arf invariant. Indeed, all the bosonic and fermionic IR endpoints can be seen to fit into Figure \ref{Bosonization-Fermionization-Map} in Appendix \ref{Fermionization}, which summarizes the relations between different bosonizations/fermionizations and topological manipulations on bosonic and fermionic CFTs.

\acknowledgments
We thank J. Harvey, C. Zhang and D. Delmastro for conversations. CC and DGS acknowledge support from the Simons Collaboration on Global Categorical Symmetries, the US Department of Energy Grant 5-29073, and the Sloan Foundation. DGS is also supported by a Bloomenthal Fellowship in the Enrico Fermi Institute at the University of Chicago.

\appendix

\section{A Summary on the Algebraic Theory of Anyons} \label{MTCsection}

In this appendix we provide a summary of the main tools regarding Modular Tensor Categories (MTC) that will be used in the main text to characterize topological cosets via anyon condensation. The content of this first section is well explained in many other references (see e.g. \cite{Kitaev:2005hzj, Benini:2018reh}), so the main point is to introduce notation and the necessary quantities of interest.

In the algebraic formulation of 3D TQFTs, one is interested in studying a set of topological line operators, called anyons, that are mathematically described by the objects of a MTC $\mathcal{C}$. An arbitrary anyon can always be expanded in a non-negative integer linear combination of a finite set of elementary anyons referred to as the \textit{simple anyons} of the theory. These are the simple objects of the MTC $\mathcal{C}$. The simple anyons of $\mathcal{C}$ are denoted as $a,b,c, \mathrm{etc.}$, while the set of all of the simple anyons in $\mathcal{C}$ is denoted by $\mathcal{I}$. The simple anyons in a MTC can fuse according to the fusion rules:
\begin{equation}
    a \otimes b = b \otimes a = \bigoplus_{c \in \mathcal{I}} N_{ab}^{c} \, c \ ,
\end{equation}
where the $N_{ab}^{c}$ are non-negative integers known as the fusion coefficients. The fusion product in an MTC is commutative and associative: $(a \otimes b) \otimes c = a \otimes (b \otimes c)$.

By definition, there exists a unique trivial anyon $0$ (called variously the vacuum, the identity anyon, etc.) such that $a \otimes 0 = 0 \otimes a = a$ for any $a \in \mathcal{I}$. Furthermore, given this identity anyon, for any simple $a \in \mathcal{I}$, there exists a unique simple $\bar{a}$ such that $a \otimes \bar{a} = 0 \oplus \cdots$ and thus we call $\bar{a}$ the conjugate of $a$. A simple anyon $a$ is said to be \textit{abelian} if and only if 
\begin{equation}
    \forall b \in \mathcal{I}, \quad a \otimes b = b \otimes a = c
\end{equation}
for some $c \in \mathcal{I}$. That is, for any simple anyon $b$, fusion with $a$ returns a single simple anyon $c$. Simple abelian anyons in a MTC always generate some abelian group, from which their name is derived. Otherwise, if there is more than one summand on the right-hand side for at least one simple $b$:
\begin{equation}
    a \otimes b = b \otimes a = c_{1} + c_{2} + \cdots,
\end{equation}
the simple anyon $a$ is called \textit{non-abelian}, or \textit{non-invertible}.

Importantly, the abstract data of a MTC can be represented by well-known diagrammatic expressions, whereby anyons are represented by lines extending through time:
\begin{equation} \label{aline}
    \includegraphics[scale=0.06, valign=c]{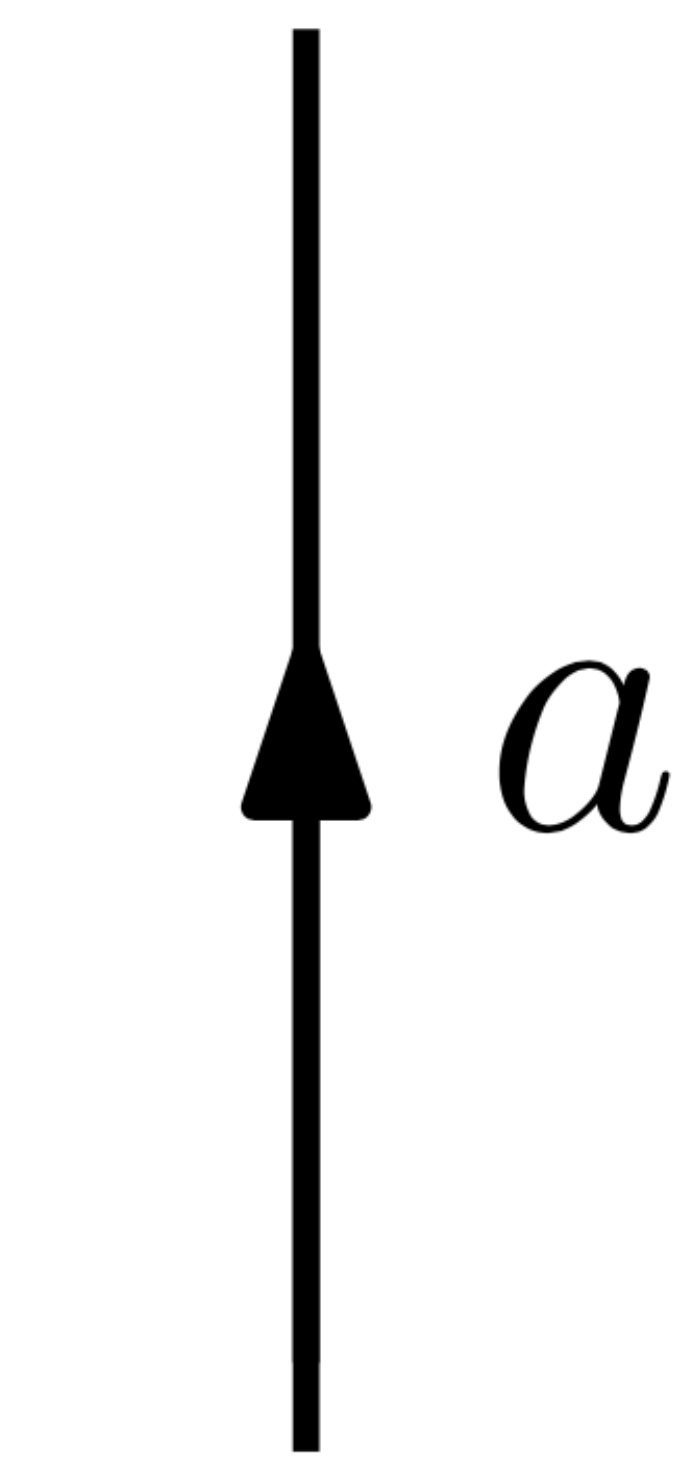} \, .
\end{equation}
In terms of these diagrams, the fusion product just introduced corresponds to the parallel fusion of the lines. Fusion also allows us to construct splitting vector spaces $V^{ab}_{c}$ and fusing vector spaces $V^{c}_{ab}$ of dimension $N_{ab}^{c}$ associated with trivalent junctions of the anyons. Equivalently, these are the vector spaces assigned to the two-sphere with three punctures in radial quantization. Choosing a basis in these vector spaces allows us to define the following vertex diagrams, from which we can construct more complex diagrams:
\begin{equation} \label{Splitting-and-Fusion}
    \Big( \frac{d_{c}}{d_{a} d_{b}} \Big)^{1/4}  \includegraphics[scale=0.06, valign=c]{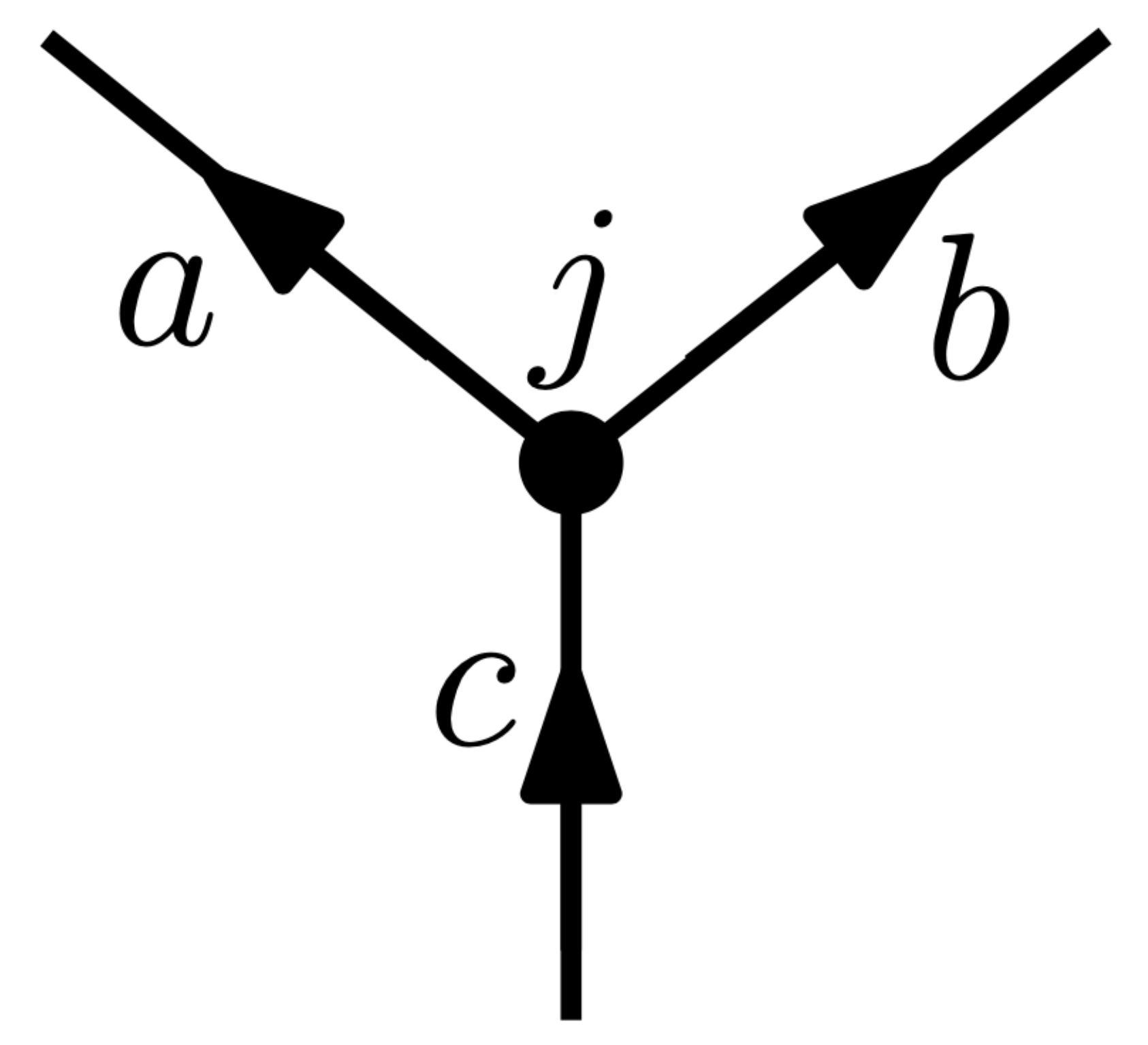} =: |a,b,c;j\rangle \in V^{ab}_{c} \,, \ \ \ \ \Big( \frac{d_{c}}{d_{a} d_{b}} \Big)^{1/4}  \includegraphics[scale=0.06, valign=c]{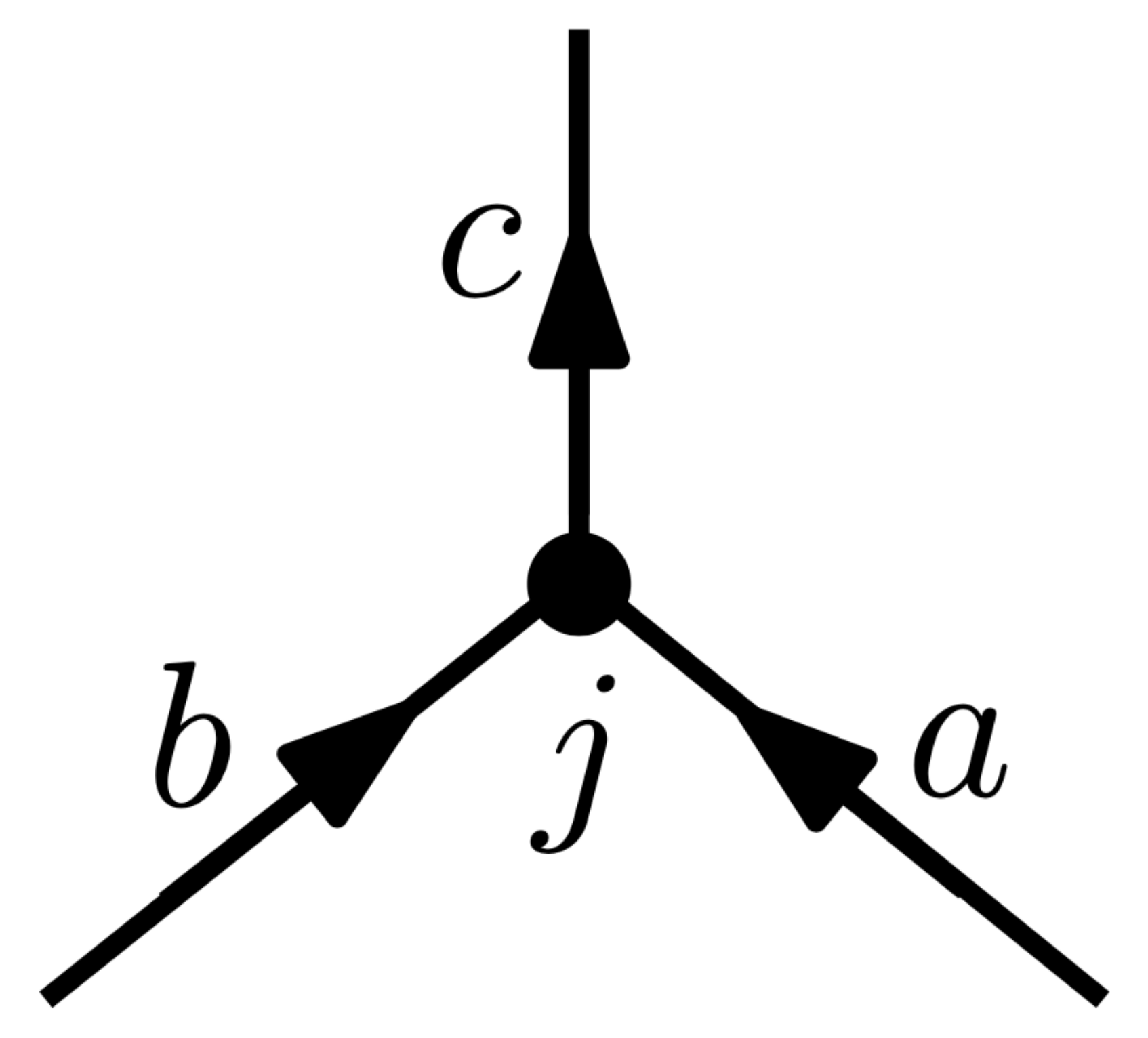} =: \langle a,b,c;j| \in V_{ab}^{c} \, .
\end{equation}
In these expressions, we have used the quantum dimension $d_{a}$ of a simple anyon $a$, defined as the largest eigenvalue of the matrix $\mathrm{\mathbf{N}}_{a}$ with entries defined by the fusion coefficients $(\mathrm{\mathbf{N}}_{a})_{bc} \coloneqq N_{ab}^{c}$. The dimension of a non-simple anyon $\oplus_{a} a$ is then defined merely as the addition of the quantum dimensions of each component: $d_{\oplus_{a} a} = \sum_{a} d_{a}$. An important property of the quantum dimensions is that they satisfy the fusion algebra: $d_{a} d_{b} = \sum_{c \in \mathcal{I}} N_{ab}^{c} d_{c}$. It is also important to define the dimension of the MTC, which in turn is given by the quantum dimensions of the individual simple anyons as
\begin{equation} \label{totalquantumdimension}
    \mathrm{dim}(\mathcal{C}) = \sum_{a \in \mathcal{I}} d_{a}^{2} \, .
\end{equation}

We may construct further diagrams using the trivalent vertices. An important one in the following is the bubble diagram:
\begin{equation} \label{Bubble}
    \includegraphics[scale=0.10, valign=c]{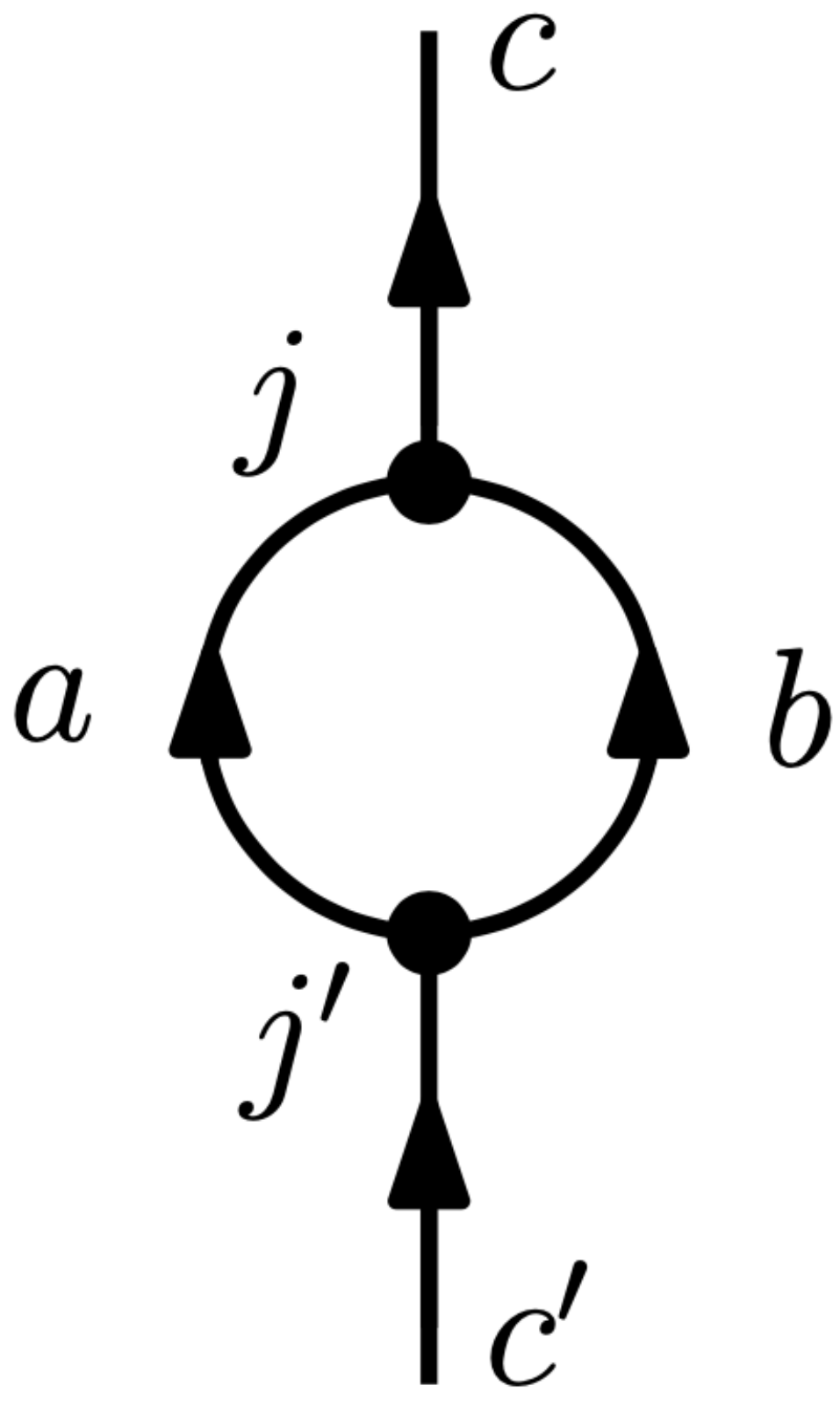} = \delta_{c,c'} \, \delta_{j,j'} \, \sqrt{\frac{d_{a} d_{b}}{d_{c}}}  \ \includegraphics[scale=0.10, valign=c]{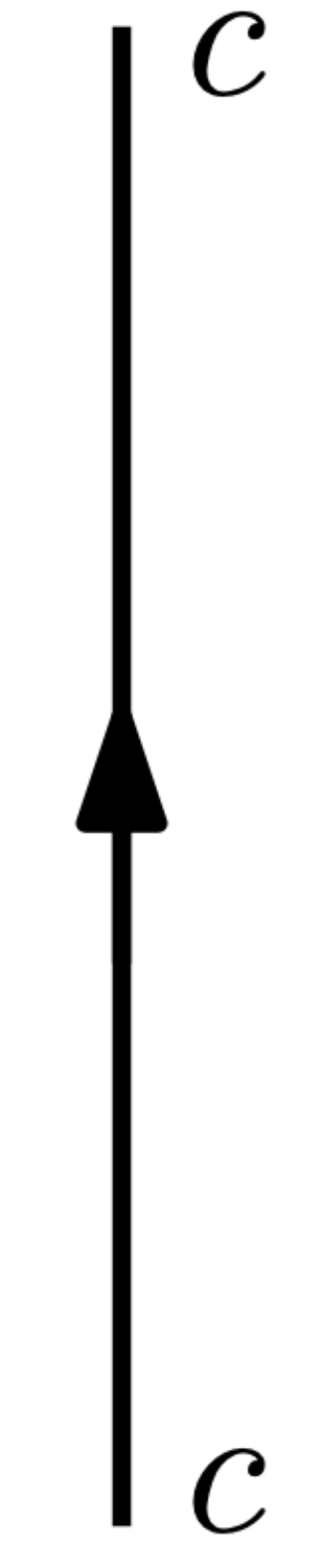} \ ,
\end{equation}
from which, taking $c = c'= 0, a=b$ we recover an expression for the quantum dimension $d_{a}$ of a simple anyon $a$ as the expectation value of an unknot of $a$:
\begin{equation} \label{QuantumDimension}
    d_{a} = \includegraphics[scale=0.05, valign=c]{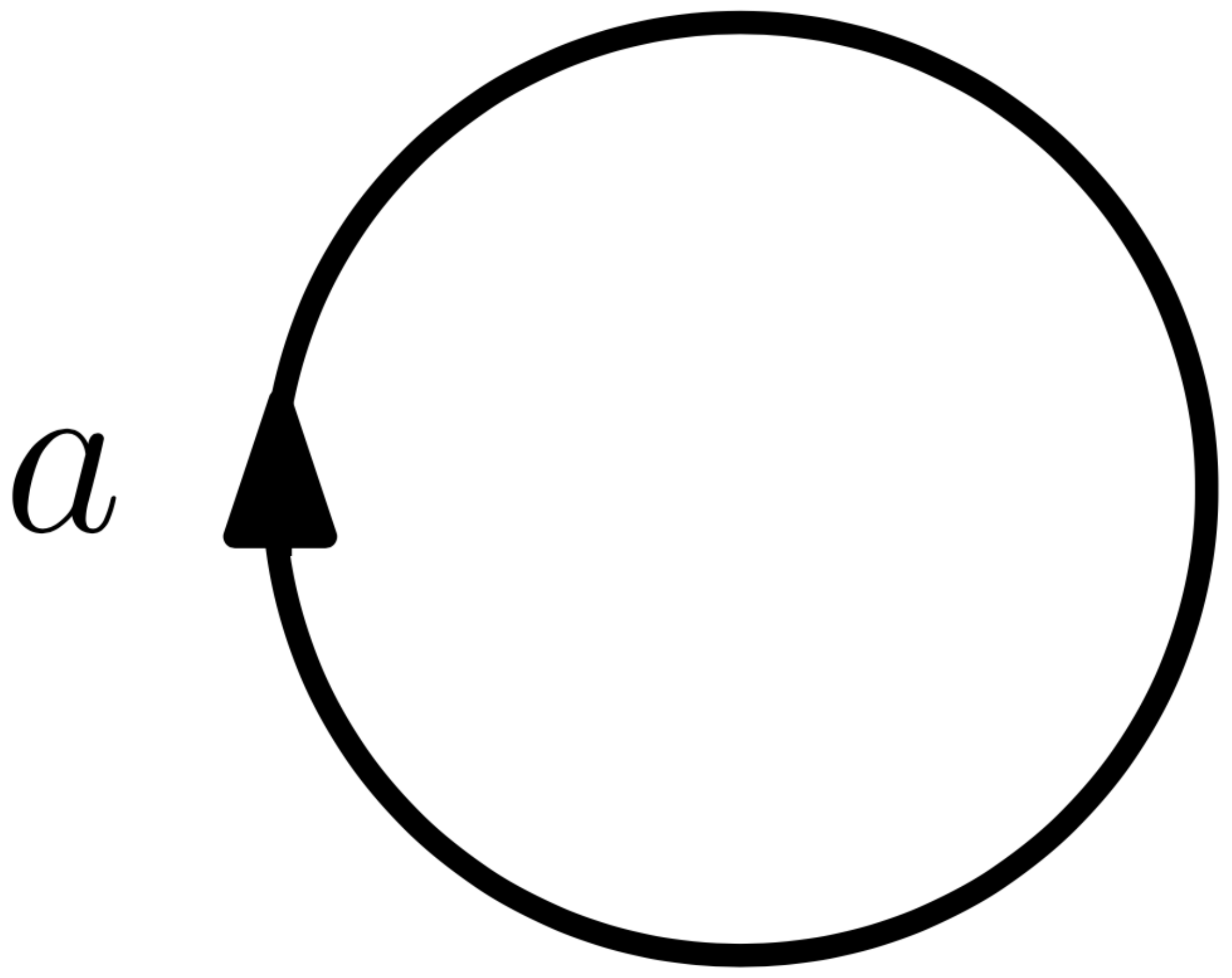} \ .
\end{equation}

We can now compose trivalent junctions to draw a diagram splitting a simple anyon $d$ into three $a,b,c$. There are two ways to do this: either fusing $a$ with $b$ first, or $b$ with $c$ first. As is well-known, the order is immaterial since it corresponds to the same splitting written in two different bases related by a unitary transformation. This unitary transformation is implemented by the $F$-symbols, which encode the associativity of splitting a single simple anyon into three. Diagrammatically:
\begin{equation} \label{FSymbolDef}
    \includegraphics[scale=0.10, valign=c]{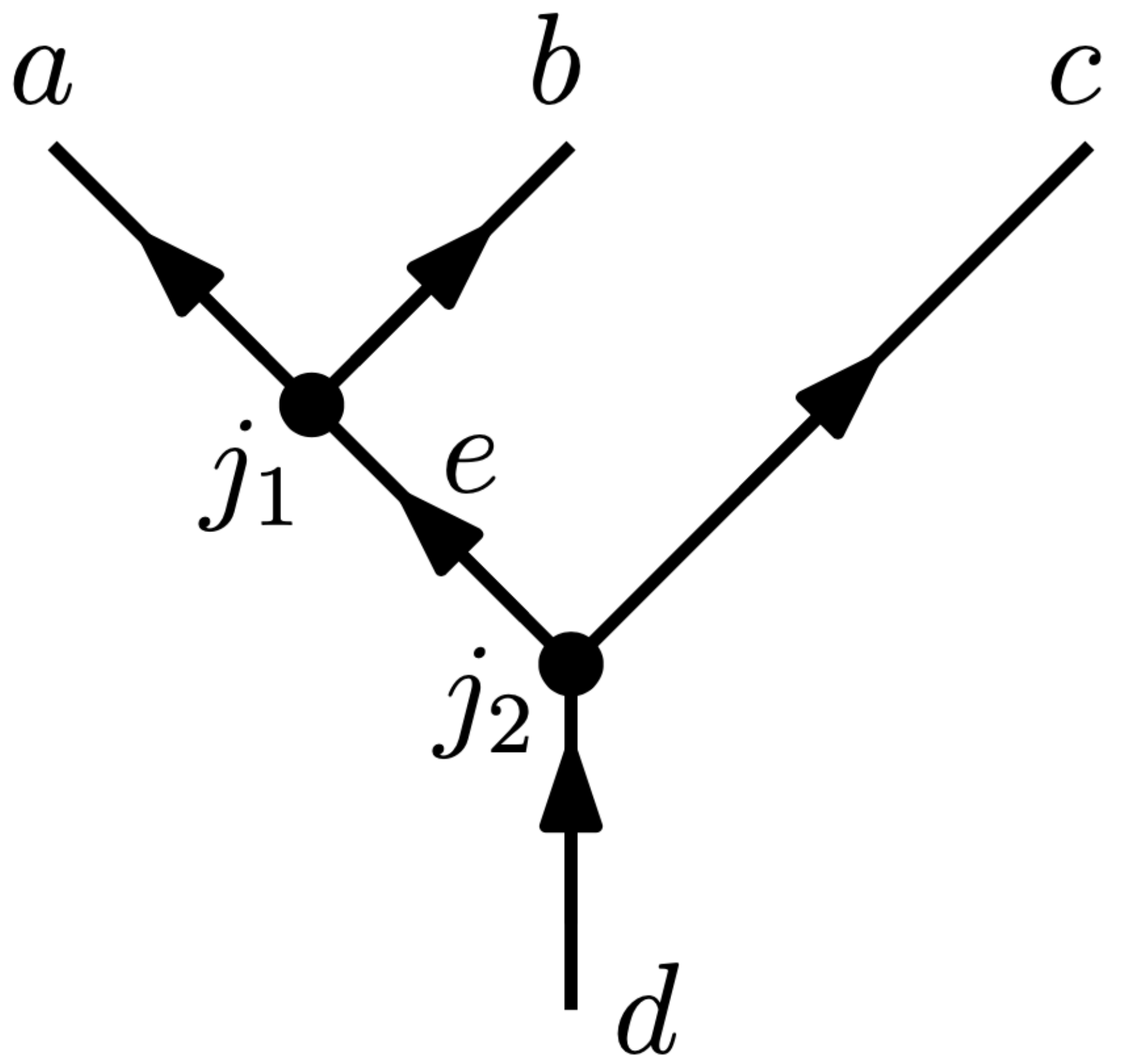} = \sum_{(f, k_{1}, k_{2})}[F^{abc}_{d}]_{(e, j_{1}, j_{2}), (f, k_{1}, k_{2})} \includegraphics[scale=0.10, valign=c]{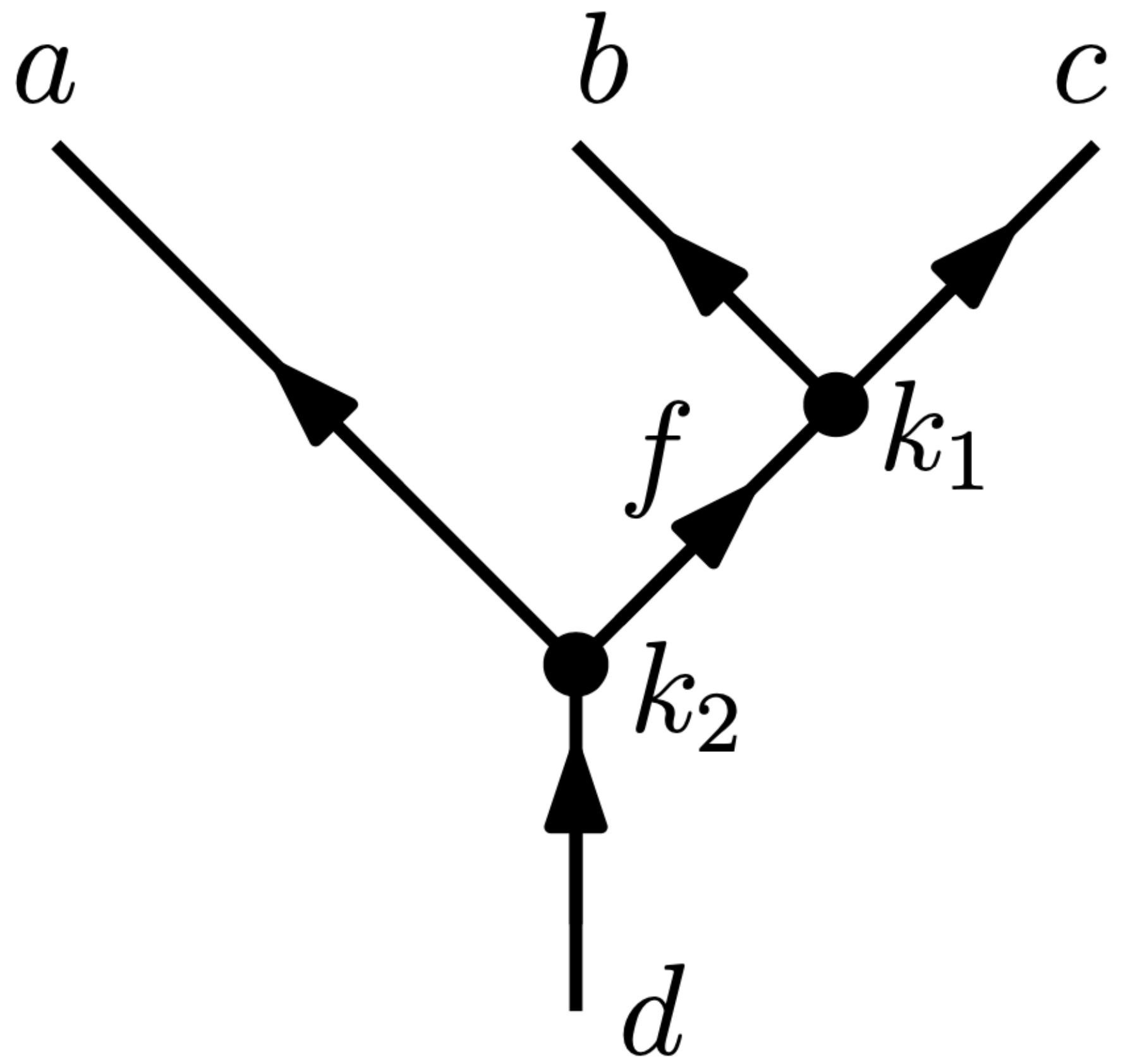},
\end{equation}
where the diagrams on the left and on the right are the different representations of the same splitting just mentioned. In a unitary theory, there always exists a basis where the matrices $[F^{abc}_{d}]$ are unitary (for a more in-depth discussion of unitarity in the context of ribbon categories, see \cite{galindo2014braided}).

Rather than being arbitrary numbers, the $F$-symbols defining an MTC are constrained by the so-called pentagon equations, which guarantee consistency of the theory when applying $F$-symbols on higher-point diagrams. We will not need these equations in the following. See the original references \cite{Moore:1988uz, Moore:1988qv} for details.

Famously, anyons satisfy non-trivial braiding statistics. Algebraically, this fact is encoded in the $R$-symbols defining an MTC. In components, this is given by the diagrammatic expression
\begin{equation} \label{RSymbolDef}
    \includegraphics[scale=0.08, valign=c]{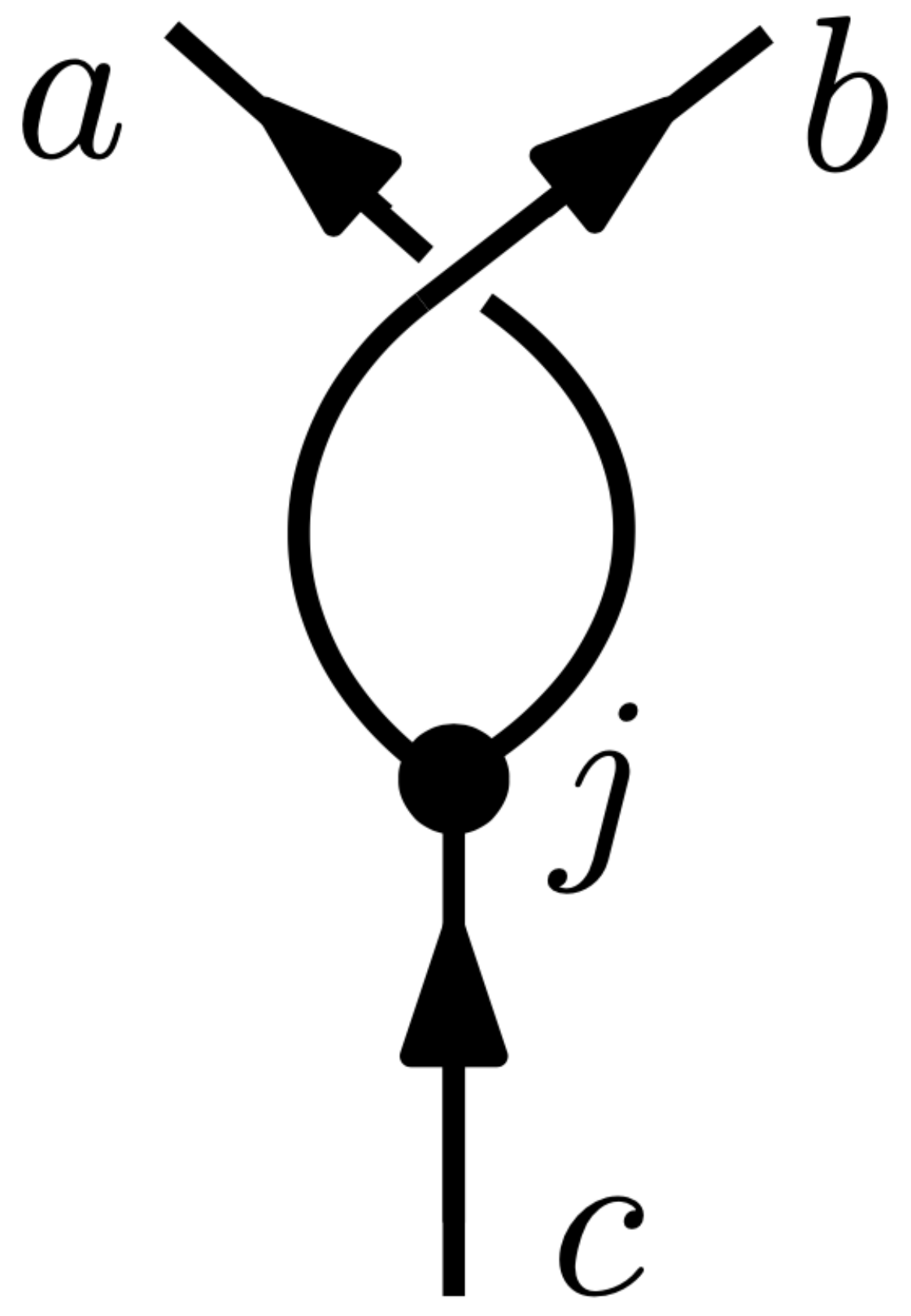} = \sum_{ k }[R^{ab}_{c}]_{j k} \includegraphics[scale=0.075, valign=c]{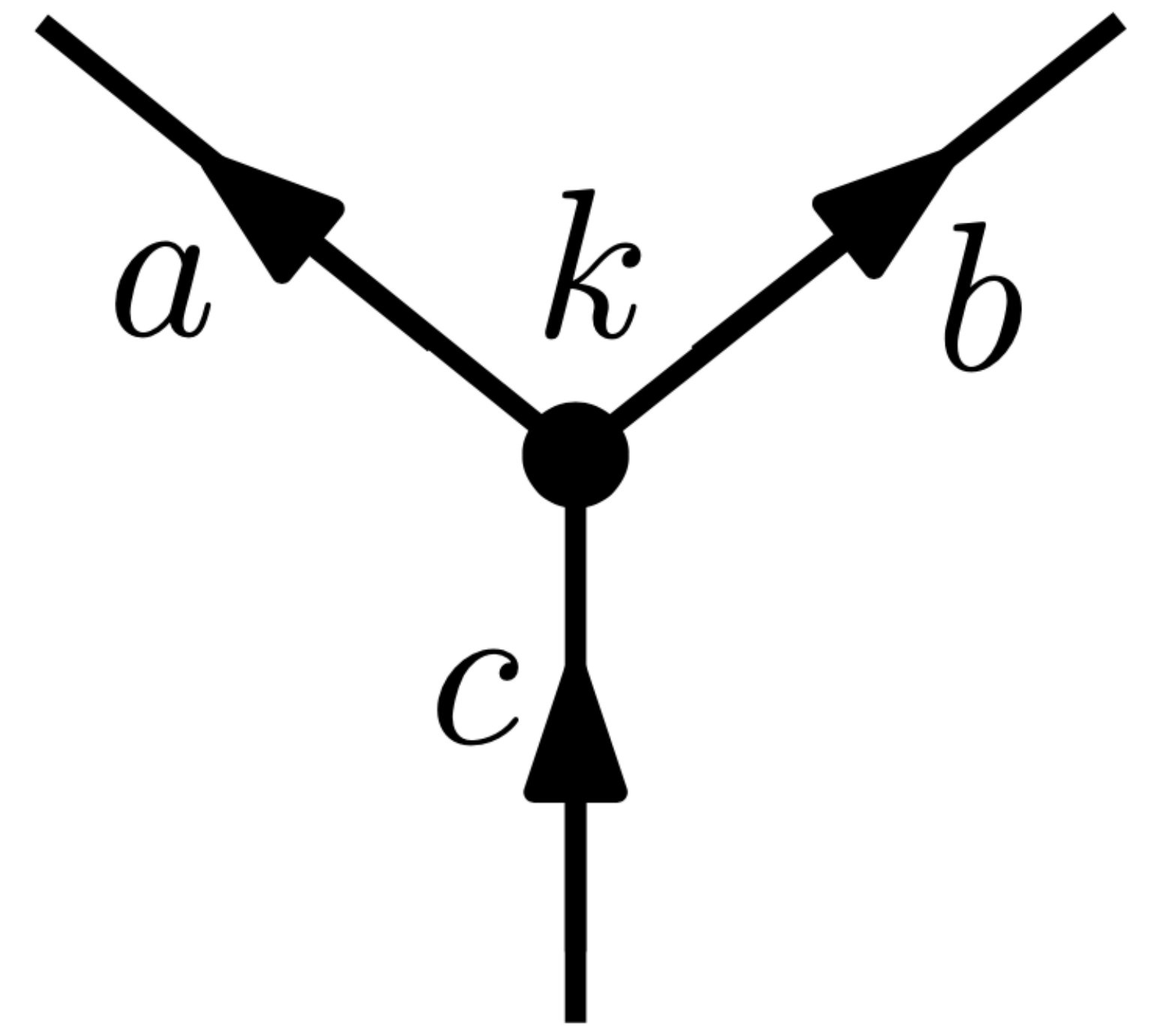}.
\end{equation}
Similar to the $F$-symbols, the $R$-symbols are also constrained by analogous consistency conditions called hexagon equations. However, we will also not need these relations in the following. See  \cite{Moore:1988uz, Moore:1988qv} for details.

Straightening a simple line $a$ that ``braids with itself'' under a $2\pi$ rotation gives the definition of a phase known as the topological spin $\theta_{a}$:
\begin{equation} \label{TopologicalTwist}
    \includegraphics[scale=0.09, valign=c]{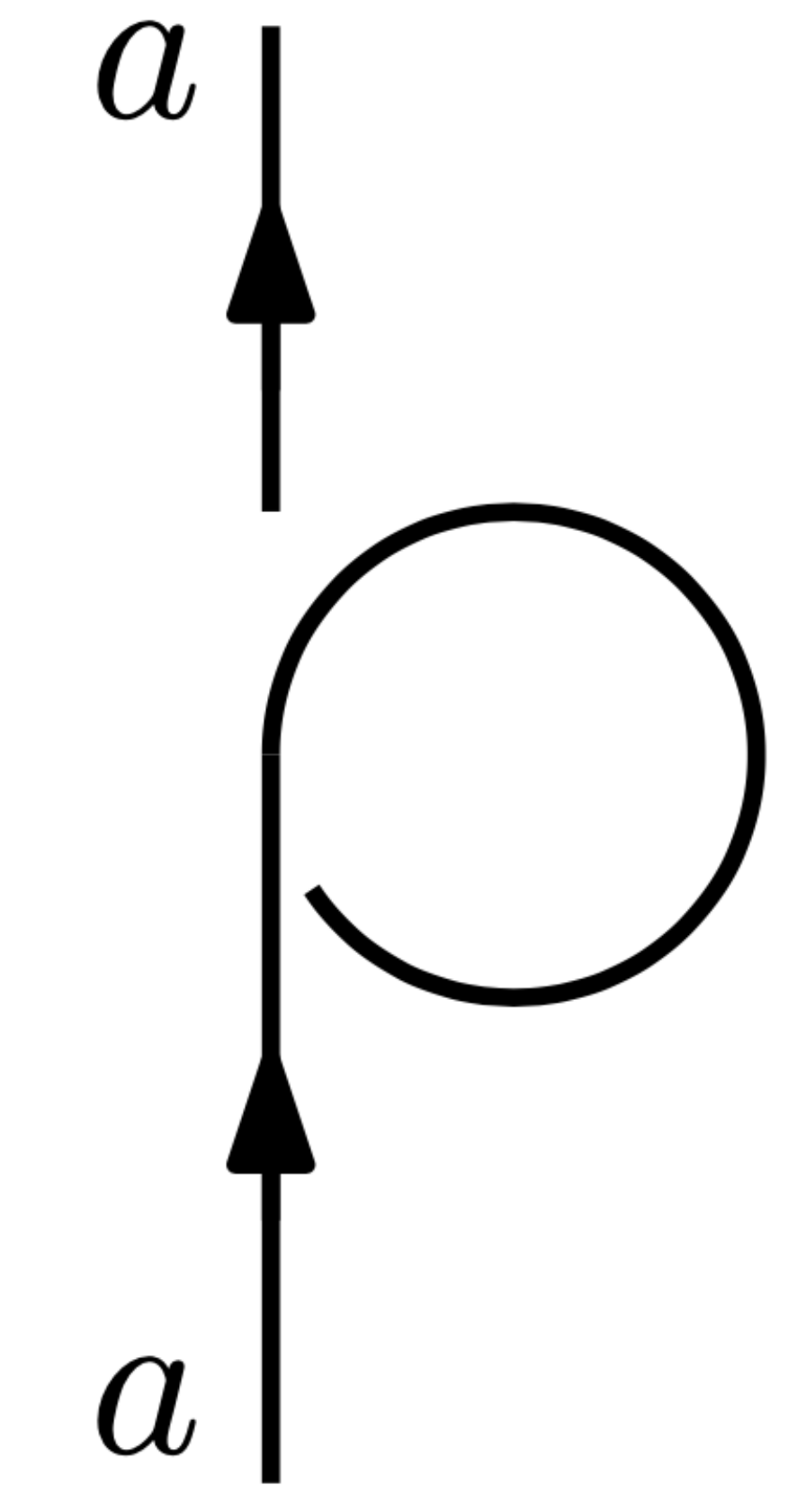} = \theta_{a} \includegraphics[scale=0.09, valign=c]{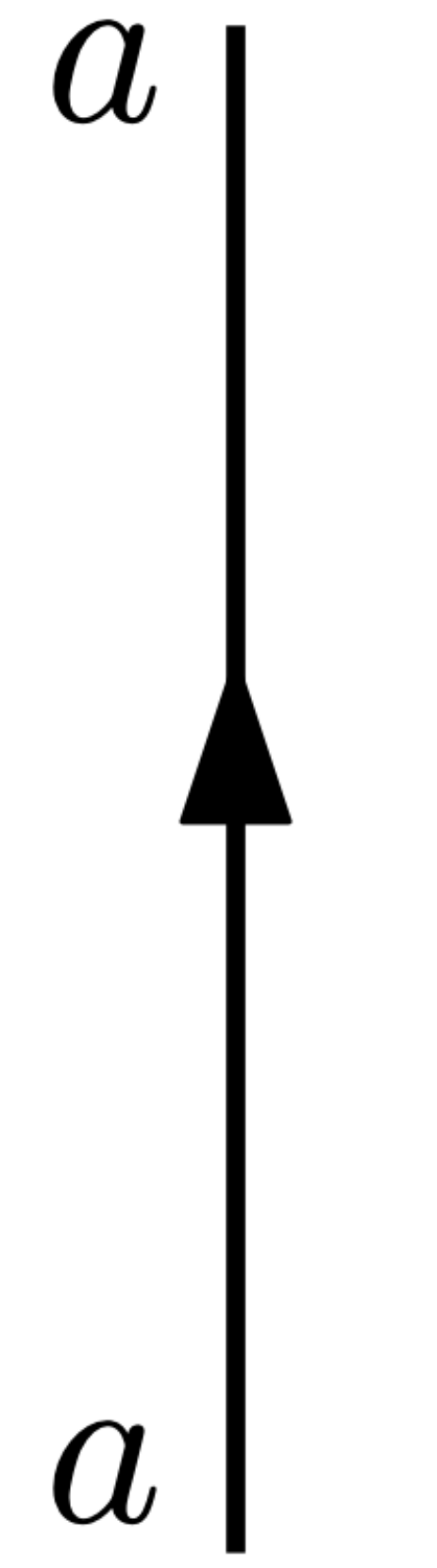}.
\end{equation}
When meaningful, the topological spin is given by $\theta_{a} = e^{2 \pi i h_{a}}$, with $h_{a}$ the conformal weight of some 2D RCFT primary $a$ related to the 3D TQFT by bulk-boundary correspondence. Often, $h_{a}$ defined mod 1 is used to refer to the topological spin of a line instead of using $\theta_{a}$ directly. It is a result of \cite{Anderson:1987ge, VAFA1988421} that $h_{a}$ is always a rational number. If $\theta_{a} = 1$, we say that the simple anyon $a$ is a \textit{boson}.

We define a matrix $S$ with entries valued on the simple anyons of the MTC as: 
\begin{equation} \label{modularS}
    \frac{S_{ab}}{S_{00}} = \includegraphics[scale=0.07, valign=c]{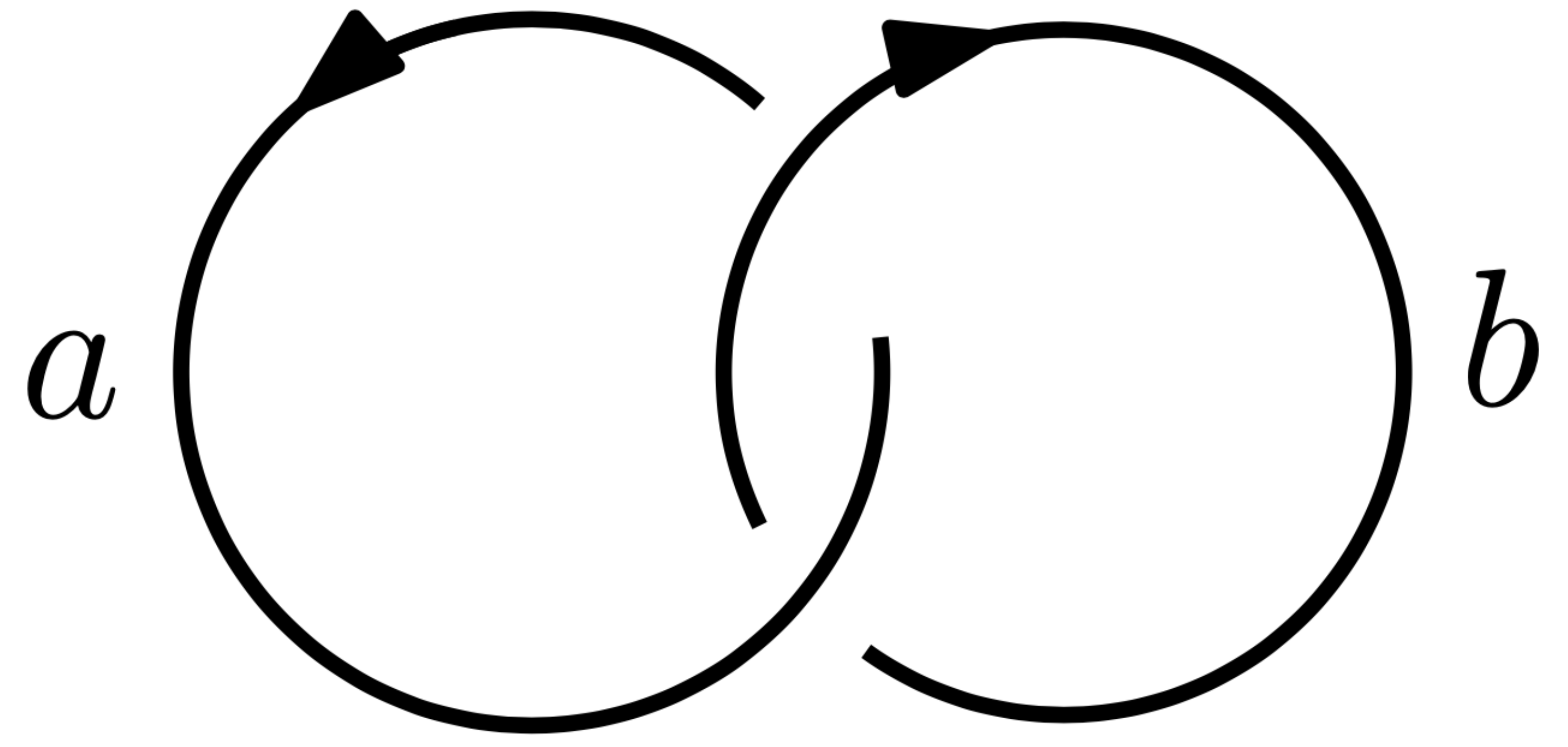},
\end{equation}
where $S_{00} = 1/\sqrt{\mathrm{dim}(\mathcal{C})}$. This is the standard modular $S$-matrix. The statement that this matrix is non-degenerate provides the final requirement in the definition of a MTC. In passing, from the modular $S$-matrix we can define the braiding phase:
\begin{equation} \label{BraidingPhase}
    \includegraphics[scale=0.08, valign=c]{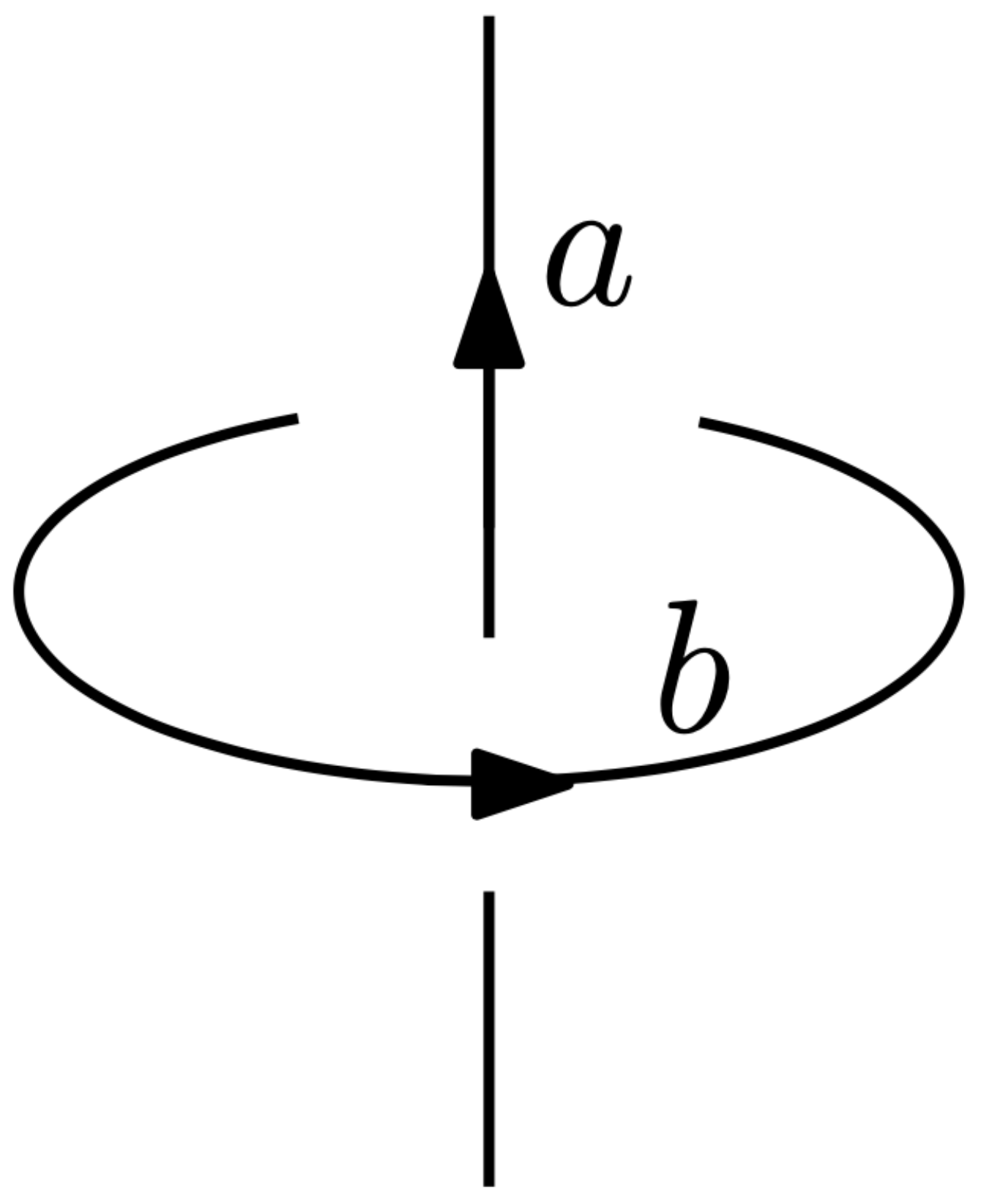} = \frac{S_{ab}}{S_{a0}} \includegraphics[scale=0.08, valign=c]{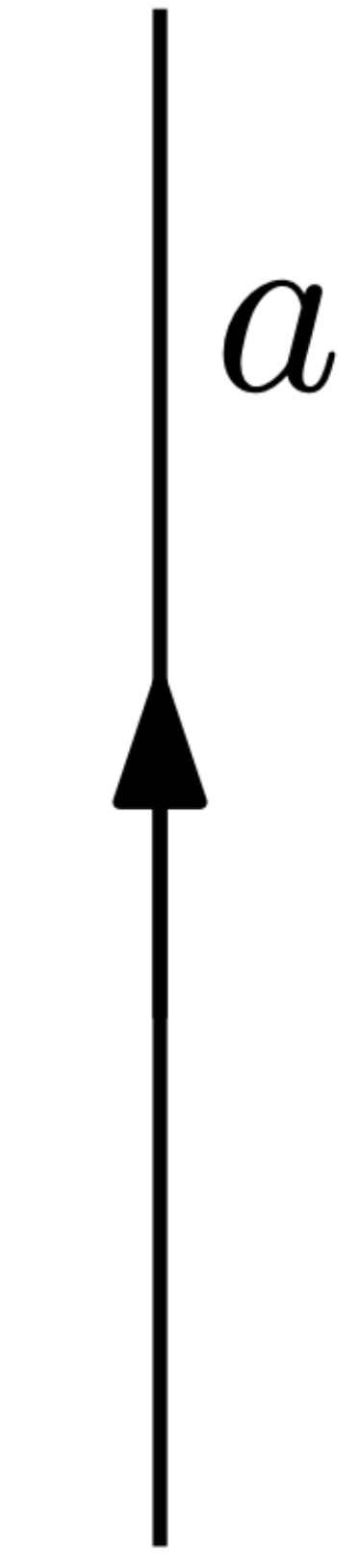},
\end{equation}
which allows us to unlink anyons circling around each other. Setting $a=0$ we recover the well-known expression for the quantum dimension for the simple anyon $b$ in terms of the modular $S$-matrix: $d_{b} = S_{0b}/S_{00}$.

\begin{figure}[!b]  
        \centering
        \includegraphics[scale=1.2]{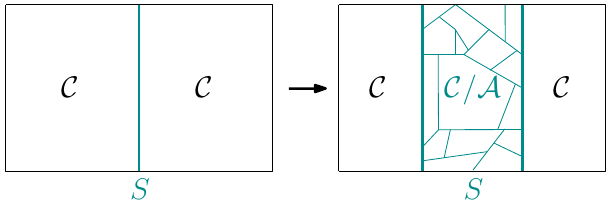} 
        \caption{When the surface $S$ in a MTC $\mathcal{C}$ is generated by a commutative Frobenius algebra $\mathcal{A}$ along a surface, we can think of the surface as a small sliver of volume with topological interfaces enclosing the topological order $\mathcal{C}/\mathcal{A}$.} \label{BlowingUp-CommutativeSurface}
\end{figure}

Finally, given a 3D TQFT described by a MTC $\mathcal{C}$, we write $\bar{\mathcal{C}}$ for the orientation-reversal of $\mathcal{C}$. In practice, this amounts to a MTC with the same number of simple anyons and the same fusion ring, but with all quantities defined above (e.g. topological spins, or $F$-symbols) differing by a complex conjugation. For instance, for simple $a \in \mathcal{C}$, call the associated anyon after orientation-reversal $a_{\mathrm{rev}} \in \bar{\mathcal{C}}$, then $\theta_{a_{\mathrm{rev}}} = \theta_{a}^{*}$ (for the precise mathematical definition of orientation-reversal, see Def. 6.13. in \cite{Frohlich:2003hm}). For a Chern-Simons theory $G_{k}$, this operation amounts to changing the sign of the level $G_{k} \to G_{-k}$. Below, unless we wish to emphasize the orientation-reversal, we will abuse language and repeat the notation for anyons in the original theory and the one obtained after orientation-reversal, as it will be clear from context with which theory we are working with.

\section{Circle and Interval Constructions of Massless $\mathrm{\bf{QCD_{2}}}$} \label{CircleInterval2DQCD}

\begin{figure}[t]  
        \centering
        \includegraphics[scale=1.4]{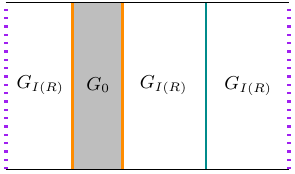} 
        \caption{Circle compactification construction of 2D QCD \cite{Komargodski:2020mxz}. The purple dotted lines are identified to form a circle geometry. The topological order $G_{I(R)}$ is glued along the circle by a $G_{0}$ Yang-Mills term. A topological surface in the $G_{I(R)}$ topological order (in green) is inserted to induce the embedding $G_{I(R)} \hookrightarrow Spin(\mathrm{dim}(R))_{1}$ capturing the free fermions in the UV.} \label{2DQCD-CircleCompactification}
\end{figure}

\begin{figure}[!b] 
        \centering
        \includegraphics[scale=1.4]{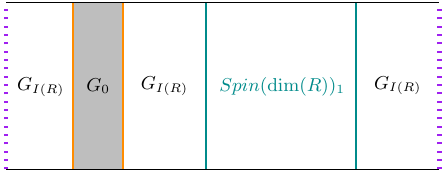} 
        \caption{We expand the topological surface into a non-trivial volume enclosing the $Spin(\mathrm{dim}(R))_{1}$ topological order. } \label{2DQCD-CircleCompactification-BlownUp}
\end{figure}

In this appendix, we show how the 3D circle compactification construction of (bosonic) 2D QCD from \cite{Komargodski:2020mxz} (see also \cite{Gaiotto:2020iye}) is equivalent to the interval compactification described in Appendix A of \cite{Delmastro:2022prj}. We make sure to keep track of all appropriate bulk topological orders, which is important to set boundary conditions.

To start, notice that a way to construct a surface in a MTC $\mathcal{C}$ consists in performing non-abelian anyon condensation along a volume enclosing a MTC $\mathcal{C}/\mathcal{A}$ and collapsing the volume into a surface. Importantly, for non-abelian condensation to take place in a region, the Frobenius algebra $\mathcal{A}$ must be commutative (see Section \ref{GappedBoundariesandTopologicalCosets} for precise definitions). This implies that if we start with the operator generated by $\mathcal{A}$ along a surface, we can braid the algebra element outside of the locus of the surface and proliferate the surface into a volume that encloses $\mathcal{C}/\mathcal{A}$, as shown in Fig. \ref{BlowingUp-CommutativeSurface}. (For a related discussion of these type of defects, see Section 5.2 of \cite{Buican:2023bzl}.)

\begin{figure}[t]  \hspace{-0.5cm}
        \includegraphics[scale=1.2]{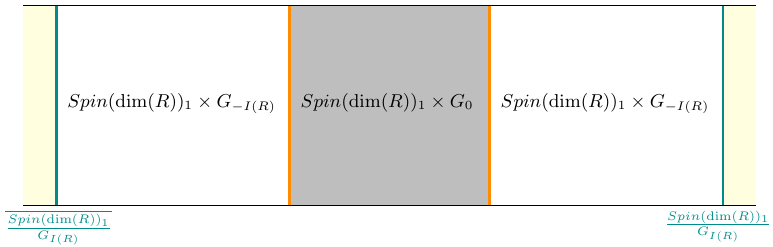} 
        \caption{After using the folding trick we obtain the interval compactification construction of 2D QCD.  Both the topological boundary conditions at the ends of this figure, and the topological interfaces connecting $G_{I(R)}$ with $Spin(\mathrm{dim}(R))_{1}$ in Fig. \ref{2DQCD-CircleCompactification-BlownUp} are due to a non-abelian anyon condensation $G_{I(R)}/\mathcal{A} = Spin(\mathrm{dim}(R))_{1}$. The interval compactification is more general, however, since it allows for the coset boundary conditions to be non-topological at the ends.} \label{2DQCD-IntervalCompactification}
\end{figure}

\begin{figure}[t]
        \centering
        \includegraphics[scale=0.4]{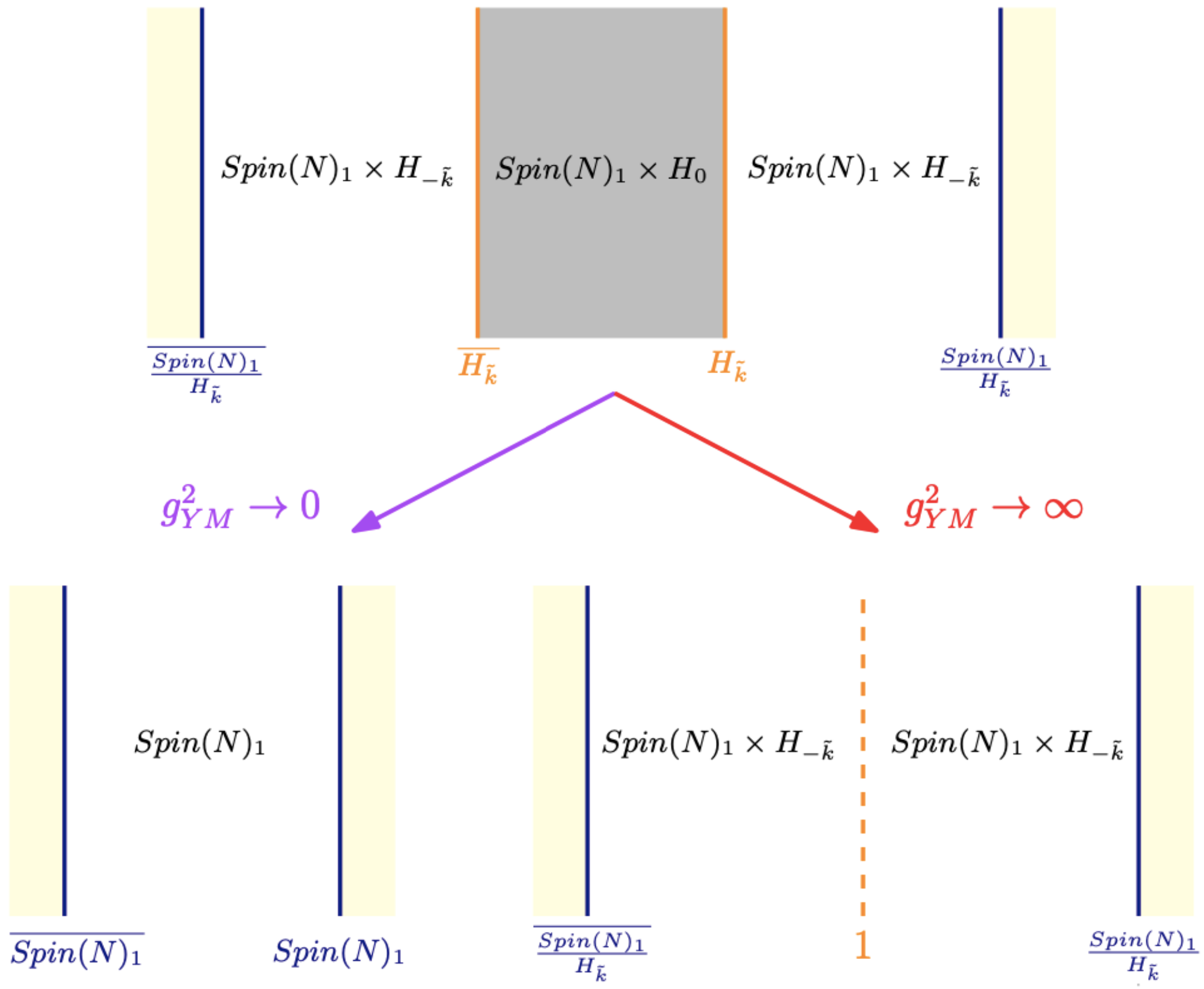} 
        \caption{Three-dimensional construction of 2D QCD upon interval compactification. We gauge the fermions and couple them to $H$ gauge fields according to the 3D realization of the coset construction, developing accordingly coset boundary conditions on the left and on the right (in blue) \cite{Moore:1989yh, Cordova:2023jip}. Meanwhile, in bulk we generate interfaces (in orange) by setting $A_{0} = 0$ boundary conditions for the $H_{\tilde{k}}$ gauge fields in the left and right regions of the bulk, which generates the standard $H_{k}$ chiral algebra boundary conditions on them \cite{Elitzur:1989nr}. We glue the interfaces via $H_{0}$ gauge fields carrying the 2D Yang-Mills kinetic term (grey region). The interfaces so generated are chosen to have transparent junction conditions for the $Spin(N)_{1}$ gauge fields. Clearly, in the UV as $g_{YM}^{2} \to 0$ we recover the expected standard 3D construction of the $Spin(N)_{1}$ theory (the bosonization of $N$ Majorana fermions), while in the IR the interfaces collapse to a defect in the $Spin(N)_{1} \times H_{-\tilde{k}}$ topological order. The statement that the infrared fixed point of massless 2D QCD is given by the topological coset is the statement that the defect so generated corresponds to the identity defect.} \label{3D-Construction-of-2DQCD}
\end{figure}

Next, we take the circle compactification construction of 2D QCD \cite{Komargodski:2020mxz}, reproduced here in Fig. \ref{2DQCD-CircleCompactification}. In this construction, the information that the gauge group can be embedded into the free fermions in the UV is encoded in the existence of a topological interface between $G_{I(R)}$ and $G_{I(R)}$ that gives rise to the modular invariant associated with the conformal embedding $G_{I(R)} \hookrightarrow Spin(\mathrm{dim}(R))_{1}$. The important point is that conformal embeddings are always implied by some non-abelian anyon condensation, or equivalently by gauging some commutative Frobenius algebra \cite{davydov2013witt, Huang:2014ixa}. In our case:
\begin{equation}
    G_{I(R)}/\mathcal{A} \cong Spin(\mathrm{dim}(R))_{1}.
\end{equation}

Using this fact, we can turn the surface into a volume containing $Spin(\mathrm{dim}(R))_{1}$, as shown in Fig. \ref{2DQCD-CircleCompactification-BlownUp}. We can now use the folding trick along the topological interfaces joining $G_{I(R)}$ with $Spin(\mathrm{dim}(R))_{1}$ to turn the circle compactification into an interval with topological boundary conditions at both ends. The result is shown in Fig. \ref{2DQCD-IntervalCompactification}, which is essentially the interval compactification of \cite{Delmastro:2022prj}, but where we have kept track of the $Spin(\mathrm{dim}(R))_{1}$ topological order to ensure appropriate coset boundary conditions at the ends. (See \cite{Cordova:2023jip} for a discussion on coset boundary conditions with a modern emphasis on anyon condensation.) Finally, notice that the interval compactification smoothly connects the gapless and gapped theories of \cite{Delmastro:2021otj}, as well as their criteria for a QCD theory to be gapped: If the coset boundary conditions support gapless degrees of freedom, the corresponding theory is gapless. On the other hand, if the coset boundary conditions are topological (in the case of the conformal embeddings), the corresponding theory is gapped. In the latter case, the construction also shows that the topological coset symmetry is present along the whole flow since the topological coset symmetry is localized at the boundary, while the flow is local in the bulk. See e.g. \cite{Kutasov:1994xq, Delmastro:2021otj, Delmastro:2022prj} for previous observations on this fact. All in all, from the 3D interval construction, the flow looks as in Fig. \ref{3D-Construction-of-2DQCD}.

\section{Bulk-to-Boundary Map, Quotient Category and Karoubi Envelope} \label{KaroubiEnvelope}

In Section \ref{TopologicalCosets}, the spectrum of lines in a topological coset and how such a spectrum of lines is obtained from the bulk MTC were discussed. However, in order to streamline the discussion, instead of providing a set of rigorous definitions and theorems to characterize the fusion category of line operators in the two-dimensional theory, we summarized a set of practical rules that allow one to study the spectrum of lines in a simplified manner. We provide a set of more rigorous results in this appendix. As an application of the formalism, we briefly outline an example of how to use this method to find the topological lines in a particularly pedagogical example in Section \ref{PedagogicalExample}.

As discussed in the subject of anyon condensation (see e.g. \cite{kong2014anyon}), the fusion category of line defects at the boundary is usually characterized in terms of $\mathcal{A}$-module categories. An alternative viewpoint, a bit more practical for our purposes, makes use of quotient categories and Karoubi envelopes. Specifically, we have the definitions:
\begin{definition} \label{QuotientCategoryDef}
	(Quotient Category $\mathcal{C}/\mathcal{A}$) Let $\mathcal{C}$ be a MTC,\footnote{More generally, a braided fusion category.} and let $\mathcal{A}$ be a special Frobenius algebra in $\mathcal{C}$. The quotient category $\mathcal{C}/\mathcal{A}$ consists of the category such that \vspace{0.2cm}
\begin{itemize}
    \item The objects in $\mathcal{C}/\mathcal{A}$ are the same as the objects in $\mathcal{C}$. \vspace{0.2cm}

    \item The morphisms of $\mathcal{C}/\mathcal{A}$ are given by
    \begin{equation} \label{HomsQuotientCategory}
        \mathrm{Hom}_{\mathcal{C}/\mathcal{A}}(a,b) = \mathrm{Hom}_{\mathcal{C}}(a, \mathcal{A} \otimes b).
    \end{equation}
\end{itemize}
One may add now definitions of composition and tensor product, but since we will not make use of these we refer the reader to \cite{muger2004galois} for details on the definition.
\end{definition}

The quotient category is guaranteed to be a tensor category when $\mathcal{A}$ is special Frobenius, but it is not guaranteed to be semisimple since $\mathrm{Hom}_{\mathcal{C}/\mathcal{A}}(a,a)$ could have dimension higher than one for an object $a$ that is simple in $\mathcal{C}$. In this case we make use of the following definition:

\begin{definition} \label{KaroubiEnvelopeDef}
	(Idempotent Completion or Karoubi envelope) For the quotient category $\tilde{Q} = \mathcal{C}/\mathcal{A}$, we construct the canonical idempotent completion $Q$ (or Karoubi envelope) of $\tilde{Q}$ as follows: \vspace{0.2cm}
\begin{itemize}
    \item The objects of $Q$ are pairs $(a,p)$, where $a \in \mathrm{Obj}(\tilde{Q})$, and $p = p^{2} \in \mathrm{End}_{\tilde{Q}}(a)$. \vspace{0.2cm}

    \item The morphisms of $Q$ are given by
    \begin{equation} \label{KaroubiEnvelopeHoms}
        \mathrm{Hom}_{Q}\big((a,p),(b,q) \big) = \{ f \in \mathrm{Hom}_{\tilde{Q}}(a,b) | f \circ p = q \circ f \},
    \end{equation}
    and other structures in $Q$ are inherited from $\mathcal{C}/\mathcal{A}$.
\end{itemize} 
\end{definition}

The category $Q$ so constructed is a semisimple tensor category. It is a result of M{\"u}ger that the construction mentioned above based on modules categories on the one side and quotient categories and Karoubi envelopes on the other side agree (see specifically Props. 2.11, 2.15, and 2.16 in \cite{muger2004galois}). 

The advantage of the formulation based on quotient categories and Karoubi envelopes is that it offers some computational organization without having to explicitly analyze the equations determining the modules (see e.g., eqn. (4.65) in \cite{Fuchs:2002cm} and the corresponding discussion). In Section \ref{PedagogicalExample} we illustrate this fact in a simple example, and it can of course be used in other examples (for a collection of calculations on non-invertible anyon condensation based on idempotents see for instance \cite{Yu:2021zmu, Cong:2016ayp}, or Appendix D in \cite{Putrov:2024uor} for a calculation based on idempotents applied to Haagerup fusion categories). Conceptually, it also allows for generalizations to higher dimensions \cite{Gaiotto:2019xmp}.

Following the discussion in Section \ref{TopologicalCosets}, the physical content of the idempotent completion construction is that some simple lines in the bulk become non-simple when pushed to the boundary, which happens whenever $\mathrm{Hom}_{\mathcal{C}/\mathcal{A}}(a,a)$ is of dimension greater than one. Then, the objects in Definition \ref{KaroubiEnvelopeDef} are simply the splittings of such bulk lines into the simple lines of the boundary theory. Furthermore, \eqref{HomsQuotientCategory} tells us that some lines in the bulk can descend into the same simple objects at the boundary. Thus, some of the previous splittings must be identified as the same simple lines of the boundary theory, in accordance with the common Hom-spaces as calculated from \eqref{HomsQuotientCategory}.

In principle, one could compute the fusion ring (and $F$-symbols) of the boundary fusion category using the abstract definitions above. However, in practice, once one has found the splitting and identification of the bulk lines into boundary lines, one can constrain the fusion ring by asking for consistency between bulk fusion rules and splitting, as explained in Section \ref{TopologicalCosets} \cite{Bais:2008ni, Eliens:2013epa}. This allows one to find the fusion ring exactly in many concrete cases.

\section{Fermionization and Bosonization of CFTs} \label{Fermionization}

In this appendix, we summarize a few facts about fermionization and bosonization of CFTs. We follow \cite{Thorngren:2021yso}. Recall that a CFT is said to be \textit{fermionic} if it depends on the choice of spin structure of the spacetime manifold. If not, the theory is said to be \textit{bosonic}. Let us take for instance the two-torus $T^{2}$, which has four possible choices of spin structure. If fermions are periodic along a cycle, we say we have a Ramond (R) boundary condition, and if they are antiperiodic, we say we have a Neveu-Schwarz (NS) boundary condition. We also use the notation $\mathrm{R}=+$ and $\mathrm{NS}=-$ to refer to these different choices of boundary conditions. The characters in a fermionic CFT depend on spin structure, and we define them as follows:
\begin{align}
    d^{\lambda}_{{\mathrm{NS} \text{-} \mathrm{NS}}}(\tau) &\coloneqq \mathrm{Tr}_{\mathcal{H}_{\mathrm{NS},\lambda}}\big(q^{L_{0}-c/24}\big), \\[0.1cm]
    d^{\lambda}_{{\mathrm{NS} \text{-} \mathrm{R}}}(\tau) &\coloneqq \mathrm{Tr}_{\mathcal{H}_{\mathrm{NS},\lambda}}\big((-1)^{F_{L}}q^{L_{0}-c/24}\big), \\[0.1cm]
    d^{\lambda}_{{\mathrm{R} \text{-} \mathrm{NS}}}(\tau) &\coloneqq \mathrm{Tr}_{\mathcal{H}_{\mathrm{R},\lambda}}\big(q^{L_{0}-c/24}\big), \\[0.1cm]
    d^{\lambda}_{{\mathrm{R} \text{-} \mathrm{R}}}(\tau) &\coloneqq \mathrm{Tr}_{\mathcal{H}_{\mathrm{R},\lambda}}\big((-1)^{F_{L}}q^{L_{0}-c/24}\big), 
\end{align}
where $(-1)^{F_{L}}$ is the holomorphic fermion parity operator, $q=e^{2 \pi i \tau}$ with $\tau$ the modular parameter as usual, and $\mathcal{H}_{\mathrm{NS}}$ and $\mathcal{H}_{\mathrm{R}}$ are the Hilbert spaces when we have anti-periodic and periodic boundary conditions on the circle, respectively.

As in bosonic CFTs, the torus partition function decomposes into characters labeled by the primaries of the theory:
\begin{equation}
    \mathcal{Z}_{\pm, \pm}(\tau,\overline{\tau}) = \sum_{\lambda, \overline{\lambda}} \mathcal{M}^{\pm}_{\lambda\overline{\lambda}} \, d_{\pm\pm}^{\lambda}(\tau) \, \overline{d}_{\pm\pm}^{\overline{\lambda}}(\overline{\tau})
\end{equation}
Notice that, unlike bosonic CFTs, the torus partition function now depends on spin structure, and it is not modular invariant but rather modular covariant. The rules for how the different spin structures are exchanged under modular transformations in a fermionic CFTs are easy to find keeping track of the boundary conditions for fermionic fields (the explicit exchanges may be found in \cite{Delmastro:2021otj}). As in the bosonic case, the mass matrix $\mathcal{M}^{\pm}_{\lambda\overline{\lambda}}$ is not arbitrary but is constrained to make the torus partition function modular covariant.

Recall that given a bosonic theory $B$ with a $\mathbb{Z}_{2}$ symmetry defined on a compact surface $\Sigma$ with genus $g$, it is possible to fermionize it and obtain a fermionic CFT with partition function
\begin{equation} \label{fermionization1}
    \mathcal{Z}_{F}(\rho) = \frac{1}{2^{g}}\sum_{\alpha \in H^{1}(\Sigma, \mathbb{Z}_{2})}(-1)^{\mathrm{Arf}[\alpha + \rho]} \mathcal{Z}_{B}(\alpha),
\end{equation}
where $\rho$ denotes a choice of spin structure, $\alpha$ denotes the $\mathbb{Z}_{2}$ gauge field for the $\mathbb{Z}_{2}$ symmetry in $B$ that we use to fermionize the theory, and $\mathrm{Arf}(\rho)$ denotes the Arf invariant, where $(-1)^\mathrm{Arf(\rho)} = 1$ for even spin structure and $(-1)^\mathrm{Arf(\rho)} = -1$ for odd spin structure. This map is invertible, and from a fermionic theory we can obtain a bosonic CFT by summing over spin structures:
\begin{equation} \label{bosonization1}
    \mathcal{Z}_{B}(\alpha) = \frac{1}{2^{g}} \sum_{\rho} (-1)^{\mathrm{Arf}[\alpha + \rho]} \mathcal{Z}_{F}(\rho).
\end{equation}
Indeed, it is straightforward to check that plugging \eqref{fermionization1} into \eqref{bosonization1} we obtain an identity.

\begin{figure}[t]
        \centering
        \includegraphics[scale=0.14]{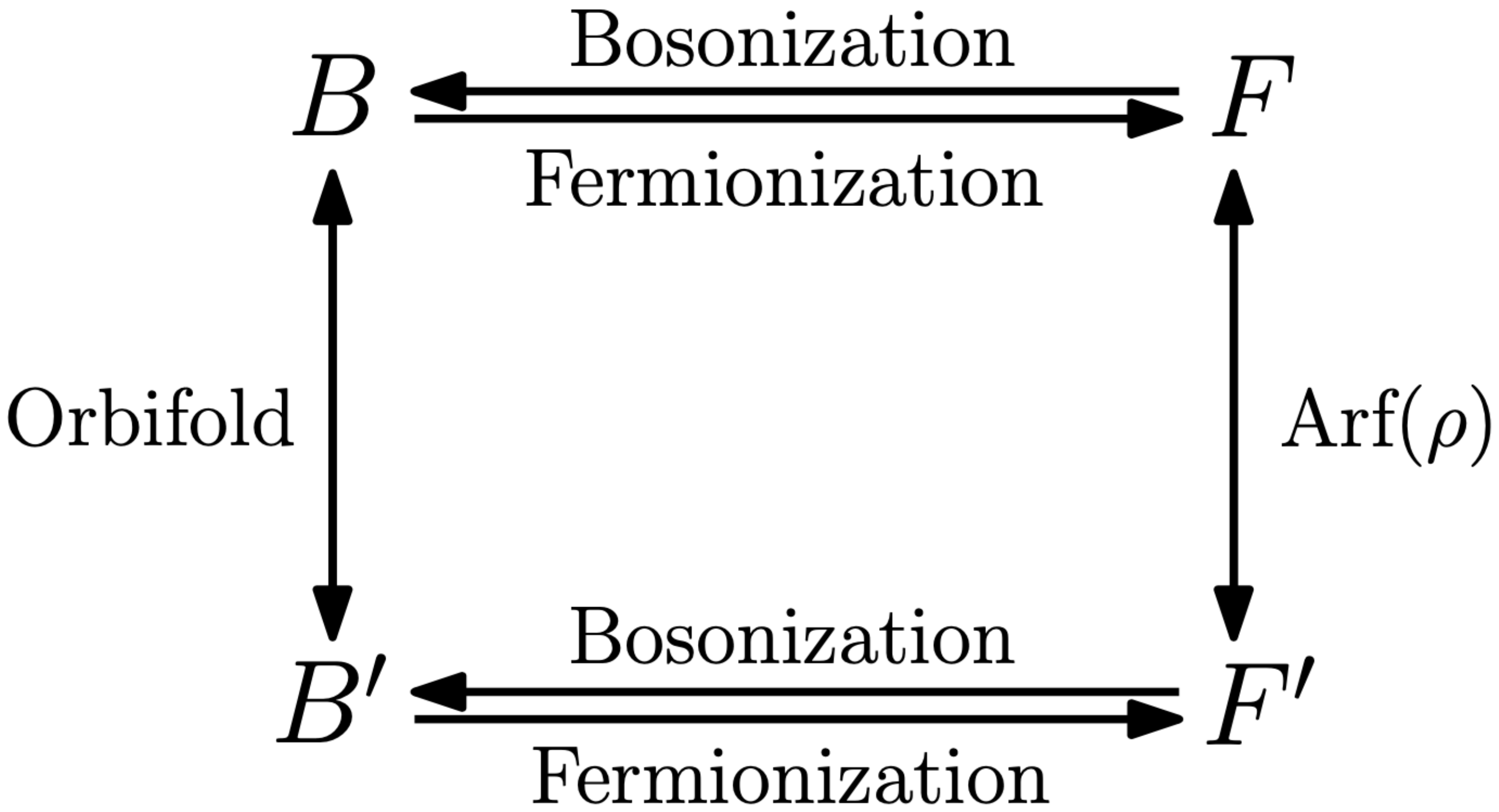} 
        \caption{Commutative diagram relating two bosonic theories $B$ and $B'$ related by a $\mathbb{Z}_{2}$ orbifold, and their corresponding fermionization differing by stacking $\mathrm{Arf}(\rho)$. The $\mathbb{Z}_{2}$ symmetry that we use to orbifold in the bosonic theories is the same $\mathbb{Z}_{2}$ symmetry used to fermionize.} \label{Bosonization-Fermionization-Map2}
\end{figure}

It is well-known that bosonization is not unique, as we can always stack a 2D Spin SPT --the Arf invariant-- and find a generically different bosonic theory with partition function:
\begin{equation} \label{bosonization2}
    \mathcal{Z}_{B'}(\alpha) = \frac{1}{2^{g}} \sum_{\rho} (-1)^{\mathrm{Arf}[\alpha + \rho] + \mathrm{Arf}[\rho]} \mathcal{Z}_{F}(\rho).
\end{equation}
Alternatively, we can think of $B'$ as the first defined bosonization \eqref{bosonization1}, but starting from a different fermionic theory which differs from the original one by stacking with the Arf invariant:
\begin{equation}
    \mathcal{Z}_{F'}(\rho) = \mathcal{Z}_{F}(\rho)(-1)^{\mathrm{Arf}[\rho]}
\end{equation}
Then, the bosonic theories $B$ and $B'$ differ by a $\mathbb{Z}_{2}$ orbifold by the same $\mathbb{Z}_{2}$ symmetry that we have used to fermionize the theories.

All these properties can be summarized in a commutative diagram, shown in Fig. \ref{Bosonization-Fermionization-Map2}.

\section{Lagrangian Algebra Multiplications for $Spin(5)_{1}/SU(2)_{10}$} \label{Spin5multiplications}

In order to verify the formalism as described in Section \ref{GappedBoundariesandTopologicalCosets}, this appendix provides a summary of the specific Lagrangian algebra multiplications for the example $Spin(5)_{1}/SU(2)_{10}$. Specifically, one may readily check that the coefficients presented in \eqref{coef1}-\eqref{coef2} are obtained via the Lagrangian algebra multiplications using Eqn. \eqref{OPEcoeffs}. The data here can be obtained directly from the MTC data provided in Appendix \ref{MTCdataAppendix}.
\begin{flalign}
    \hspace{1cm}  m_{(0,0) (a,b ) }^{(a,b)} & = m_{(a, b) (a, b) }^{(0,0)} = 1,  \ \mathrm{for} \ (a,b) \in \mathcal{A} && \\[0.3cm]
    \hspace{1cm}  m_{(0, 6) (0, 6) }^{(0, 6)} & =2^{1/4} i , && \\[0.3cm]
    \hspace{1cm}  m_{(0, 6) (v, 4) }^{(v, 4)} & = m_{(v, 4) (0, 6) }^{(v, 4)} = -2^{1/4} i , && \\[0.3cm]
    \hspace{1cm}  m_{(0, 6) (v, 4) }^{(v, 10)} & = - m_{(v, 4) (0, 6) }^{(v, 10)} = - i , && \\[0.3cm]
    \hspace{1cm}  m_{(0, 6) (v, 10) }^{(v, 4)} & = - m_{(v, 10) (0, 6) }^{(v, 4)} =  i , && \\[0.3cm]
    \hspace{1cm}  m_{(0, 6) (\sigma, 3) }^{(\sigma, 3)} & = m_{(\sigma, 3) (0, 6) }^{(\sigma, 3)} =  i/2^{1/4} , && \\[0.3cm]
    \hspace{1cm}  m_{(0, 6) (\sigma, 3) }^{(\sigma, 7)} & = - m_{(\sigma, 3) (0, 6) }^{(\sigma, 7)} = ( 3/2 )^{1/4} , && \\[0.3cm]
    \hspace{1cm}  m_{(\sigma, 7) (0, 6) }^{(\sigma, 3)} & = - m_{(0, 6) (\sigma, 7) }^{(\sigma, 3)} = ( 3/2 )^{1/4} , && \\[0.3cm]
    \hspace{1cm}  m_{(0, 6) (\sigma, 7) }^{(\sigma, 7)} & = m_{(\sigma,7) (0, 6) }^{(\sigma,7)} = - i/2^{1/4} , && \\[0.3cm]
    \hspace{1cm}  m_{(v, 10) (\sigma, 3) }^{(\sigma, 7)} & = m_{(\sigma, 3) (v, 10) }^{(\sigma, 7)} =  i , && \\[0.3cm]
    \hspace{1cm}  m_{(v, 10) (\sigma, 7) }^{(\sigma, 3)} & = m_{(\sigma, 7) (v, 10) }^{(\sigma, 3)} =  i , && \\[0.3cm]
    \hspace{1cm}  m_{(v, 4) (v, 4) }^{(0, 6)} & = - 2^{1/4} i , && \\[0.3cm]
    \hspace{1cm}  m_{(v, 4) (v, 10) }^{(0, 6)} & = - m_{(v, 10) (v, 4) }^{(0, 6)} =  -  i , && \\[0.3cm]
    \hspace{1cm}  m_{(v, 4) (\sigma, 3) }^{(\sigma, 3)} & = m_{(\sigma, 3) (v, 4) }^{(\sigma, 3)} = ( 3/2 )^{1/4}, && \\[0.3cm]
    \hspace{1cm}  m_{(v, 4) (\sigma, 3) }^{(\sigma, 7)} & = - m_{(\sigma, 3) (v, 4) }^{(\sigma, 7)} = i/2^{1/4}, && \\[0.3cm]
    \hspace{1cm}  m_{(v, 4) (\sigma, 7) }^{(\sigma, 3)} & = - m_{(\sigma, 7) (v, 4) }^{(\sigma, 3)} = - i/2^{1/4}, && \\[0.3cm]
    \hspace{1cm}  m_{(v, 4) (\sigma, 7) }^{(\sigma, 7)} & = m_{(\sigma, 7) (v, 4) }^{(\sigma, 7)} = - ( 3/2 )^{1/4}, && \\[0.3cm]
    \hspace{1cm}  m_{(\sigma, 3) (\sigma, 3) }^{(0, 6)} & = i/2^{1/4}, && \\[0.3cm]
    \hspace{1cm}  m_{(\sigma, 7) (\sigma, 7) }^{(0, 6)} & = - i/2^{1/4} , && \\[0.3cm]
    \hspace{1cm}  m_{(\sigma, 3) (\sigma, 3) }^{(4, v)} & = + ( 3/2 )^{1/4} , && \\[0.3cm]
    \hspace{1cm}  m_{(\sigma, 7) (\sigma, 7) }^{(4, v)} & = - ( 3/2)^{1/4}, && \\[0.3cm]
    \hspace{1cm}  m_{(\sigma, 3) (\sigma, 7) }^{(v, 10)} & = m_{(\sigma, 7) (\sigma, 3) }^{(v, 10)} =  i , && \\[0.3cm]
    \hspace{1cm}  m_{(\sigma, 3) (\sigma, 7) }^{(v, 4)} & = - m_{(\sigma, 7) (\sigma, 3) }^{(v, 4)} = + i/2^{1/4} , && \\[0.3cm]
    \hspace{1cm}  m_{(\sigma, 3) (\sigma, 7) }^{(0, 6)} & = - m_{(\sigma, 7) (\sigma, 3) }^{(0, 6)} = + ( 3/2 )^{1/4} , &&
\end{flalign}

\section{Explicit MTC Data} \label{MTCdataAppendix}

For self-containment, in this appendix we summarize the MTC data of $Spin(\nu)_{1}$ for $\nu$ odd and $SU(2)_{k}$ for integer $k$, which are cases of interest in this work.

\subsection{Fibonacci}

This MTC has two simple lines called $0$ and $\tau$ with topological spins $\theta_{0} = 1$, $\theta_{\tau} = e^{is \frac{4\pi}{5}}$ and quantum dimensions $d_{0}=1$, $d_{\tau}=\phi$. The only non-trivial fusion rule is
\begin{equation}
    \tau \times \tau = 0 + \tau.
\end{equation}
The non-trivial $F$-symbols are given by
\begin{equation}
    [F^{\tau \tau \tau}_{\tau}]_{ef} = \begin{bmatrix}
\phi^{-1} & \phi^{-1/2} \\
\phi^{-1/2} & -\phi^{-1}
\end{bmatrix}_{ef},
\end{equation}
while the non-trivial $R$-symbols are given by
\begin{equation}
    R^{\tau \tau}_{0} = e^{-is \frac{4\pi}{5}}, \quad R^{\tau \tau}_{\tau} = e^{i s \frac{3 \pi}{5}}.
\end{equation}

Above, $\phi$ is the golden ratio and $s = \pm 1$ corresponds to the chirality of the two MTCs with this data and central charges $c_{-} = 14 s / 5 \ \mathrm{mod} \ 8$. For instance, the $(G_{2})_{1}$ and $(F_{4})_{1}$ Chern-Simons theories are described by this MTC data with $s=+1$ and $s=-1$ respectively.

\subsection{$Spin(\nu)_{1}$}

We take $\nu$ an odd integer. In this case the MTC has three simple lines labeled $0$, $v$ and $\sigma$ with fusion rules
\begin{equation}
    v \times v = 0, \quad \sigma \times v = \sigma, \quad \sigma \times \sigma = 0 + v.
\end{equation}
The topological spins are $\theta_{0} = 1$, $\theta_{v} = -1$, $\theta_{\sigma} = e^{2 \pi i \nu/16}$, and the modular S-matrix is that of Ising:
\begin{equation}
    S = \frac{1}{2}\begin{pmatrix}
1 & 1 & \sqrt{2} \\
1 & 1 & -\sqrt{2} \\
\sqrt{2} & -\sqrt{2} & 0
\end{pmatrix}.
\end{equation}
The quantum dimensions are $d_{0} = d_{v} = 1$, $d_{\sigma} = \sqrt{2}$. Meanwhile, the non-trivial $F$-symbols are given by
\begin{equation}
    F^{v \sigma v}_{\sigma} = F^{\sigma v  \sigma}_{v} = -1, \quad [F^{\sigma \sigma \sigma}_{\sigma}]_{ef} = \frac{\kappa_{\sigma}}{\sqrt{2}} \begin{bmatrix}
1 & \ 1 \\
1 & -1
\end{bmatrix}_{ef},
\end{equation}
where $\kappa_{\sigma}$ is the Frobenius-Schur indicator $\kappa_{\sigma} = (-1)^{(\nu^{2}-1)/8}$, while the $R$-symbols are
\begin{equation}
    R^{v v}_{0} = -1, \quad R^{v \sigma}_{\sigma} = R^{\sigma v}_{\sigma} = (-i)^{\nu}, \quad R^{\sigma \sigma}_{0} = \kappa_{\sigma} e^{-i \frac{\pi}{8}\nu}, \quad R^{\sigma \sigma}_{v}=\kappa_{\sigma} e^{ i \frac{3\pi}{8}\nu}.
\end{equation}

\subsection{$SU(2)_{k}$}

The MTC $SU(2)_{k}$ with integer $k$ consists of $k+1$ simple lines labeled from $0$ to $k$, with fusion rules
\begin{equation}
    \Lambda_{1} \times \Lambda_{2} = \sum_{\Lambda = |\Lambda_{1}-\Lambda_{2}|}^{\mathrm{min}(\Lambda_{1}+\Lambda_{2},2k-\Lambda_{1}-\Lambda_{2})}  \Lambda \ , \label{SU2kFusionRules}
\end{equation}
where the sum is restricted such that $\Lambda_{1} + \Lambda_{2} - \Lambda$ is even. 

The topological spins are given by $\theta_{j} = \exp(2 \pi i \frac{j(j+2)}{4(k+2)})$ with $j = 0, 1, \ldots, k$, while the modular $S$-matrix is:
\begin{equation}
    S_{j_{1} j_{2}} = \sqrt{\frac{2}{k+2}}\sin{\bigg( \frac{\pi(j_{1} + 1)(j_{2} + 1)}{(k+2)} \bigg)}, \quad j_{1},j_{2} = 0, 1, \ldots, k,
\end{equation}
from which we obtain the quantum dimensions $d_{j} = \sin{\big( \frac{(j+1)\pi}{k+2} \big)}/\sin{\big( \frac{\pi}{k+2} \big)}$.

In the following $q = e^{2 \pi i/(k+2)}$. Then, the $F$-symbols have entries
\begin{equation}
    [F^{abc}_{d}]_{ef} = i^{a+b+c+d} \sqrt{[e+1]_{q} [f+1]_{q}} \begin{Bmatrix}
a & \ \ b & \ \ e \\
c & \ \ d & \ \ f
\end{Bmatrix},
\end{equation}
where
\begin{align}
& \begin{Bmatrix}
a & \ \ b & \ \ e \\
c & \ \ d & \ \ f
\end{Bmatrix} = \Delta(a,b,e) \Delta(e,c,d) \Delta(b,c,f) \Delta(a,f,d) \nonumber  \\[0.3cm] & \times \sum_{z} \bigg[ \frac{(-1)^{z} [z+1]_{q}!}{ [z - \frac{a+b+e}{2}]_{q}! [z - \frac{e+c+d}{2}]_{q}! [z - \frac{b+c+f}{2}]_{q}!} \frac{1}{[z - \frac{a+f+d}{2}]_{q}! [\frac{a+b+c+d}{2} - z]_{q}!} \nonumber \\[0.3cm] & \hspace{8cm} \times \frac{1}{[\frac{a+e+c+f}{2} - z]_{q}! [\frac{b+e+d+f}{2} - z]_{q}!} \bigg]
\end{align}
with
\begin{equation}
    \Delta(a,b,c) = \sqrt{\frac{[\frac{-a + b + c}{2}]_{q}! [\frac{a - b + c}{2}]_{q}! [\frac{a + b - c}{2}]_{q}!}{[\frac{a + b + c}{2} + 1]_{q}!}}
\end{equation}
and
\begin{equation}
    [n]_{q}! = \prod_{m=1}^{n}[m]_{q}, \quad [n]_{q} = \frac{q^{n/2} - q^{-n/2}}{q^{1/2} - q^{-1/2}}.
\end{equation}
Finally, the $R$-symbols are given by:
\begin{equation}
    R^{ab}_{c} = i^{c-a-b} q^{\frac{1}{8}[c(c+2) - a(a+2) - b(b+2)]}.
\end{equation}

\bibliographystyle{JHEP}
\bibliography{references}

\end{document}